\documentclass[%
 aip,jap,
 amsmath,amssymb,
 reprint,%
]{revtex4-1}
\usepackage{graphicx}
\usepackage{dcolumn}
\usepackage{bm}
\usepackage[utf8]{inputenc}
\usepackage[T1]{fontenc}
\usepackage{mathptmx}
\usepackage[load=prefixed]{siunitx}
\usepackage{multirow}

\usepackage[utf8]{inputenc}
\usepackage[T1]{fontenc}
\usepackage{mathptmx}

\usepackage{color}
\newcommand{\un}[1]{\ensuremath{\ \mathrm{#1}}} 

\begin{document}
\preprint{AIP/123-QED}

\title{A next-generation inverse-geometry spallation-driven ultracold neutron source}

\author{K.K.H.~Leung}
\email{kkleung@ncsu.edu}
\affiliation{Department of Physics, North Carolina State University, Raleigh, NC 27695, USA}
\affiliation{Triangle Universities Nuclear Laboratory, Durham, NC 27708, USA}
\author{G.~Muhrer}
\email{Gunter.Muhrer@esss.se}
\affiliation{European Spallation Source, Lund 22592, Sweden}
\affiliation{Los Alamos National Laboratory, Los Alamos, NM 87545, USA}
\author{T.~H{\"u}gle} 
\affiliation{Los Alamos National Laboratory, Los Alamos, NM 87545, USA}
\affiliation{Oak Ridge National Laboratory, Oak Ridge, TN 37830, USA}
\author{T.M.~Ito} 
\affiliation{Los Alamos National Laboratory, Los Alamos, NM 87545, USA}
\author{E.M.~Lutz}
\affiliation{Department of Physics, North Carolina State University, Raleigh, NC 27695, USA}
\affiliation{Triangle Universities Nuclear Laboratory, Durham, NC 27708, USA}
\author{M.~Makela} 
\affiliation{Los Alamos National Laboratory, Los Alamos, NM 87545, USA}
\author{C.L.~Morris}
\affiliation{Los Alamos National Laboratory, Los Alamos, NM 87545, USA}
\author{R.W.~Pattie, Jr.} 
\affiliation{Los Alamos National Laboratory, Los Alamos, NM 87545, USA}
\affiliation{Department of Physics, East Tennessee State University, Johnson City, TN 37614, USA}
\author{A.~Saunders} 
\affiliation{Los Alamos National Laboratory, Los Alamos, NM 87545, USA}
\author{A.R.~Young}
\email{aryoung@ncsu.edu}
\affiliation{Department of Physics, North Carolina State University, Raleigh, NC 27695, USA}
\affiliation{Triangle Universities Nuclear Laboratory, Durham, NC 27708, USA}

\date{\today}
             
\begin{abstract}
The physics model of a next-generation spallation-driven high-current ultracold neutron (UCN) source capable of delivering an extracted UCN rate of around an-order-of-magnitude higher than the strongest proposed sources, and around three-orders-of-magnitude higher than existing sources, is presented. This UCN-current-optimized source would dramatically improve cutting-edge UCN measurements that are currently statistically limited. A novel ``Inverse Geometry'' design is used with 40 liters of superfluid $^4$He (He-II), which acts as the converter of cold neutrons (CNs) to UCNs, cooled with state-of-the-art sub-cooled cryogenic technology to $\sim 1.6\un{K}$. Our source design is optimized for a 100~W maximum heat load constraint on the He-II and its vessel. In this paper, we first explore modifying the Lujan Center Mark-3 target for UCN production as a benchmark. In our Inverse Geometry, the spallation target is wrapped symmetrically around the cryogenic UCN converter to permit raster scanning the proton beam over a relatively large volume of tungsten spallation target to reduce the demand on the cooling requirements, which makes it reasonable to assume that water edge-cooling only is sufficient. Our design is refined in several steps to reach $P_{\rm UCN} = 2.1\times 10^{9}\un{s^{-1}}$ under our other restriction of 1\un{MW} maximum available proton beam power. We then study effects of the He-II scattering kernel used as well as reductions in $P_{\rm UCN}$ due to pressurization to reach $P_{\rm UCN} = 1.8\times 10^{9}\un{s^{-1}}$. Finally, we provide a design for the UCN extraction system that takes into account the required He-II heat transport properties and implementation of a He-II containment foil that allows UCN transmission. We estimate a total useful UCN current from our source of $R_{\rm use} \approx 5 \times 10^{8}\un{s^{-1}}$ from a 18\un{cm} diameter guide $\sim 5\un{m}$ from the source. Under a conservative ``no return'' (or ``single passage'') approximation, this rate can produce an extracted density of $> 1\times 10^4\un{UCN\,cm^{-3}}$ in $< 1000\un{L}$ external experimental volumes with a $^{58}$Ni (335\un{neV}) cut-off potential.
\end{abstract}

\maketitle

\section{Introduction}

Ultracold neutrons (UCNs) are an important tool in experiments with impact in nuclear physics, particle physics, astrophysics, cosmology,\cite{Nico2005,Abele2008a,Dubbers2011} as well as condensed matter physics.\cite{Golub1996} The kinetic energy of UCNs are less than the neutron optical potential of materials $U_{\rm opt}$ so that they can reflect at all incident angles via the nuclear strong force. $U_{\rm opt}$ is determined by the coherent neutron scattering length of nuclei making up the material and their number density. Materials commonly used for UCN reflection include: $^{58}$Ni (335\un{neV}), Be (250\un{neV}), deuterated plastics ($\sim 160 - 200\un{neV}$), or fluoropolymers ($\sim 100 - 130\un{neV}$). This kinetic energy range ($\lesssim 335\un{neV}$) corresponds to neutron wavelengths $> 490\un{\text{\AA}}$ or speeds $<8\un{m\,s^{-1}}$. This allows UCNs to be stored in an experimental volume for times limited by their $\beta$-decay lifetime of $\sim 15\un{min}$, a weak-interaction process. The UCN storage times typically achieved are shorter than this due to a loss probability per material reflection of $10^{-3} - 10^{-5}$ caused by absorption or upscattering. The latter is caused by scattering off phonon excitations in a material, which transfers kinetic energy to a UCN so that its final energy is above the storable range. The size of the neutron's magnetic moment ($\mu_{\rm n}  = - 60\un{neV\,T^{-1}}$) allows UCNs to be polarized or spin-analyzed by passage through a region with a magnetic field of several Tesla, e.g. achievable with an electromagnetic coil or a ferromagnetic foil. This effect also allows polarized UCNs to be reflected by ``magnetic walls'' to avoid material loss. Furthermore, the neutron's mass in Earth's gravitational field generates a potential gradient of $102\un{neV\,m^{-1}}$, which can be used favorably in an experiment.

UCNs allow sensitive studies over hundreds of seconds on slow-moving, electrically-neutral, spin-polarizable, free nucleons that participate in all four fundamental forces. Examples of experiments that benefit from the unique properties of UCNs include: high-precision measurements of the neutron lifetime and $\beta$-decay correlation parameters,\cite{Wietfeldt2011,Nico2009,Young2014b} searches for a beyond standard model neutron permanent electric dipole moment,\cite{Lamoreaux2009,Chupp2019} hunts for new short range forces,\cite{Nesvizhevsky2003, Baessler2007,Jenke2011,Jenke2014} and tests of Lorentz violation.\cite{Altarev2009} A limiting factor in all these experiments is the limited rate or densities of UCNs available. A dramatic increase would improve these measurements as well as create new opportunities for next-generation experiments. 

A range of UCN sources are currently in operation or in development world-wide. They use either solid deuterium (sD$_2$) or superfluid $^4$He (He-II) as the cold neutron (CN) to UCN converter material (except for Ref.~\onlinecite{Steyerl1986}). CNs can downscatter to UCN energies by scattering off phonon excitations in these cryogenic materials.\cite{Golub1977} UCN sources are located either ``in-pile'' (i.e. close to a reactor core or spallation target) or at the end of a CN guide. sD$_2$ has a larger UCN conversion cross-section than He-II and also converts CNs from higher energies where the incident flux is also higher. Furthermore, from a cryogenic perspective, sD$_2$ sources can be operated $\sim 4\,{\rm K}$, where ample cooling power is available to ensure the thermal properties of sD$_2$ are compatible with the requirements of heat removal for practical sources. However, after UCNs are created in the sD$_2$ they become lost quickly\cite{Morris2002} ($\sim 30-60\un{ms}$) via upscattering or nuclear absorption. The loss mean free path for UCN in sD$_2$ can be as small as $1-2\un{cm}$, reducing the UCN extraction efficiency from thick sD$_2$ converters. sD$_2$ sources have been constructed at Los Alamos National Laboratory (LANL),\cite{Saunders2013,Ito2018} at the Paul Scherrer Institute,\cite{Anghel2009,Becker2015} and at the University of Mainz.\cite{Frei2007,Kahlenberg2017} There are also several others under construction at North Carolina State University\cite{Korobkina2014} and at the Technical University of Munich.\cite{Trinks2000a,Frei2012} The latter, which is mostly constructed, is the most ambitious of these sD$_2$ sources. It has a predicted useful UCN current or rate (i.e. the integrated UCN flux) of $R_{\rm use} \approx 5\times 10^7 \un{s^{-1}}$ and a density in the source delivery system of $\sim 5,000\un{cm^{-3}}$. In this paper, we denote the total produced UCN rate by $P_{\rm UCN}$ and the total useful UCN current rate out of the source by $R_{\rm use}$.

He-II offers much higher thermal conductivities and a UCN upscattering time constant given by $\tau_{\rm up} \approx (100 \un{s\,K^{7}}) / T^7$ (see Sec.~\ref{sec:UCNextraction}), resulting in UCN loss times $\gtrsim 3\un{s}$ for a He-II bath at a temperature $T < 1.6\un{K}$. At these temperatures, the total scattering mean-free-path for $8.9\un{\text{\AA}}$ (1\un{meV}) neutrons, the primary UCN producing CN component, is $\approx 18\un{m}$.\cite{Sommers1955} This permits the design of a simple, large volume source where the converter material also serves as the coolant. 

There are two ``in-pile'' He-II sources currently under construction. One is at TRIUMF, Canada, where the source is coupled to a 500\un{MeV} proton beam with 20\un{kW} incident beam power.\cite{Masuda2012,Martin2013,Masuda2014,Picker2017,Ahmed2019} The next-generation version of this source will have an expected $R_{\rm use} \sim 10^{7}\un{s^{-1}}$ and UCN density deliverable to an experimental volume of $\sim (0.7-6) \times 10^3 \un{cm^{-3}}$ (both polarized). The other He-II UCN source is situated in the thermal column of the WWNR reactor at Petersburg Nuclear Physics Institute with a predicted $R_{\rm use} = (6-8)\times 10^{7}\un{s^{-1}}$ and a UCN density delivered to an experiment of $\sim 13,000\,{\rm cm^{-3}}$.\cite{Serebrov2009b,Serebrov2010,Serebrov2011,Serebrov2015a,Serebrov2018} The former will use evaporative $^3$He refrigeration to cool the He-II converter to 0.8\un{K} via a heat exchanger, while the latter will directly evaporate the He-II converter to reach 1.2\un{K} (corresponding to 0.8\un{mbar} vapor pressure). There are also He-II sources located at the end of CN beams to profit from the long $8.9{\;\rm\text{\AA}}$ neutron mean-free-path and reduced heat loads, which allows temperatures of $\sim 0.5\,{\rm K}$ to be reached. Using a valve on the He-II converter volume, these sources are optimized for accumulating high UCN densities inside the source. These sources are in operation,\cite{Zimmer2007a,Piegsa2014,Leung2016} with UCN densities of $\sim 120\,{\rm cm^{-3}}$ observed inside $\sim 4\,{\rm L}$ volumes of He-II after direct extraction to a detector.\cite{Zimmer2018} Future versions under construction have expected UCN densities of $\sim 2000\,{\rm cm^{-3}}$ in larger ($\sim 20\un{L}$) volumes.\cite{Zimmer2015}


In the present work, we develop the physics model for a next-generation He-II UCN source located at a spallation target optimized for high UCN current. The geometry, which was briefly outlined in Ref.~\onlinecite{Young2014}, is based on using a 40~L cylindrical volume of He-II for UCN production. Our design philosophy is to optimize the UCN production rate with the following constraints: (1) allow up to time-averaged 1~MW spallation proton beam power with a fixed $800\un{MeV}$ proton energy, (2) use of state-of-the-art sub-cooled He-II technology to providing a cooling power of $100\un{W}$ at $T\approx 1-2\un{K}$, and (3) operate with edge-cooling of a tungsten spallation target with water. 

The design constraint of only 100~W heat load in the He-II converter is a critical departure from the figure-of-merit for a typical CN source, where more than an order of magnitude more cooling power is available for a cold moderator operated near 20\un{K}. A CN source design hinges on coupling the cold moderator (typically liquid LH$_2$ or LD$_2$) as closely as possible to the spallation target. This can be seen, for example, in the Lujan Center's Mark-3 CN source at LANSCE of LANL\cite{Mocko2013}, which combined a close-coupled target and pre-moderator with 20\un{K} Be reflectors to optimize the CN flux. For lower temperature sources, such as very-cold neutron or UCN sources, one must manage the heat loads, thus resulting in different design strategies.\cite{Carpenter2005,Micklich2007} In evaluating the effectiveness of our conceptual UCN source design, we begin with a Lujan Mark-3-inspired design to make contact with a conventional CN source (see Sec.~\ref{sec:lujan}) before investigating our specific source geometry.

In order to achieve the goal of edge water cooling for the high proton beam power, the beam is rastered so the heating is distributed. This leads to a design of having the pre-moderator, moderator, and spallation target wrapped axial-symmetrically around the He-II UCN converter in order to have symmetric heat loads and UCN production as the proton beam is scanned. This symmetry leads to a quasi-constant CN flux into the He-II from different azimuthal-angle directions. By distributing the heat load in the spallation target, as pointed out in Ref.~\onlinecite{Carpenter2005}, the design constraints will be limited by the heating on the cryogenic components. In our design, the heat load on the He-II converter and its vessel is dominated by fast neutrons. To have maximal UCN production, the CN flux spectrum in the He-II needs to have strong overlap with the UCN production function in He-II, which has a large peak for 1\un{meV} ($\sim 10~{\rm K}$ or $8.9\,\text{\AA}$) CNs.

The choice of $\sim 1.6\un{K}$ for the He-II converter, warmer than other He-II UCN source designs, is a result of the large cooling powers offered by newly-developed sub-cooled He-II cooling systems. These systems are made possible by technology that allows efficient compression and heat exchange of circulating helium coolant at cryogenic temperatures. Systems have been developed for the Large Hadron Collider's magnet systems at CERN,\cite{LeBrun1994, Lebrun2014} and are now common place at facilities like Thomas Jefferson National Accelerator Facility and the Spallation Neutron Source. By using a series of five cold compressors, a nominal 100\un{W} of cooling power can be provided at $\sim 1.6\un{K}$.

To reduce UCN absorption loss, the $^3$He concentration in the converter He-II must be restricted to $\lesssim 10^{-9}$, a level that is routinely achieved with superleak technology. Modest pressurization of the He-II converter to $\sim 1\un{bar}$ will be used to suppress formation of localized helium gas bubbles that can upscatter UCNs,\cite{Gudkov2007} as well as improve heat removal. Pressurization can be achieved with a capillary connected to a pressure-regulated helium gas bottle, which is routinely done in cryogenic experiments. The $1\un{bar}$ pressure does not impose any additional engineering safety constraints as most cryogenic vessels will need to be pressure-rated for several times higher pressure. Our total He-II volume of $\sim 90\un{L}$ is smaller than or comparable to typical volumes routinely used for superconducting cavities and superconducting magnets.

The removal of the 100\un{W} of heating from the static 40\un{L} He-II converter requires a large diameter (chosen to be $18\un{cm}$) He-II filled conduit around $2\un{m}$ long (adding a further $\sim 50\un{L}$ to the total He-II volume). A part of this conduit will be used for transporting UCNs out of the converter volume also. The conductive heat transport in He-II for our heat flux ($\gtrsim 1\un{kW\,m^{-2}}$) falls in the Gorter-Mellink regime, which is well-studied experimentally.\cite{Lebrun2014,Van-Sciver2012} The 100\un{W} heating is then removed by a flowing He-II line from a sub-cooled He-II refrigeration system.

Heat exchange with the flowing He-II line will need to be efficient. Kapitza resistance usually dominates at temperatures below 2\un{K}, which has a theoretical $T^{-3}$ dependence. For a typical ``dirty surface'' at 1.6\un{K} the thermal conductance is $\sim 1\un{kW\,m^{-2}\,K^{-1}}$ for a copper to He-II interface.\cite{Lebrun2014} Thus, 5\un{m^2} of area is needed to keep the temperature drop $< 10\un{mK}$. This is routinely achieved with sintered-copper heat exchangers, where typical areas per mass of Cu of $\sim 1\un{m^2\,g^{-1}}$ are attainable.\cite{Pobell1996} An important design feature for He-II thermal conductivity is maintaining large cross-section areas in the conduit leading to the heat exchanger, avoiding any narrow apertures (see Eq.~\ref{eq:conductivityConduit}). Since the large surface areas and materials used for heat exchangers are lossy to UCNs, a high heat throughput system will result in UCN loss. The efficiency of the UCN extraction system will need to take this loss into account, which is considered in our design.

UCN source designs can fall on a spectrum between being optimized for high UCN-current or high UCN-density. This is determined by the lifetime of UCNs in the source and the UCN extraction system.\cite{Young2014b} The short $\tau_{\rm up}$ at our chosen He-II temperature, which permits us to tolerate higher He-II heat loads, pushes our design towards being current-optimized. The UCN density extractable to an external experimental volume at a current-optimized source is less sensitive to losses in the UCN transport system compared to a density-optimized source. This is because UCNs only make very few passages between source and volume to reach equilibrium density. Indeed, when estimating our achievable densities we use a conservative ``no return'' approximation (or ``single passage'' approximation), which assumes UCN only make a single passage from source to external volume. This reduced sensitivity is particularly advantageous for cryogenic UCN transport systems since special care is required to avoid build-up of frozen contamination on UCN reflecting components. The UCN loss rates in the UCN transport system will also be typically much smaller than the rapid loss of UCNs at our He-II temperature. Current-optimized sources are also particularly advantageous for UCN experiments requiring large volumes or a high time-averaged UCN throughput.\cite{Zimmer2014}


We will use a 800~MeV proton beam for our analysis. This is similar to the high-current medium-energy accelerator at LANSCE.\cite{Lisowski2006} This will permit us to compare our expected CN flux to LANSCE's Lujan-Center Mark-III target, for which a detailed performance assessment exists. The LANSCE accelerator has recently been refurbished and is capable of delivering up to 1~MW of proton beam power at 800~MeV.\cite{Clendenin1996,Sinnis2019} Our upper limit of~1 MW is thus consistent with available accelerator technology.\cite{Bauer2001} There are also other powerful proton beams available for spallation targets world-wide including: 1.4~MW at the Spallation Neutron Source (with a planned 2.8~MW upgrade \cite{Howell2017}), 5~MW at the European Spallation Source, 1~MW at the Japan Proton Accelerator Research Complex, and 1.3~MW at the Paul Scherrer Institute.

For the 1\un{MW} proton power and 100\un{W} He-II heat load design restrictions, in this paper we describe optimizations with individual improvements of $\gtrsim 10\%$. This allowed us to reach $P_{\rm UCN} = 2.1 \times 10^{9}\un{s^{-1}}$. We then further studied subtle effects due to the He-II temperature, pressure, and its quantum fluid properties that reduce this to $P_{\rm UCN} = 1.8 \times 10^{9}\un{s^{-1}}$. In the final section, we perform UCN extraction simulations to give an estimate for $R_{\rm use} \approx 5\times 10^{8}\un{s^{-1}}$ from a 18\un{cm} diameter UCN guide at a position of $\sim 5\un{m}$ from our source. These UCN rate numbers assume a $^{58}$Ni coating in the UCN transport system, which has a potential $U_{\rm ^{58}Ni} = 335\un{neV}$.

\begin{table}
\caption{\label{tab:sourceDensity} The filling of an external spherical UCN bottle with an opening diameter $D_{\rm open} = 18\un{cm}$ matched to the UCN extraction guide diameter. The UCN rate entering the volume is $R_{\rm use} = 5 \times 10^{8}\un{UCN\,s^{-1}}$, which is based on the work in this paper. The UCN rate exiting the bottle is calculated assuming kinetic theory. The ``no return'' approximation is used (see text). $V_{\rm bottle}$ is the volume of the bottle, $\rho_{\rm bottle}$ is the equilibrium UCN density, and $\tau_{\rm bottle}$ is the build-up time constant. Other parameters used are: average UCN speed $\bar{v} = 7\un{m\, s^{-1}}$ (equal to 255\un{neV} kinetic energy), average wall loss per bounce $\bar{\mu} = 5 \times 10^{-4}$, and the bottle coated with $^{58}$Ni. Details about this calculation are given in Sec.~\ref{sec:UCNdensities}.}
\begin{ruledtabular}
\begin{tabular}{ c c c c c c}
$V_{\rm bottle}$ [L]                           & 5         & 50     & 500        & $5 \times 10^3$ & $5\times 10^{4}$ \\\hline
$\rho_{\rm bottle}$ [$\times 10^4\, {\rm UCN\, cm^{-3}}$] & $1.12$ & $1.11$ & $1.05$ & $0.80$ & $0.31$ \\\hline 
$\tau_{\,\rm bottle}$ [s]                        & 0.11       & 1.1        & 10         & 80  & 315 \\
\end{tabular}
\end{ruledtabular}
\end{table}

With this useful UCN current entering an external volume (or ``bottle''), the maximum UCN density that can be achieved can be estimated. The equilibrium densities $\rho_{\rm bottle}$ for different sized bottles are shown in Table~\ref{tab:sourceDensity}. Equilibrium is reached when the UCN rate entering and exiting the bottle through the 18\un{cm} guide are equal, where the latter is estimated using kinetic theory. The equilibrium build-up time constant is denoted by $\tau_{\rm bottle}$. As mentioned earlier, the conservative ``no return'' approximation is applied for our current-optimized source, which assumes that once a UCN leaves the external bottle it is lost and has zero probability of returning again. Therefore, we only use the single-passage useful UCN current $R_{\rm use}$ for estimating the density. These densities are not specific to our source design, and are valid for any UCN filling process where this approximation is made, which is particularly relevant for current-optimized sources. The details of this calculation are described more in Sec.~\ref{sec:UCNdensities}. For scale, most neutron electric dipole moment experiments have volumes $< 20\un{L}$ and neutron lifetime experiments $< 1000\un{L}$. It should be noted that the design leading to the above UCN current and density is in a physics model stage only. When moving to an engineering design a reduction in performance is expected. 

\section{Strategy of source evaluation and optimization \label{sec:strategy}}

In this section we describe our strategy for developing and optimizing the physics model of our source design. We first optimize the CN flux for a given configuration (which includes the geometry and material choice) that produces the highest $P_{\rm UCN}$ under the 100~W He-II heat load and 1~MW proton power constraints. This is done by calculating with MCNP the track-length weighted, energy-differential CN flux in the He-II converter, as well as the heating of the He-II converter and its vessel walls. When the terms ``differential CN flux'' and ``He-II heat load'' are used in this paper, these refer to the quantities described in this paragraph (i.e. for the latter it is the combined heat load on the He-II and its vessel). 

The differential CN flux used are from the so-called track-length (``F4'') tallies from MCNP, where the distance neutrons travel in the He-II volume are summed up and then normalized by the cell volume and number of protons (i.e. its raw units are: ${\rm cm/cm^3/proton}$). The number of protons is converted to $100\un{\mu A\cdot s}$ and the differential energy flux calculated by dividing by the energy bin width of the MCNP tally (e.g. to arrive at units of: $\un{cm^{-2}\,s^{-1}\,meV^{-1}\,(100\,\mu A)^{-1}}$). The He-II heat load is also quoted for $100\un{\mu A}$ proton current. The proton energy is fixed at 800\un{MeV} in our study, therefore $100\un{\mu A}$ of proton current corresponds to 80\un{kW} of proton power. The proton powers quoted are time-averaged values.

From the He-II heat load per proton, the maximum proton power that can be used for a given configuration is calculated by scaling this up to the constraints of maximum 100\un{W} He-II heat load. This scaling assumes that the He-II heat load increases linearly with proton power. In some cases the proton power required to reach this constraint exceeds the 1\un{MW} limit. In this case, the lower maximum proton power limit is used instead. 

The differential CN flux for a configuration is scaled linearly to the maximum proton power, and from this $P_{\rm UCN}$ is calculated. The effects of the pulse structure of the proton beam, typically $> 10\un{Hz}$ at spallation neutron facilities, are assumed to be averaged over since this is faster than the time-scales for UCN transport as well as the thermal transport in our large-volume cryogenic system.

The CN-energy-differential, volumetric UCN-production-rate function per CN flux (e.g. this quantity has units: ${\rm UCN\,cm^{-5}\,s^{-2}\,meV^{-2}}$) will be referred to as the ``UCN production function''. This is calculated following Ref.~\onlinecite{Korobkina2002} using the measured dynamic structure factor from neutron time-of-flight inelastic scattering from Refs.~\onlinecite{Andersen1994a,Andersen2015} that covers $0.2\un{meV}$ to $4\un{meV}$. This technique has become the standard for calculating UCN production rates in He-II.\cite{Masuda2012,Schmidt-Wellenburg2015,Leung2016} The width of the UCN production peak from single-phonon scattering at $\sim1\un{meV}$ calculated from this data has a width of $\sim 0.4\un{meV}$ due to the resolution of the time-of-flight spectrometer used (e.g. Fig.~\ref{fig:UCNproductionTemperature}). The ``true'' single-phonon linewidth has been measured to be $\sim 20-50\un{\mu eV}$ (FWHM) at $1\un{meV}$ in our temperature range.\cite{Mezei1983} However, the use of the broadened single-phonon peak width in the UCN production function is valid so long as the CN spectrum is broad, as is the case here. The UCN production function used is for He-II at 1.5\un{K} (e.g. see Ref.~\onlinecite{Leung2016}). The UCN production function in Ref.~\onlinecite{Korobkina2002} used previously in Ref.~\onlinecite{Young2014}, which is for He-II at 1.2\un{K}, contains an error in the binning and is thus avoided here. To reach CN-energy-differential volumetric UCN production rate ${\rm d}P_{{\rm UCN}/V}/{\rm d}E_{\rm CN}$ (e.g. this quantity has units ${\rm UCN\,cm^{-3}\,s^{-1}\,meV^{-1}}$) for a given configuration, the UCN production function is folded with the differential CN flux. Integration of this differential rate over the CN energy range and multiplication by the 40\un{L} He-II volume gives $P_{\rm UCN}$ (with units: ${\rm UCN\,s^{-1}}$).

The specific strategy for optimizing the geometry of the Inverse Geometry source design is described in Sec.~\ref{sec:optimization}, after the key components in the physics model have been introduced.

The UCN production rate depends on the choice of a UCN cut-off energy $E_c$. UCNs produced with kinetic energy above $E_c$ are considered to be lost quickly in the system so that they cannot be transported or stored. When UCNs are produced in He-II, they fill phase-space with constant density. Therefore, the produced UCN spectrum is given by ${\rm d}P_{\rm UCN} \propto v^2 {\rm d}v$, which can be expressed in terms of the UCN kinetic energy $E_{\rm UCN}$ as ${\rm d}P_{\rm UCN}/{\rm d}E_{\rm UCN} \propto \sqrt{E_{\rm UCN}} \, {\rm d}E_{\rm UCN}$. The calculations in this paper assume a $^{58}$Ni coating will be used in the He-II converter volume and in the guides for UCN reflection. This is a common coating used for neutron transport and has a potential $U_{\rm ^{58}Ni}  = 335\un{neV}$. Because of the neutron potential of He-II, $U_{\textrm{He-II}} = 18.5\un{neV}$, the cut-off UCN kinetic energy that can be reflected when they are produced inside the He-II is reduced to $E_c = 316.5\un{neV}$.

UCNs gain kinetic energy when they fall in Earth's gravity, therefore the maximum UCN kinetic energy that a surface can reflect depends on the height at which a UCN is produced and where the reflection occurs. In order to calculate the behavior of UCNs for a given configuration, Monte-Carlo based UCN trajectory tracking simulations are used. These simulations take into account effects such as the gravitational acceleration, the UCN-energy-dependent loss probability upon a reflection off a surface (which depends on the incoming UCN kinetic energy and angle, as well as the loss cross-sections of the material), geometric features, and the nature of wall reflections (i.e. specular or non-specular). After the optimization for $P_{\rm UCN}$ as described, in Sec.~\ref{sec:UCNextraction} we move onto studying a physics model of the UCN transport properties of our source design and its extraction efficiency.

The UCN extraction system also needs to provide an unobstructed, large diameter, He-II-filled conduit to the cooling system's heat exchanger. If UCNs reach the heat exchanger they will have a high probability of being lost due to the large surface area and materials typically used here. Fortunately, Earth's gravity, as well as geometric features on the conduit walls, can be used to affect UCN trajectories while having minimal influence on the thermal transport. The design goal for our heat removal is a 50\un{mK} increase from the heat exchanger to the main 40\un{L} converter volume. As described in the Introduction, the low-temperature techniques at our temperature range are well-developed, therefore this temperature drop should be attainable. However, cryogenics research and development will be useful for successfully executing an engineered design.

The overall optimization strategy of the UCN transport system is described in Sec.~\ref{sec:UCNextraction}. Some features discussed are: (1) diameter and length needed for the dual purpose UCN extraction and heat removal conduit, (2) implementation of a He-II containment foil while allowing high UCN transmission, (3) losses in additional UCN guide needed to reach an external experimental bottle 5\un{m} away from the source, and (4) shielding of the UCN extraction guide. The UCN extraction system's figure-of-merit that is optimized in our design is the single-passage extraction efficiency out of the source, since we will use the ``no return'' (or ``single passage'') approximation for estimating the extractable density. The total single-passage transport efficiency $\epsilon_{\rm tot\;single}$ is then used to calculate the useful UCN current $R_{\rm use} = \epsilon_{\rm tot\;single}\, P_{\rm UCN}$. The calculation of the accumulable UCN density that can achieved in different-sized external bottles with this $R_{\rm use}$ is shown in Table~\ref{tab:sourceDensity} and is described in Sec.~\ref{sec:UCNdensities}.

We now return to the discussion of the UCN production rate $P_{\rm UCN}$. The first geometry we analyze is the Lujan Mark-3 target modified to produce UCNs (and not the Inverse Geometry). This will provide a useful reference for our subsequent discussions.

\section{Lujan center Mark-3-inspired UCN design \label{sec:lujan}}

The production of UCN in He-II is strongly peaked for incoming CNs with a wavelength of 8.9~\text{\AA} or 1\un{meV} energy due to the kinematics of single-phonon scattering. This means that in the case of a spallation-driven UCN source, the high energy neutrons, which are produced through spallation with $\sim$ MeV energies, need to be cooled significantly to be useful for UCN production. In addition, the He-II bath needs to be shielded from $\gamma$-photon and charged particle heating.

\begin{figure}[tb!!]
\begin{center}
\includegraphics[width=0.88\columnwidth]{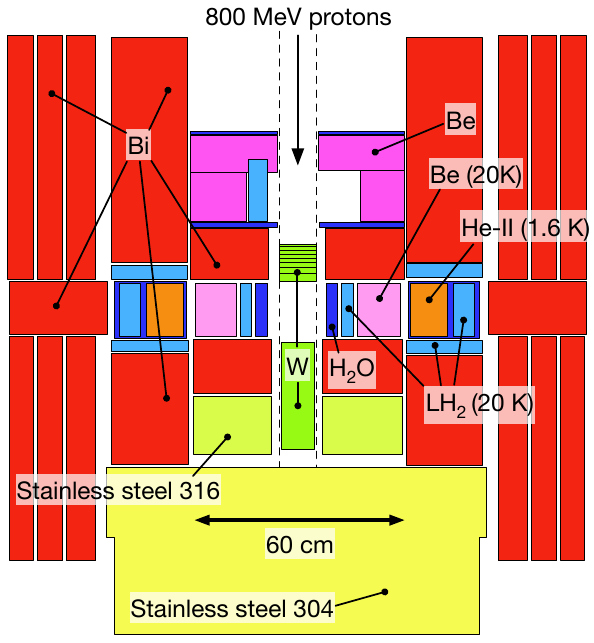}
\end{center}
\caption{The MCNP model of the Lujan Center Mark-3 target modified for UCN production by adding a 40~L annular-shaped He-II volume surrounded by LH$_2$, which we call the ``Mark-3-inspired UCN design'' in this paper (described in Sec.~\ref{sec:lujan}). The UCN extraction system from the He-II is not shown. Materials without an indicated temperature are at 300\un{K}. This design is used as a reference in our study.}
\label{fig:sourceLujan}
\end{figure}

The well-bench-marked Lujan Center Mark-3 target geometry\cite{Mocko2013} has been optimized for long-wavelength CNs using the cold beryllium reflector filter concept for small angle neutron scattering and reflectometry. Such a geometry is useful for UCN production in He-II due to the need for long-wavelength CNs. To provide a useful starting point for subsequent discussions the Lujan Center Mark-3 target is modified to produce UCNs by adding a 40~L annular volume of He-II surrounded by a liquid hydrogen (LH$_2$) moderator. Both are placed at the same height as the cold beryllium reflector. Bismuth (Bi) is used as an internal shielding component, instead of lead or iron, to reduce heat loads and to improve thermal neutron transport properties. This choice permits a straightforward comparison of the Inverse Geometry design which also utilizes Bi. We will refer to this as the ``Mark-3-inspired UCN design'', which is shown in Fig.~\ref{fig:sourceLujan}.

In Ref.~\onlinecite{Young2014}, the differential CN flux in the He-II of the Mark-3-inspired UCN design obtained using MCNPX\cite{MCNPX} calculations was shown (it is also plotted here in Fig.~\ref{fig:CNspectrumPlot}). This resulted in $P_{\rm UCN} = 9.4 \times 10^{7}\un{s^{-1}}$ for the nominal Lujan Center beam conditions of $100\un{\mu A}$ and $800\un{MeV}$ proton beam current and energy (equivalent to $80\un{kW}$ beam power). The He-II heat load was $\sim 67\un{W}$, with $\sim 67\%$ generated by neutrons (cold and fast), $\sim 28\%$ by photons and  $\sim 5\%$ by protons. Based on the assumption of 100\un{W} cooling power at $1.6\un{K}$, this calculation predicted that using a proton power of $120\un{kW}$ is possible, resulting in $P_{\rm UCN} = 1.4 \times 10^{8} \un{s^{-1}}$. This design does not utilize a full $1\un{MW}$ beam power. These results are summarized in Table~\ref{tab:OptimizeSummary}.

The Mark-3 target's neutron production efficiency has been exhaustively optimized, benchmarked  and established, providing vetted scattering kernels and a baseline for thermal and cold neutron moderator performance in its design studies reported in Secs.~\ref{sec:baselineDesign} through \ref{sec:optimization}. Next we shift our focus to the Inverse Geometry source design and explore its performance as a UCN current-optimized source. 




\section{Baseline Inverse Geometry design \label{sec:baselineDesign}}

\begin{figure*}[tb!!]
\begin{center}
\includegraphics[width=5.5in]{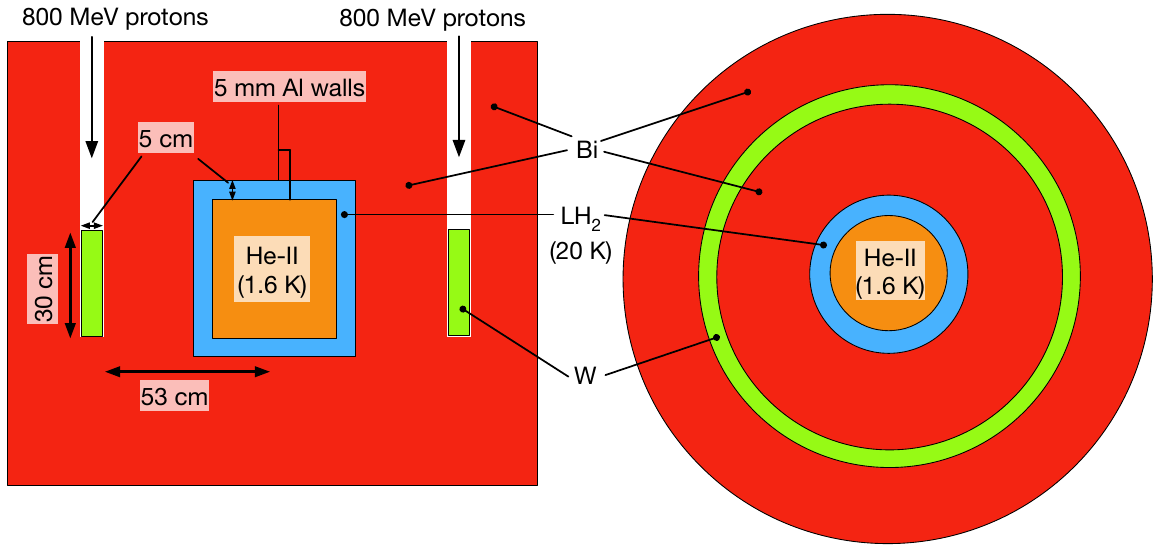}
\end{center}
\caption{MCNP model of the \emph{baseline} design of our ``Inverse Geometry'' source described in Sec.~\ref{sec:baselineDesign}. The axis of the 40\un{L} cylindrical He-II converter volume is parallel to the direction of the incoming rastered proton beam. Temperatures of the material are at room temperature unless otherwise indicated. The UCN extraction system, studied and described in Sec.~\ref{sec:UCNextraction}, is not included in this model. (\emph{Left}) Cross-section at a plane coincident with the cylindrical He-II volume's axis. (\emph{Right}) Cross-section at a plane perpendicular to the He-II volume's axis and passing through its center.}
\label{fig:sourceBaselineInverse}
\end{figure*}

Fig.~\ref{fig:sourceBaselineInverse} shows our baseline Inverse Geometry design. The spallation target is a cylindrical shell made from tungsten with an outer radius of 58\un{cm} and a wall thickness of 5\un{cm}. The target is 30\un{cm} long to the proton beam, which is longer than the 22\un{cm} stopping length for 800\un{MeV} protons in tungsten.\cite{STAR} The impinging proton beam is rastered to quasi-uniformly illuminate the front face of the target.

In the Lujan Center Mark-3 target, the front surface of the target illuminated by the proton beam is $77\un{cm^2}$, and the target can accept 160\un{kW} of proton beam heating. For the Inverse Geometry target, the front illuminated surface being considered is $1700\un{cm^2}$, with the heating being up to 1\un{MW}. As shown in Ref.~\onlinecite{Bultman1998}, the maximum volumetric heat load in the target occurs in the first tungsten plate. This means that to the first order, the volumetric heat load is proportional to the front illuminated surface area. Therefore, the spatial-averaged beam power at the front face of the Inverse Geometry target is $< 28\%$ of the Lujan Mark-3 target. Being able to adequately distributing this heat load is what determines the smallest dimensions possible for the Inverse Geometry's target. 

The rastering technique for a pulsed proton beam to quasi-evenly distribute the heating on a tungsten target will be used at the European Spallation Source (ESS).\cite{Shea2013,Shea2014,Thomsen2014} A long pulse structure is better suited for rastering. At LANSCE, this can be achieved by using the H$^+$ beam and not sending it through the proton storage ring. The LANSCE pulse will then be long enough for ESS-type rastering.\cite{Shea2019} Doing this for the LANSCE pulse would also reduce the power density deposited by a factor of two. This is because the beam profile LANSCE delivers to the Lujan target is Gaussian, which has a higher power density at the center of the beam than a flat profile.

While a more detailed analysis of the target cooling system is beyond the scope of this work, the assumption that edge cooling is possible with this target is reasonable based on these heat distribution estimates.


At the core of the model is a cylindrical 40\un{L} He-II bath with a 18\un{cm} radius and 40\un{cm} length. The He-II bath is surrounded by 5\un{cm} thick of LH$_2$ ($T=20\un{K}$) as moderator. Both are encased by 5\un{mm} wall-thickness aluminum (Al) canisters. In our baseline Inverse Geometry design, the thickness of the LH$_2$ moderator is chosen to be the same in the Mark-3-inspired UCN design. The reflector between the target and the cold source is chosen to be Bi, which is a good attenuator of $\gamma$-photons and protons, and ensures minimal neutron absorption between the target and the LH$_2$ moderator. 

In our baseline Inverse Geometry design, for a beam power of 80\un{kW} ($100\un{\mu A}$) the He-II heat load is $\sim 27\un{W}$. Therefore, we can operate the target with almost 300\un{kW} beam power to reach our 100\un{W} He-II heat load specification. For this baseline design $P_{\rm UCN} = 2.4 \times 10^8\un{s^{-1}}$ is attained, which is already 1.7 times larger than for the Mark-3-inspired UCN design. This increase is primarily due to the lower heat deposited in the He-II per incident neutron. 

The use of Bi poses additional engineering and operational requirements due to the accumulation of radio-isotopes specific to Bi at the end-of-life of the source.\cite{Carpenter2005} Although we do not address these concerns here, we hope that our design motivates engineered solutions to this problem.

One of the key ideas for our Inverse Geometry approach is based on the backscattering concept that was first proposed and implemented by Russell\cite{Russell1997}. The idea is that, in the free gas model, if a neutron scatters inelastically the energy loss increases with increasing scattering angle. To avoid ambiguity, ``inelastic scattering'' is used in this paper to refer to a loss of neutron kinetic energy in the laboratory frame. This means that the spectrum observed in the backscattering direction is colder\cite{Ino2004,Muhrer2004} and can lead to a so-called ``over-moderated spectrum''.\cite{Muhrer2012} This is advantageous due to the long $8.9\un{\text{\AA}}$ wavelength desired for UCN production in He-II. In the next section where we optimize this geometry, it will be desirable to increase the distance between the He-II volume and the tungsten target to improve shielding of fast neutrons, since these neutrons produce a He-II heat load for negligible UCN production. In this case, moving the tungsten target towards the downstream direction of the proton beam (keeping the He-II volume position fixed) decreases the apparent temperature of the CN differential flux, which is advantageous for UCN production.

\section{Inverse Geometry design modifications \label{sec:modification}}

In this section we describe key modifications of the material choice in our baseline Inverse Geometry design that will improve $P_{\rm UCN}$ further. Table~\ref{tab:OptimizeSummary} provides a summary of these incremental improvements.

\subsection{Beryllium canister}

The neutron absorption cross-section of Al is fairly low at the thermal energy (25.8\un{meV}), $\sigma_{\rm abs} = 0.23 {\,\rm b}$. However, it is well known that the absorption cross-section scales by the inverse of the neutron velocity. For our particularly long wavelength 1\un{meV} CNs, $\sigma_{\rm abs} = 1.17\un{b}$. A 5\un{mm} thick Al wall would thus cause a $\sim 30~\%$ reduction for 1\un{meV} CNs for a single-passage.

On the other hand, $\sigma_{\rm abs}$ of beryllium (Be) at 1 meV is 40.6\un{mb} only. The reduction of 1~meV neutrons through the same wall thickness would only be 2 \%. We therefore replace the Al canisters with Be canisters. This not only increased the CN flux efficiency of the system per proton but also slightly reduced the heat load. The CN spectrum temperature in the He-II was also reduced to 24~K due to the absorption cross-section scaling with the inverse of the neutron velocity. Both effects combine to give a $P_{\rm UCN} = 4 \times 10^8 \un{s^{-1}}$ for the 100\un{W} He-II heat load.

\subsection{Heavy water (D$_2$O) pre-moderator}

Many of the designs that use LH$_2$ moderators also use thermal pre-moderators. This comes from the fact that, as shown in many nuclear physics textbooks (e.g. see Ref.~\onlinecite{Kaplan1963}), the slowing-down power of a moderator is proportional to the macroscopic scattering cross-section of its media. LH$_2$ has only about two thirds of the hydrogen density of water, and therefore only about 2/3 of its slowing-down power. Even though, as shown in Ref.~\onlinecite{Muhrer2012}, this ``golden rule'' from reactor physics is not exactly applicable to moderator design at spallation sources because these moderators are smaller and therefore challenge the boundary conditions of the ``infinite'' media used in the standard textbook theory, it is usually still of benefit to use a pre-moderator in combination with a LH$_2$ moderator.

The most commonly used materials for pre-moderation are light and heavy water. As mentioned above, we chose Bi as the reflector material to minimize the absorption between the tungsten target and the He-II. The neutron absorption by the deuteron is significantly lower than by protons. Based on this consideration, we add a heavy water (D$_2$O) pre-moderator. The thickness of the pre-moderator is chosen to be 5~cm in our baseline design. The walls of the D$_2$O vessel are made from Al. While this modification did not significantly increase the efficiency of the target system, even though the spectral temperature was reduced to 23 K, it did reduce the He-II heat load per proton by more than 20 \%. This leads to
$P_{\rm UCN} = 5.4 \times 10^{8} \un{s^{-1}}$ for 100\un{W} He-II load with 425~kW proton beam power.

\subsection{Liquid deuterium (LD$_2$) moderator}

Even though LH$_2$ is more effective in moderating neutrons compared to liquid deuterium (LD$_2$), its thermal absorption cross-section is $\sim 640$ times higher. This causes the ${\rm (n, \gamma)}$ production rate in LH$_2$ to be significantly higher than in LD$_2$. This means the expected particle heating in the He-II bath should be lower if the bath is surrounded by a deuterated material compared to a protonated material. In a final step of modifying the baseline concept, we replace the LH$_2$ moderator with a LD$_2$ moderator of the same size. While this led to a small reduction in the efficiency of the system, because the CN spectrum temperature increased to 26\un{K}, the reduction in the He-II heat load per proton by more than 25\% leads to $P_{\rm UCN} = 7.1 \times 10^{8}\un{s^{-1}}$ for 100\un{W} He-II heat load with a 680\un{kW} proton beam.

\section{Optimizing the Inverse Geometry design \label{sec:optimization}}

In this section we optimize the Inverse Geometry geometry after applying the modifications described in Sec.~\ref{sec:modification}. The dimensions of the thickness of the LD$_2$ moderator, the thickness of the D$_2$O pre-moderator, and the position of the tungsten target relative to the He-II converter volume are studied. The dimensions of the tungsten target are kept constant is this study.

When increasing the D$_2$O thickness, the tungsten target diameter could have been increased to maintain the gap between target and the outer wall of the D$_2$O vessel. However, if the target diameter is increased, the neutron \emph{density} inside the target would decrease, making the source less effective. The biggest trade-off of having the distance between the target and D$_2$O reduced is an increase in radiation-induced heat load. However, the cooling of the D$_2$O is not deemed to be problematic here.

The strategy used is to optimize each of the three parameters independently by finding the maximum $P_{\rm UCN}$ using the procedure described in Sec.~\ref{sec:strategy}, while keeping the other two at their baseline values. To study correlations in the optimization, the target position is scanned by alternatively setting the LD$_2$ thickness or the D$_2$O thickness at their previously determined optimum values.

\begin{figure*}
\begin{center}
\includegraphics[width=5.5in]{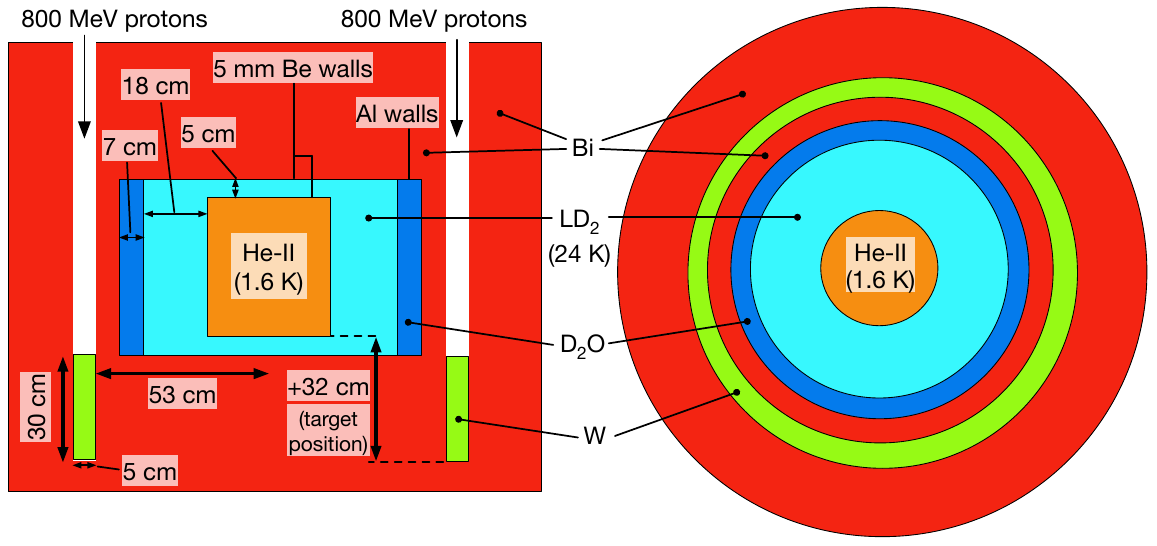}
\end{center}
\caption{MCNP model of the \emph{optimized} Inverse Geometry design after the modifications described in Sec.~\ref{sec:modification} and optimizations described in Sec.~\ref{sec:optimization}. Temperatures of the material are at room temperature unless otherwise indicated. The UCN extraction system, studied and described in Sec.~\ref{sec:UCNextraction}, is not included in this model. The two cross-section views (\emph{left} and \emph{right}) are the same as in Fig.~\ref{fig:sourceBaselineInverse}.}
\label{fig:inverseGeometryOptimized}
\end{figure*}

A final optimization step is made by setting both the LD$_2$ moderator and D$_2$O pre-moderator thicknesses at their optimized values and then scanning the tungsten target position once more. Table~\ref{tab:OptimizeSummary} gives the key parameters concluded from each step of the optimization study. The results are summarized and discussed in Sec.~\ref{sec:summaryMCNP}, and the final optimized geometry is shown in Fig.~\ref{fig:inverseGeometryOptimized}. In Sec.~\ref{sec:incorrectHe}, the effects of the available scattering kernel in MCNP to model the behavior of our He-II converter are studied. Finally, in Sec.~\ref{sec:heliumTempPress}, the effects of the He-II temperature and pressure are discussed and incorporated.

\begin{table*}
\caption{Summary of key parameters from different geometries and configurations studied. Column 2: ``proton power at 100\un{W} He-II'' refers to the proton power that can be applied to reach the 100\un{W} He-II heat load constraint. The asterisk (*) indicates that increasing the target position further would increase the proton power beyond the 1.0\un{MW} constraint. Columns 3-5: the percentage of the total He-II heat load due to neutrons, photons, or protons, respectively. Column 6: the CN flux at 1\un{meV} per 100\un{\mu A} proton current. Column 7: the energy of the peak in the differential CN flux. Column 8: the optimized $P_{\rm UCN}$. Before the last two rows these $P_{\rm UCN}$ values are before the reduction assigned to the effects of the He-II scattering kernel (Sec.~\ref{sec:incorrectHe}) and the He-II temperature and pressure (Sec.~\ref{sec:heliumTempPress}). The last two rows show these reductions.}
\begin{center}
\begin{ruledtabular}
\begin{tabular}{c c c c c c c c}
Geometry/Configuration/Effect & \parbox[c]{1.8cm}{\footnotesize proton power at 100~W He-II} & \parbox[c]{0.9cm}{\footnotesize neutron heating} & \parbox[c]{0.9cm}{\footnotesize photon heating} & \parbox[c]{0.9cm}{\footnotesize proton heating}& \parbox[c]{3.2cm}{{\footnotesize CN flux at 1{\,\rm meV} per $100\un{\mu A}$ proton} {\scriptsize [${\rm cm^{-2}\,s^{-1}\,meV^{-1}(100\mu A)^{-1}}$]}}&  \parbox[c]{1.3cm}{\footnotesize peak CN flux [meV]}  & \parbox[c]{1cm}{$P_{\rm UCN}$ [${\rm s^{-1}}$]} \\\noalign{\smallskip}\hline 
{\footnotesize Mark-3-inspired UCN source} & 120 kW & 67 \% & 28 \% & 5 \% & $5.8\times10^{10}$ & 2.6 & $0.1 \times 10^9$\\
{\footnotesize Baseline Inverse Geometry (before modifications)} & 300$\,$kW &  &  &  & $4.5\times10^{10}$ & 2.2 & $0.2 \times 10^9$  \\
{\footnotesize \emph{Inverse Geometry after modifications:}} \\
{\footnotesize D$_2$O = 5$\,$cm (base), LD$_2$ = 5$\,$cm (base), target = 0$\,$cm (base)} & 680$\,$kW & 77 \% & 19 \% & 4 \% &  $7.4\times10^{10}$ & 2.0 & $0.7 \times 10^9$\\
{\footnotesize D$_2$O = 5$\,$cm (base), LD$_2$ = 18$\,$cm (opt), target = 0$\,$cm(base)} &  700$\,$kW& 63 \% & 30 \% & 7 \% &  $11\times10^{10}$ & 2.0 & $1.3 \times 10^9$ \\
{\footnotesize D$_2$O = 7$\,$cm (opt), LD$_2$ = 5$\,$cm (base), target = 0$\,$cm (base)} & 710$\,$kW& 74 \% & 21 \% & 5 \% &  $6.0\times10^{10}$ & 2.0 & $0.8 \times 10^9$ \\
{\footnotesize D$_2$O = 7$\,$cm (opt), LD$_2$ = 18$\,$cm (opt), target = 0$\,$cm (base)} &  600$\,$kW& 69 \% & 24 \% & 7 \% & $17\times10^{10}$ & 1.7 & $1.6 \times 10^9$\\
{\footnotesize D$_2$O = 5$\,$cm (base), LD$_2$ = 5$\,$cm (base), target = 29$\,$cm (opt)*} &  1.0$\,$MW*& 76 \% & 23 \% & 1 \% & $6.3\times10^{10}$ & 2.0 & $1.1 \times 10^9$ \\
{\footnotesize D$_2$O = 7$\,$cm (opt), LD$_2$ = 5$\,$cm (base), target = 25$\,$cm (opt)*} & 1.0$\,$MW* & 74 \% & 24 \% & 2 \% & $5.3\times10^{10}$ & 2.1 &$0.9 \times 10^9$\\
{\footnotesize D$_2$O = 5$\,$cm (base), LD$_2$ = 18$\,$cm (opt), target = 26$\,$cm (opt)*} & 1.0$\,$MW* & 62 \% & 36 \% & 2 \% & $9.9\times10^{10}$ & 2.0 & $1.7 \times 10^9$\\
{\footnotesize D$_2$O = 7$\,$cm (opt), LD$_2$ = 18$\,$cm (opt), target = 32$\,$cm (opt)*} & 1.0$\,$MW* & 67 \% & 32 \% & 1 \% & $14\times10^{10}$ & 1.7 & $2.1 \times 10^9$ \\
\\
{\footnotesize \emph{MCNP He-II scattering kernel (10\% reduction)}} & & & & & & & $1.9 \times 10^9$ \\
{\footnotesize \emph{He-II pressurization to $1{\,\rm bar}$ (3\% reduction)}} & & & & & & & $1.8 \times 10^9$ \\
\end{tabular}
\end{ruledtabular}
\end{center}
\label{tab:OptimizeSummary}
\end{table*}

The simulations in this section were performed using MCNP 6.1\cite{Goorley2012}, whereas those in the previous sections and previous publications\cite{Young2014} were done using MCNPX. The neutron energy bin sizes were reduced from $0.04-0.10\un{meV}$ in the earlier simulations (the smaller size was used over the single-phonon UCN production peak) to a constant 0.02\un{meV}. Furthermore, the lowest CN energy used is reduced from 0.9\un{meV} to 0.03\un{meV}. This allows the full-width of the single-phonon UCN production peak to be covered and increases the number of points over the peak. For the different CN flux plots, these smaller bins are averaged over six bins.

The MCNP scattering kernel used for Bi at ambient temperature was generated using NJOY.\cite{MacFarlane2010} In order to do so, the released version of NJOY was modified to include the crystal structure of Bi (as shown in Ref.~\onlinecite{Cho2002}) in the LEAPR subroutine. The He-II in our MCNP calculations is treated as a free-gas scatterer at 1.6~K with an approximately constant cross-section of $0.76-0.78 \un{b}$ per atom for $1-5\un{meV}$ CNs, our energy range of interest. This does not fully describe neutron inelastic scattering in He-II. The impact of this is studied in Sec.~\ref{sec:incorrectHe}.

For the other materials used, we have restricted ourselves to those with known and vetted characteristics. For further improvements of our source design we are also exploring novel moderators, such as triphenylmethane \cite{Young2014}, and high-albedo CN reflectors, such as diamond nanoparticles.\cite{Cubitt2010}

\subsection{LD$_2$ moderator and D$_2$O pre-moderator thicknesses}

The LD$_2$ moderator thickness was scanned between 1\un{cm} to 25\un{cm}, while keeping the D$_2$O pre-moderator thickness and the target location at their baseline values. The $P_{\rm UCN}$ for 100\un{W} He-II heat load has a peak value of $1.3 \times 10^{9} \un{s^{-1}}$ when the LD$_2$ moderator thickness is 18\un{cm} with a full-width half maximum of $\sim 20\un{cm}$. This peak is caused by the beam power per 100\un{W} He-II decreasing with LD$_2$ thickness, while the CN flux per proton increases and essentially flattens off at $\sim 16\un{cm}$. At this LD$_2$ thickness, a maximum proton beam power of 700\un{kW} can be used. We will refer to this 18\un{cm} D$_2$O thickness as the optimized value.

The D$_2$O pre-moderator thickness was scanned between 1\un{cm} to 9\un{cm}, while keeping the LD$_2$ moderator thickness and the target position at their baseline values. The baseline 5\un{cm} thickness for the LD$_2$ is close to the optimum of 7\un{cm}, which gives the highest $P_{\rm UCN} = 8 \times 10^{8} \un{s^{-1}}$ for 100~W He-II heat load. $P_{\rm UCN}$ decreases to around a half only when the D$_2$O thickness is decreased to $\sim 2.5\un{cm}$. This peak in $P_{\rm UCN}$ is due to the monotonically decreasing beam power per He-II heat load with D$_2$O thickness, while the CN flux per proton increases at small thicknesses but flattens off after $\sim 6\un{cm}$. At the optimized D$_2$O thickness, a maximum proton beam power of 710\un{kW} can be used. We will refer to this 7\un{cm} D$_2$O thickness as the optimized value.

\subsection{Tungsten target position \label{sec:tungstenPosition}}

The final parameter to optimize is the position of the spallation tungsten target. The baseline position of 0\un{cm} is defined for when the downstream (relative to the proton beam) end of the tungsten target is coincident with the back wall (also relative to the proton beam) of the cylindrical He-II vessel. An increasing target position is defined as moving the target towards the proton beam's downstream direction.

\begin{figure}
\begin{center}
\includegraphics[width=0.95\columnwidth]{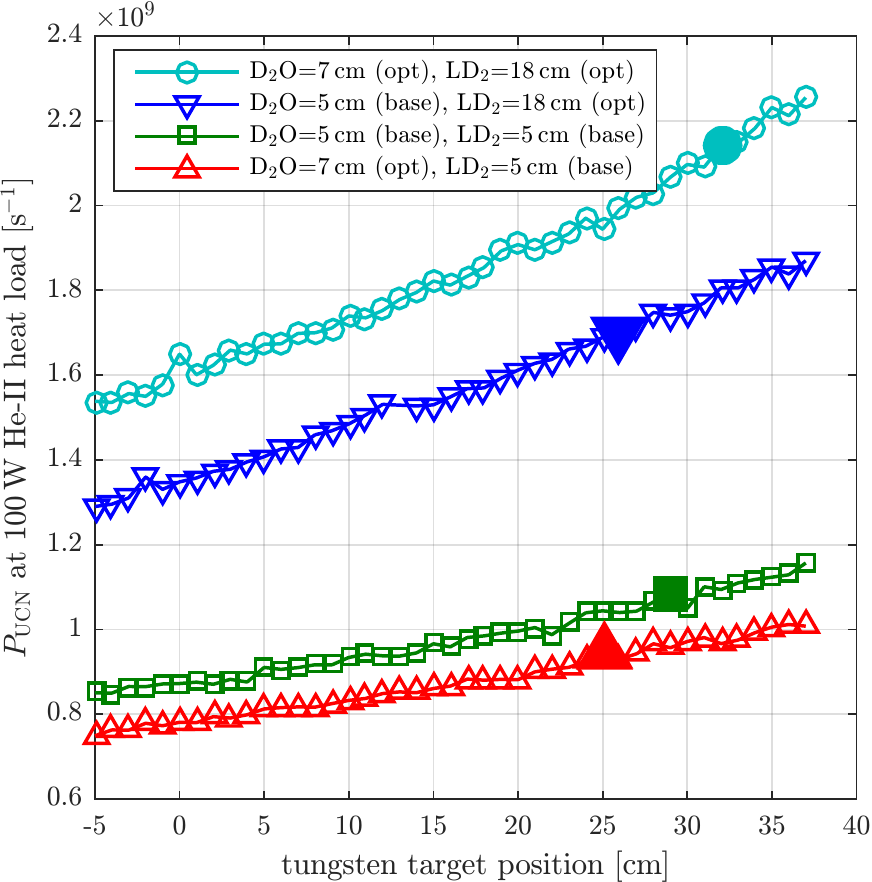}
\end{center}
\caption{The total UCN production rate $P_{\rm UCN}$ at the 100~W He-II heat load constraint. The tungsten target position is scanned with the LD$_2$ moderator thickness and/or the D$_2$O pre-moderator thickness with the original baseline (``base'') values compared with their optimized (``opt'') values. The larger solid marker on each line indicates when the 1\un{MW} proton beam power restriction has been reached (see Fig.~\ref{fig:beamPowerPerHeatLoad}).}
\label{fig:targetPositionScans}
\end{figure}

The target position is scanned to determine its optimized value while setting the D$_2$O thickness and the LD$_2$ thickness at their optimized values (described previously). For completeness, this scan is also performed while having both at their baseline values, as well as alternatively having either at their optimized value. Plots of $P_{\rm UCN}$ for 100~W He-II heat load of these target position scans are shown in Fig.~\ref{fig:targetPositionScans}.
 
\begin{figure}
\begin{center}
\includegraphics[width=0.95\columnwidth]{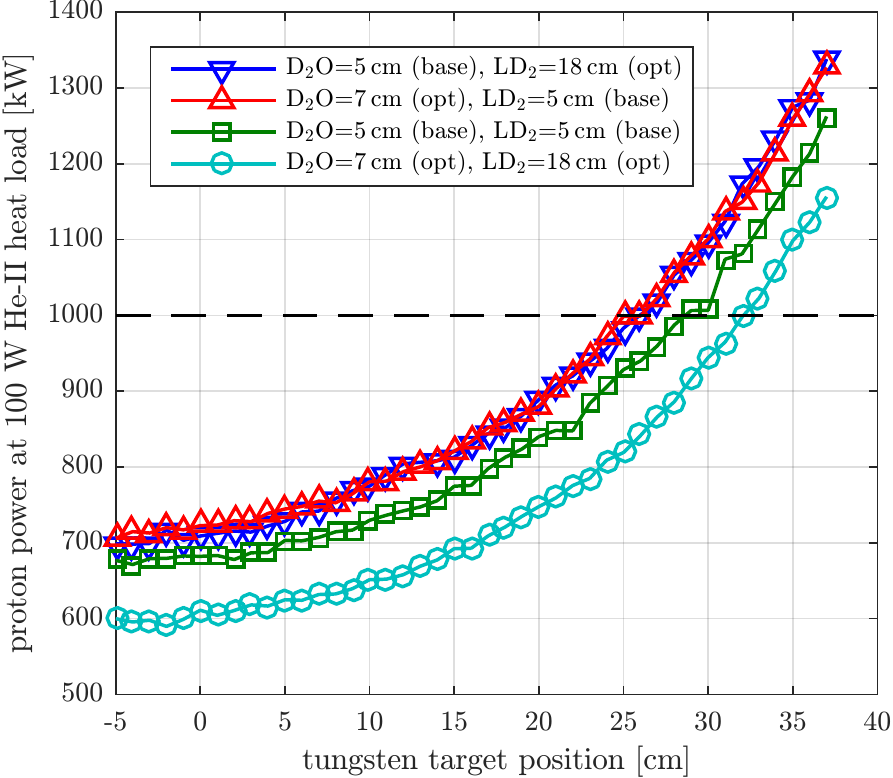}
\end{center}
\caption{The proton beam power that can be used at the 100~W He-II heat load constraint while scanning the tungsten target position for optimized (``opt'') or baseline (``base'') values for the D$_2$O pre-moderator and LD$_2$ moderator thicknesses. The horizontal dashed line marks the 1\un{MW} proton beam power restriction.}
\label{fig:beamPowerPerHeatLoad}
\end{figure}

\begin{figure}
\begin{center}
\includegraphics[width=0.95\columnwidth]{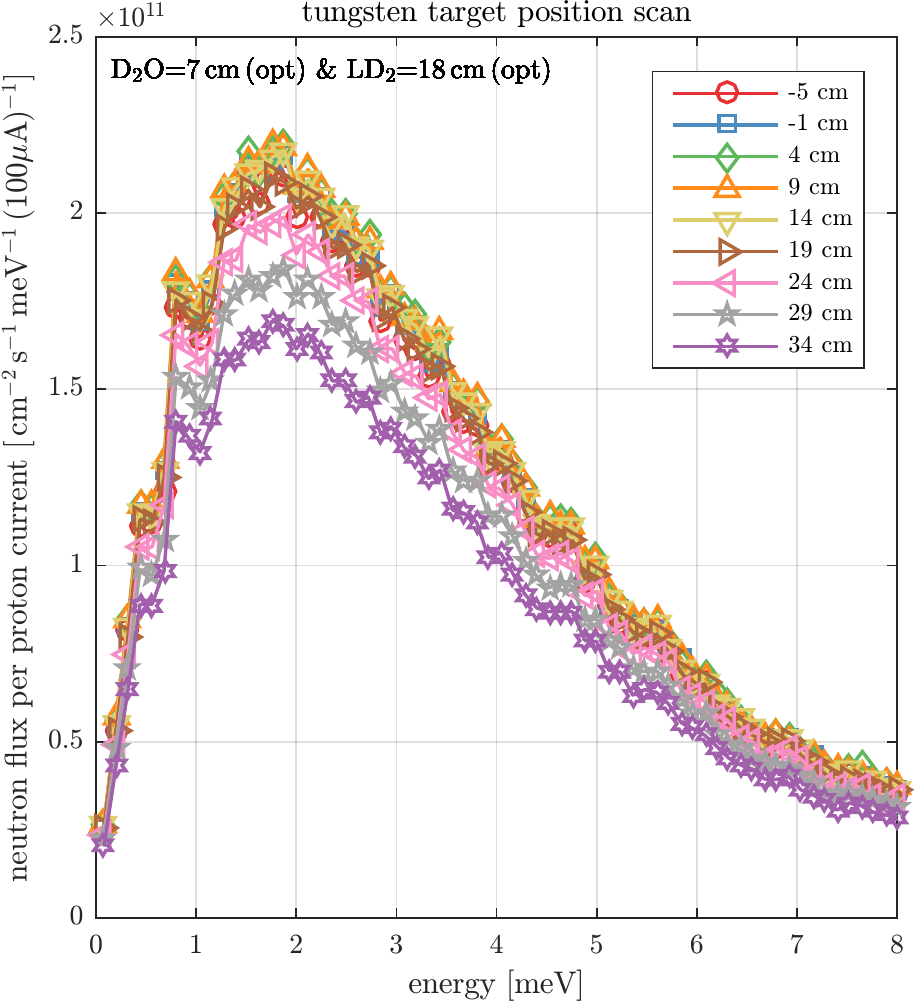}
\end{center}
\caption{The differential CN flux (see Sec.~\ref{sec:strategy}) while scanning the tungsten target position and keeping the thicknesses of D$_2$O and LD$_2$ at their optimized values (denoted by ``opt'').}
\label{fig:CNFluxTargetPos}
\end{figure}

There is no peak in $P_{\rm UCN}$ per 100~W He-II heat load for all these cases for the range of target positions studied, rather it continually increases when the target position is increased. This is because the beam power per He-II heat load increases with target position to approximately the 2nd to 3rd power (see Fig.~\ref{fig:beamPowerPerHeatLoad}), while the CN flux incident on the He-II is roughly constant from $-5\un{cm}$ to 20\un{cm} (see Fig.~\ref{fig:CNFluxTargetPos}), before starting to decrease approximately linearly. This behavior is consistent with that seen in Ref.~\onlinecite{Micklich2007}.

When increasing the tungsten target position while keeping the 100\un{W} He-II heat load constraint, there is a position at which the maximum 1\un{MW} proton beam power is also reached that depends on the D$_2$O and LD$_2$ thicknesses. This occurs for a tungsten target position of 32\un{cm} for the D$_2$O thickness at 8\un{cm} and LD$_2$ thickness at 18\un{cm}, their optimized values. This corresponds to $P_{\rm UCN} = 2.1 \times 10^{9} \un{s^{-1}}$, which is the largest value obtained so far. However, later we look into effects that could decrease this value.

It is worth noting that this behavior for the tungsten target position implies two things. Firstly, if more proton beam power were to be available, $P_{\rm UCN}$ can be increased further even under the 100\un{W} He-II heat load restriction by further increasing the target position. Secondly, if the maximum proton beam power available is less than 1\un{MW}, the corresponding reduction in the optimized $P_{\rm UCN}$ will be less than a proportionality decrease because with this new restriction the target position (as well as the other parameters) can be re-optimized. 


\subsection{Summary and interpretation of improvements \label{sec:summaryMCNP}}

\begin{figure}
\begin{center}
\includegraphics[width=0.95\columnwidth]{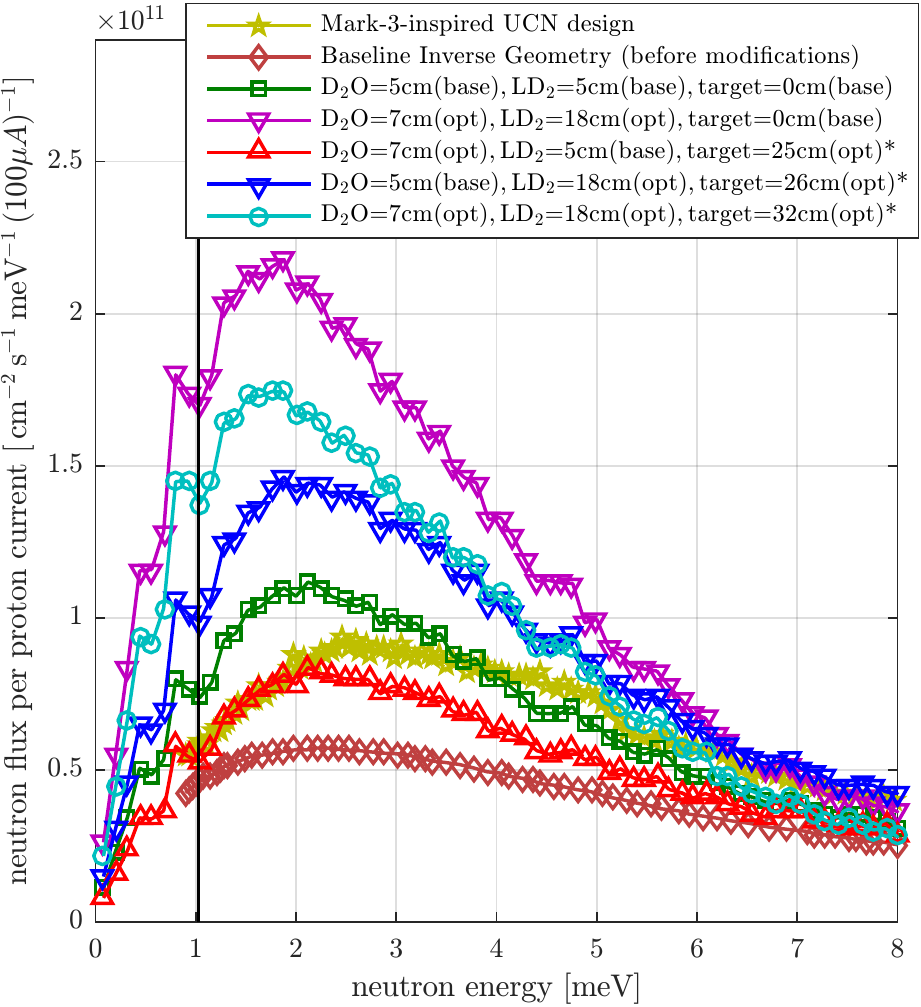}
\end{center}
\caption{The differential CN flux (see Sec.~\ref{sec:strategy}) for the various designs and configurations discussed: the Mark-3-inspired UCN design (described in Sec.~\ref{sec:lujan}), the baseline Inverse Geometry (Sec.~\ref{sec:baselineDesign}), and the Inverse Geometry after the modifications of Sec.~\ref{sec:modification}. For the latter, different combinations of the D$_2$O pre-moderator thickness, LD$_2$ moderator thickness, and tungsten target position at their baseline (``base'') and optimized (``opt'') values are shown to illustrate the effects of each parameter. The vertical solid line marks the 1.04\un{meV} single-phonon UCN production peak in He-II.}
\label{fig:CNspectrumPlot}
\end{figure}

\begin{figure}
\begin{center}
\includegraphics[width=\columnwidth]{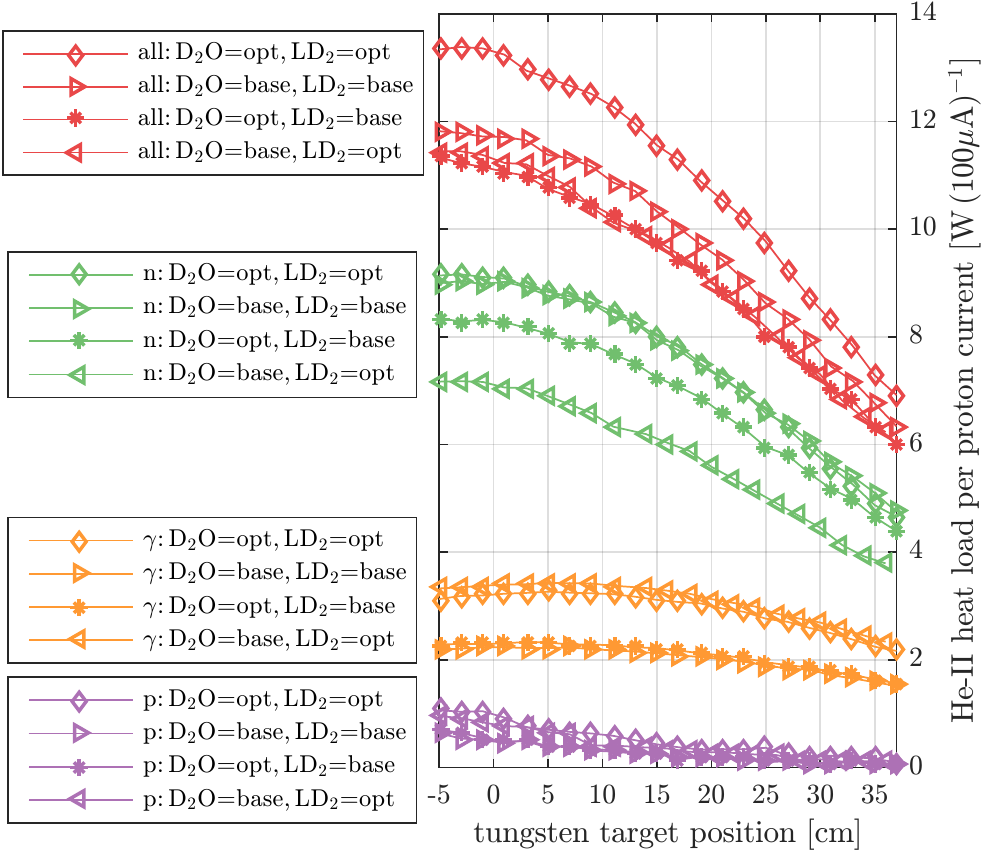}
\end{center}
\caption{Contributions to the He-II heat load per $100\un{\mu A}$ proton beam current as the tungsten target position is scanned for different configurations. In the legend, the total heat load is denoted by `all', while the heating from neutrons (both fast and cold) are denoted by `n', from gamma-photons by `$\gamma$', and protons by `p'.}
\label{fig:sourceOfHeat}
\end{figure}

To understand how the Inverse Geometry, along with its modifications and optimizations produce an increased $P_{\rm UCN}$ for 100~W He-II heat load, it is useful to look at plots of the different differential CN flux in the He-II per proton shown in Fig.~\ref{fig:CNspectrumPlot}, and the He-II heat load per proton shown in Fig.~\ref{fig:sourceOfHeat}. The latter is broken up into contributions by neutrons (both fast and cold), $\gamma$-photons, and protons. These should be combined with the previously mentioned Figs.~\ref{fig:targetPositionScans} and \ref{fig:beamPowerPerHeatLoad}. Table~\ref{tab:OptimizeSummary} provides a summary of these key parameters.

In general, relative to our baseline Inverse Geometry after the modifications described in Sec.~\ref{sec:modification}, the optimized Inverse Geometry source featured a reduction in the effective CN temperature (with the peak in the differential CN flux now at $\sim 1.7\un{meV}$) and an improved CN flux per proton (e.g. 1.9 times larger at 1~meV). This combined with a 1.5 times higher proton beam power per He-II heat load to produce an increase of $P_{\rm UCN}$ at 1~MW proton power and 100~W He-II heat load by a factor of 2.8, or to $P_{\rm UCN} = 2.1 \times 10^9\un{UCN\,s^{-1}}$. For reference, this CN flux per proton is around a factor of 2.4 better than the Mark-3-inspired UCN design. The $P_{\rm UCN}$ for the same 100~W He-II heat load is now around 15 times higher in our optimized Inverse Geometry design. Next we describe the incremental improvements coming from each step.

The transition from the Mark-3-inspired UCN design to the baseline Inverse Geometry design (before the modifications described in Sec.~\ref{sec:modification}) produced a colder CN spectrum, with the peak shifting from $\sim 2.6\un{meV}$ to $\sim 2.2\un{meV}$. This transition increased $P_{\rm UCN}$ from $1.4\times10^{8}\un{s^{-1}}$ to $2.4\times10^{8}\un{s^{-1}}$. This CN temperature decrease, if the integrated CN flux per proton remained constant, would only provide a modest improvement in UCN production. The gain in $P_{\rm UCN}$ predominantly came from an increase in the maximum proton power that can be used. The CN flux per proton was actually reduced to $0.7$ times at $1\un{meV}$. The He-II heat load per $100{\rm \mu A}$ of proton beam was 67\un{W} for the Mark-3-inspired UCN design, whereas for the baseline Inverse Geometry design (before modifications) it is only 27\un{W}.

The modifications described in Sec.~\ref{sec:modification} replaced the Al canisters with Be, added a D$_2$O pre-moderator, and changed the LH$_2$ moderator to LD$_2$. These modifications increased the CN flux per proton by $1.6$ times as well as decreased the He-II heat load per beam current to 12\un{W} per $100\un{\mu A^{-1}}$. The peak in the differential CN flux shifted from 2.2\un{meV} to 2.0\un{meV}. This combined to give the $\sim 3.5$ times gain in $P_{\rm UCN}$ to $7\times 10^{8}\un{s^{-1}}$.

The final optimization of the D$_2$O thickness, LD$_2$ thickness, and tungsten target position gave a further factor of $\sim 3$ times gain to reach $P_{\rm UCN} = 2.1\times10^9 \un{s^{-1}}$. The CN temperature reduced slightly with the peak in the spectrum shifting from 2.0\un{meV} to 1.7\un{meV}. To understand this large gain, it is useful to look at the various contributions to the He-II heat load (Fig.~\ref{fig:sourceOfHeat}) for different cases of optimization. For reference, the Mark-3-inspired UCN design's heat load per 100\un{\mu A} proton current was 67~W total (44.7~W from neutrons, 18.7~W from $\gamma$, and 3.3~W from protons), essentially an order of magnitude greater than the optimized geometry.

When the LD$_2$ thickness was increased from 5\un{cm} to 18\un{cm}, the 1\un{meV} CN flux per proton increased by $2.5$ times (keeping the D$_2$O thickness and target position optimized). This can be attributed to better matching of the moderator thickness before neutron loss kicks in. The increase in the CN flux overwhelmed the improvement in He-II heat load per proton, which only increased by $1.2$ times. This gave the total $2$ times increase in $P_{\rm UCN}$. The He-II heat load per proton decrease came from a combination of reduced neutrons, $\gamma$, and protons. The heat load due to neutrons is dominated by thermal and fast neutrons ($> 80\un{meV}$). The contribution from CNs (defined between $0-12\un{meV}$) is $\sim 10^{-3}\un{W}$ per 100\un{\mu A} beam and from thermal to fast neutrons (defined between $12-80\un{meV}$) it is $\sim 10^{-4}\un{W}$ per 100\un{\mu A} beam.

When the D$_2$O thickness was increased from 5\un{cm} to 7\un{cm} (keeping LD$_2$ and target position optimized), the increase in the CN flux at 1\un{meV} of $1.5$ times can be attributed to a better optimized pre-moderation. The He-II heat load per proton increased by $1.2$ times so overall there was a slight increase in $P_{\rm UCN}$ by a factor $\sim 1.3$. The increase in the heat load was due to an increase in fast neutrons, whereas the $\gamma$ and proton heating went down.

The optimal thicknesses of the D$_2$O pre-moderator and LD$_2$ moderator are correlated. This can be seen in Fig.~\ref{fig:targetPositionScans} for the case of D$_2$O optimized and LD$_2$ at baseline (with the target position at the optimized value), where the $P_{\rm UCN}$ is actually reduced to a factor $0.9$ compared to the case with D$_2$O at baseline and LD$_2$ at baseline. However, the effect is small $<10\%$, and with the optimization steps that we performed, we should be close to the optimum value for the precisions considered in this work.

The optimization from shifting the target position from 0\un{cm} to 32\un{cm} (at D$_2$O and LD$_2$ optimized) was interesting as the CN flux per proton was reduced to $0.8$ times. However, the He-II heat load per proton was reduced to 0.6 times, which permitted our design goal of applying the full 1\un{MW} proton beam power on the target. This results in an increase of $P_{\rm UCN}$ per He-II heat load by $1.3$ times. Since the He-II heat load is dominated by fast neutrons, the effect of moving the He-II volume further upstream towards the beam away from the tungsten target causes a suppression of the fast neutrons more quickly than compared to the loss of CN flux.

The trend of increasing $P_{\rm UCN}$ per 100\un{W} He-II with tungsten target position appears to continue beyond the largest $37\un{cm}$ studied. This indicates that if more powerful proton beams beyond the 1\un{MW} considered here were to be available, it could be utilized to increase $P_{\rm UCN}$ further while still satisfying the 100\un{W} He-II heat load constraint. But since the UCN production per proton falls as this distance is increased, 32~cm is the optimal position for the 1\un{MW} proton beam power and 100\un{W} of He-II cooling constraints considered here.

\subsection{Effects of incorrect He-II scattering in MCNP \label{sec:incorrectHe}}

As mentioned earlier the He-II was treated as a free gas scatterer in MCNP using the correct density for 1.6\un{K} He-II (0.145\un{g/cm^3}) and with an approximately constant cross-section of $0.78-0.76 \un{b}$ per atom for $1-5\un{meV}$ CNs. This corresponds to a neutron scattering mean-free-path of $\sim 60\un{cm}$. 

Liquid helium cooled below $T_\lambda = 2.17\un{K}$ starts exhibiting properties of a quantum fluid with unique phonon and roton quasi-particle excitations. Neutrons inelastically scatter off these excitations and thus both the cross-sections and energy loss in the free gas scatterer model are incorrect. The total scattering cross-section, which is predominantly inelastic scattering, decreases with temperature with the drop being largest for lower CN energies. At 2.2\un{meV}, approximately the peak of the CN spectrum incident on the He-II, this cross-section has been measured to be $0.24\un{barns}$ per atom at 1.6\un{K} ($1.9\un{m}$ mean-free-path).\cite{Sommers1955} At 1\un{meV} the measured cross-section is 25\un{mb} (18\un{m} mean-free-path).

The cross-sections used in the MCNP simulations are larger than the measured values. This will affect the UCN production in the following ways. Firstly, in the simulations more 1\un{meV} neutrons entering the He-II volume undergo inelastic scattering. Since any scattering of these neutrons will take them out of the single-phonon peak, the UCN production in our simulations from these neutrons is underestimated. Secondly, in the simulations a large fraction of the incoming $1.3-8\un{meV}$ neutrons undergo scattering (or multiple scattering) in the 40~L He-II volume. This scattering increases their path lengths in the He-II (e.g. an albedo effect) and results in an overestimate of UCN production for neutrons in this energy class. Thirdly, in the simulations the additional inelastic scattering in the He-II moderates the neutron spectrum to lower energies where the UCN production cross-sections are larger. This effect causes an overestimate of the UCN production.

\begin{figure}
\begin{center}
\includegraphics[width=0.95\columnwidth]{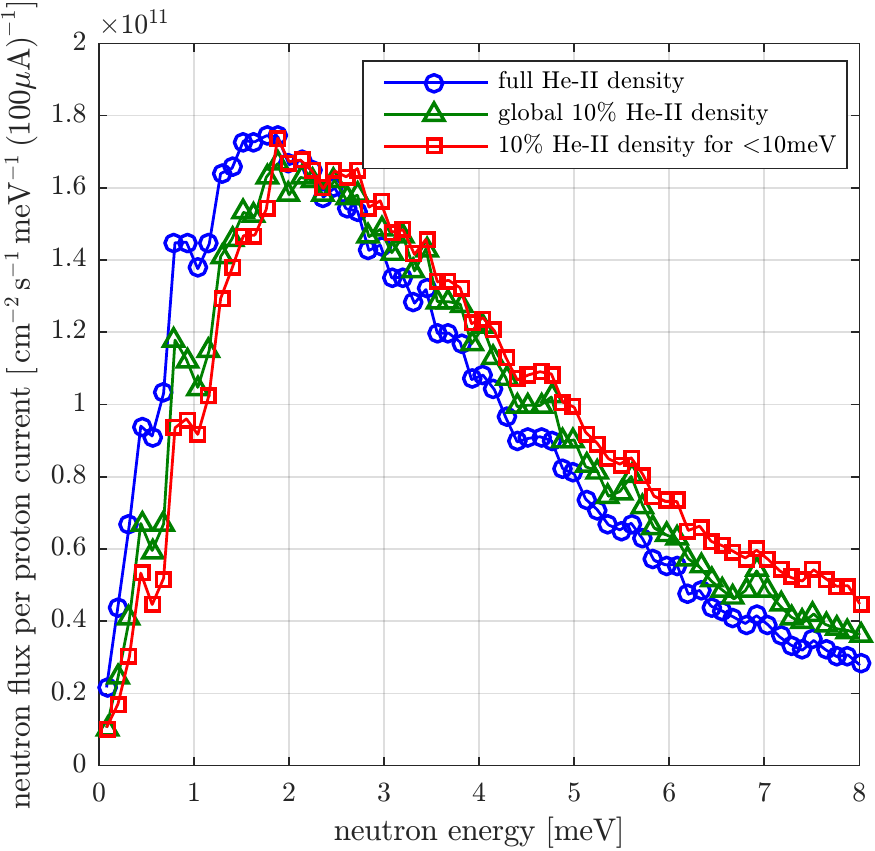}
\end{center}
\caption{Impact on the CN flux from reducing the He-II density to study the effects of the incorrect MCNP He-II scattering kernel. The spectra are for the optimized geometry (i.e. D$_2$O=7cm, LD$_2$=18cm, target=32cm).}
\label{fig:reducedHeDensity}
\end{figure}

MCNP scattering kernels for He-II do not exist to our knowledge. Nevertheless, we study the impact of the incorrect kernel. First, the He-II density is reduced to 10\% for a simulation run for the optimized parameters case. This causes the track-length weighted CN flux at 1\un{meV} to decrease by $\sim 20\%$ and the peak of the CN spectrum shifts from $1.8\un{meV}$ to $2.2\un{meV}$. This is shown in Fig.~\ref{fig:reducedHeDensity}. However, the calculated total UCN production per proton, which comes from integrating over the CN differential spectrum folded with the UCN production function, the decrease is only 10\%. The He-II heating per proton from the full He-II density simulations is used since the heating is dominated by fast neutrons and gammas, which are not affected by the He-II quantum liquid behavior. In this case, $P_{\rm UCN}$ is reduced to $1.9 \times 10^{9}\un{s^{-1}}$.

The He-II scattering cross-section reduction due to the quantum collective excitations occurs only for neutrons with energies $\lesssim 10\un{meV}$. The global He-II density reduction of the previous study affected all neutrons. To check that the $P_{\rm UCN}$ decrease was not caused by the lack of moderation of thermal and fast neutrons in the He-II, a MCNP with perturbation capability (i.e. a ``PERT card'') simulation at the optimized parameters was performed. The He-II density was reduced only for neutrons with energy $<10\un{meV}$. The spectrum and $P_{\rm UCN}$ were the same to $< 5\%$ compared to the global He-II density reduction study. As can be seen in Fig.~\ref{fig:reducedHeDensity}, this decrease is only slightly larger than the fluctuations in the simulations.

With these studies, we have shown that the impact of having the incorrect He-II scattering kernel in our simulations is around $10\%$. This is not a large effect at this stage of our physics design for the source.

\subsection{Effects of He-II temperature and pressure \label{sec:heliumTempPress}}

The He-II operating temperature of this UCN source design is higher than the $\lesssim 1.2\un{K}$ typically used. In this section, we study the effects on $P_{\rm UCN}$ by the warmer He-II temperature. The amplitude of the single-phonon peak of the dynamic structure function in He-II $S(q,\omega)$, where $\hbar q$ and $\hbar \omega $ are the momentum and energy transfer, decreases with increasing temperature and becomes more broad. However, $S(q,\omega)$ obeys the conservation (``sum'') rules of the zeroth and first frequency moments, which can be expressed as \cite{Turchin1965,Berk1993} (to some normalization constants):
\begin{equation}
\int_{-\infty}^{\infty} S(q,\omega) {\rm d}\omega  = 1 \textrm{ and } \int_{-\infty}^{\infty} \omega S(q,\omega) {\rm d}\omega  = \frac{\hbar^2 q^2}{2 m} \;,
\end{equation}
where $m$ is the neutron mass and $S(q,\omega)$ is in units of ${\rm meV^{-1}}$. These can be compared to the expression for the UCN production rate\cite{Korobkina2002}:
\begin{equation}
P_{\rm UCN} \propto \int \frac{{\rm d}\Phi(E)}{{\rm d}E}\,S\left(q = \frac{k}{\hbar},\omega = \frac{\hbar k^2}{2m}\right) \frac{{\rm d}E }{\sqrt{E}} \;,
\label{eq:UCNproduction}
\end{equation}
where ${\rm d}\Phi(E)/{{\rm d}E}$ is the differential CN flux and $k$ is the initial CN momentum. The  arguments to $S(q,\omega)$ in Eq.~\ref{eq:UCNproduction} indicate that the integral is evaluated along the free neutron ``dispersion'' $E = \hbar^2 k^2/(2m)$. Therefore, so long as the CN spectrum changes slowly over the width of the features in $S(q,\omega)$, $P_{\rm UCN}$ should remain relatively constant. This should be generally satisfied for $T < 2\un{K}$ and a broad CN differential flux from our target (as opposed to a CN flux after a neutron monochromator). This lack of change in $P_{\rm UCN}$ is demonstrated in the calculations described in this section.

\begin{figure}
\begin{center}
\includegraphics[width=0.9\columnwidth]{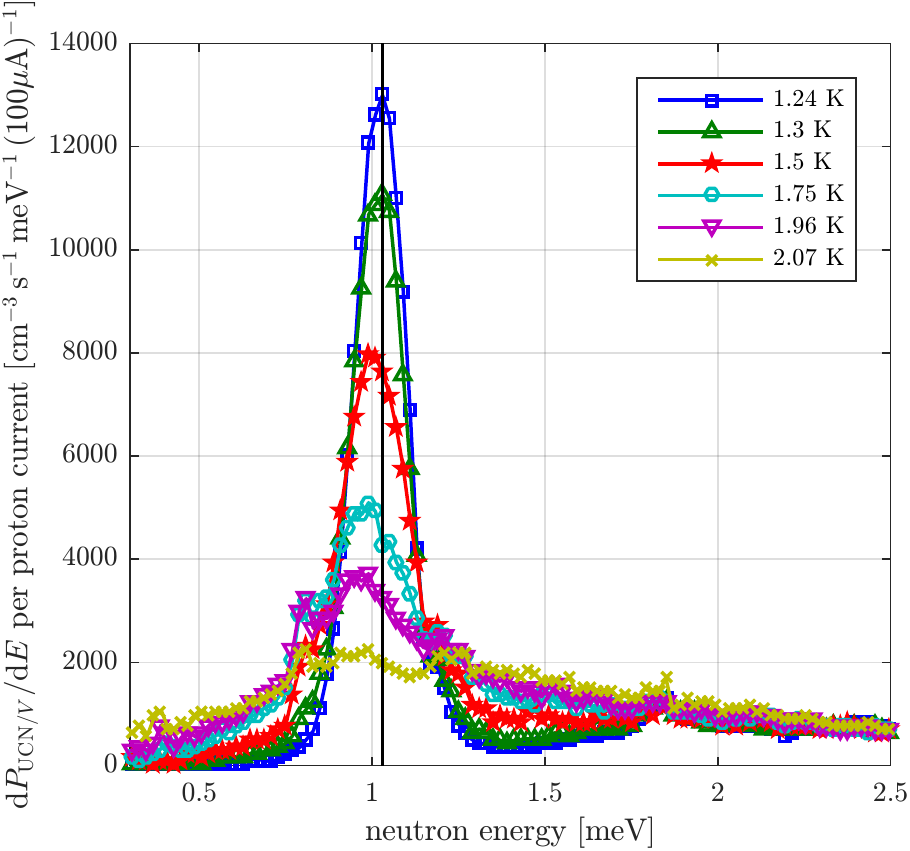}
\includegraphics[width=0.9\columnwidth]{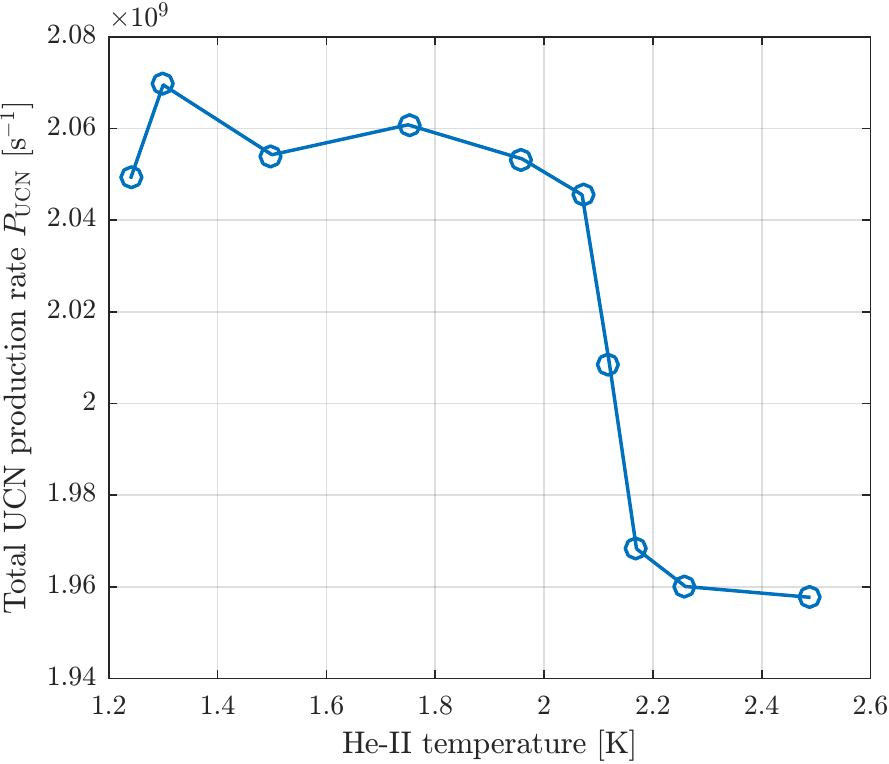}
\end{center}
\caption{Changes in the UCN production function (\emph{top}) and total UCN production rate (\emph{bottom}) with He-II temperature at saturated vapor pressure for the optimized Inverse Geometry design. ${\rm d}P_{{\rm UCN}/V}/{\rm d}E$ is the energy-differential volumetric UCN production rate per $100\un{\mu A}$ proton beam current. The CN energy bin size shown in this plot is finer than previous plots to increase detail over the 1\un{meV} single-phonon UCN production peak.}
\label{fig:UCNproductionTemperature}
\end{figure}

To verify $P_{\rm UCN}$ does not decrease significantly at 1.6\un{K} for our CN spectrum, the differential UCN production is folded with the differential CN flux for the optimized geometry. Experimental studies of UCN production have been performed between $1.1-2.4\un{K}$ in Ref.~\onlinecite{Leung2016}. The calculation here follows the appendix of Ref.~\onlinecite{Leung2016}, which used neutron inelastic scattering data between 1.2\un{K} and 2.5\un{K} from Ref.~\onlinecite{Andersen1994a,Andersen2015}. $P_{\rm UCN}$ is then calculated by integrating the differential production. As shown in Fig.~\ref{fig:UCNproductionTemperature}, it can be seen that $P_{\rm UCN}$ is approximately constant until $\sim 2\un{K}$ after which it drops, though only by $\sim 6\%$. The largest effect of increasing the temperature is the increase of the upscattering of UCNs and thus reduces the extraction efficiency (described in Sec.~\ref{sec:HeUpscatterUCN}).

We also note that experimental studies of UCN production in pressurized He-II have been performed.\cite{Schmidt-Wellenburg2015} These measurements were made at $T= 1.1\un{K}$ at SVP ($\sim 0.3\un{mbar}$), 6\un{bar}, 11\un{bar}, 16\un{bar}, and 20\un{bar}. In this experiment, the He-II was pressurized with a capillary connected to a helium gas bottle, the same technique as that proposed for pressurizing our He-II to 1\un{bar}. The UCN production rate due to single-phonons is expected and observed to decrease with increasing pressure. The measured decrease in the UCN production rate was slightly higher than expected. However, at our low 1\un{bar} pressure, interpolation of the more pessimistic experimental results give a $6\%$ reduction only. The UCN production due to multi-phonon production is expected to increase with pressure, but was experimentally observed to remain constant. We assume the more pessimistic (i.e. remain constant) experimental result. Because the UCN production rate due to multi-phonons is around 50\% at our temperature (see Fig.~\ref{fig:UCNproductionTemperature}), the reduction in the total UCN production due to pressurization to 1\un{bar} is expected to be $\sim 3\%$ only. Including this pressurization loss and the effect of the incorrect He-II scattering kernel, $P_{\rm UCN} = 1.8 \times 10^{9} \un{UCN\, s^{-1}}$ is expected from our optimized source design.

\section{UCN extraction geometry \label{sec:UCNextraction}}

In this section we discuss UCN extraction geometries and make estimates of the extraction efficiency at the level of a physics model. Once UCNs are produced they can be lost due to upscattering in the He-II (Sec.~\ref{sec:HeUpscatterUCN}) or upscattering or absorption losses at the walls of the volumes and guides. The losses of the latter are small compared to the former for our He-II temperature. Vital performance parameters for a UCN source are the UCN current and/or density that can be delivered to an external experimental volume. Our source design will be optimized for high-current output due to the short $\tau_{\rm up} \approx 3\un{s}$ for our 1.6\un{K} He-II converter. The goal of the UCN extraction system is to get the UCNs out of the He-II quickly.

\begin{figure*}
\begin{center}
\includegraphics[width=1.8\columnwidth]{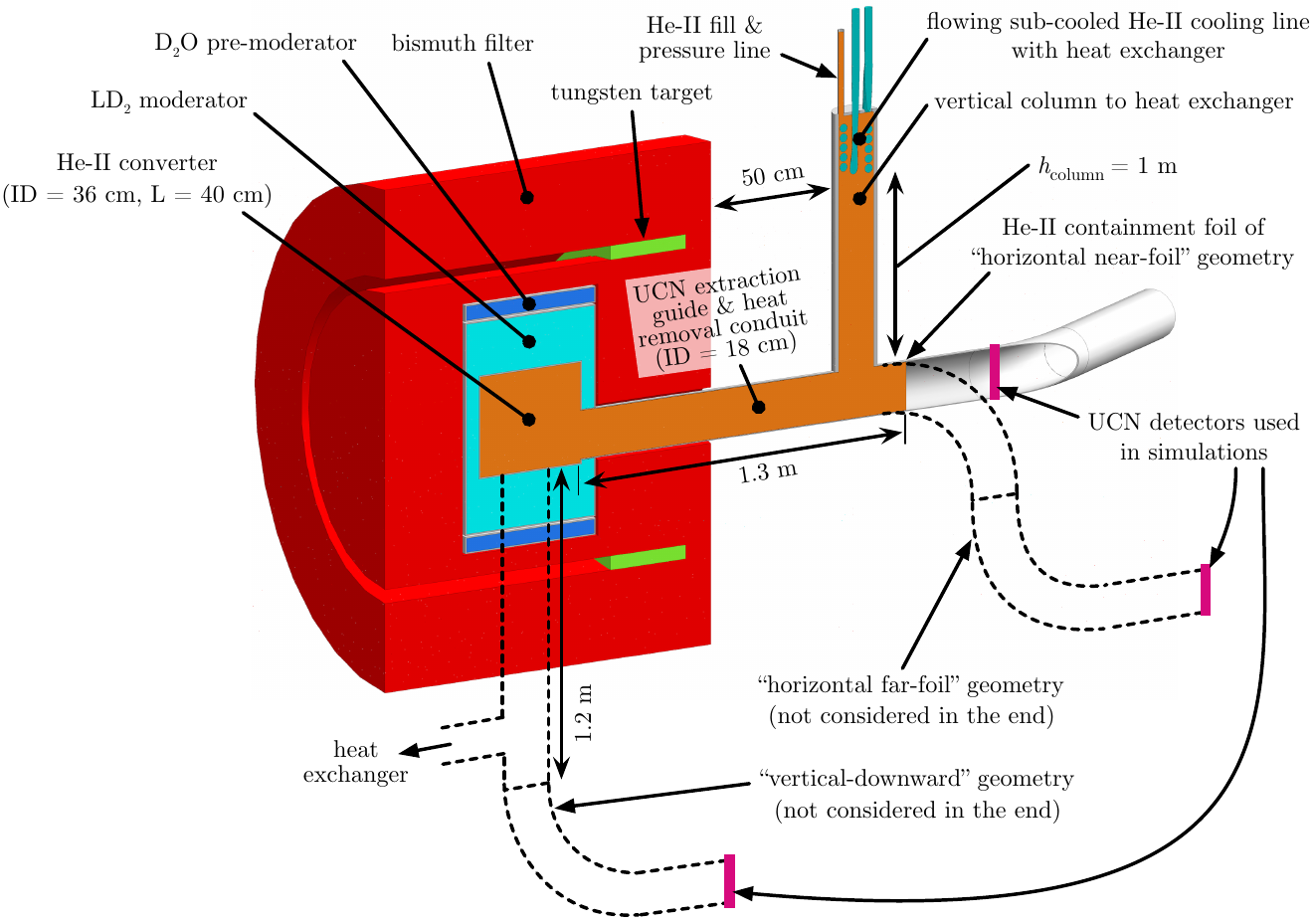}
\end{center}
\caption{A 3D model of the optimized Inverse Geometry source design with UCN extraction system (compare with Fig.~\ref{fig:inverseGeometryOptimized}). A vertical cross-section at the center of the He-II volume is shown. The horizontal proton beam (since this is easier to implement than a vertical beam) arrives from the left. The proton beam is rastered to strike the tungsten target at different positions to distribute the heat load. The D$_2$O pre-moderator, LD$_2$ moderator, and He-II converter volumes (with inner diameter ``ID'' and length ``L'') are nestled radially inwards. In the primary UCN extraction geometry (``horizontal near-foil''), a horizontal conduit (with inner diameter ``ID'') filled with He-II is used to transport both UCNs and heat out of the He-II converter volume. A vertical column, with a height denoted by $h_{\rm column}$ and also filled with He-II, is used to transport heat to the heat exchanger of the flowing cooling line from a sub-cooled He-II cooling system. UCNs exit the He-II out of the containment foil and into guides connected to externally-located experiments. Other extraction geometries briefly studied, shown as dashed-lines, but not considered further are the ``horizontal far-foil'' and the ``vertical-downward''. The position of the UCN simulation detectors used in the UCN trajectory simulations are also shown.}
\label{fig:inverseGeometry3D}
\end{figure*}

A brief overview of the UCN extraction system was given at the end of Sec.~\ref{sec:strategy}. Fig.~\ref{fig:inverseGeometry3D} shows the UCN extraction geometries we will study. So far it has been more convenient to show the Inverse Geometry source's axis in a vertical orientation. However, due to the engineering simplifications of having a horizontal proton beam, it will be easier to orient the source's axis horizontally.

The ``horizontal near-foil'' extraction geometry will be the primary scheme considered. Here, a straight UCN guide section filled with He-II and with diameter $D_{\rm guide}$ is connected to the 40\un{L} He-II converter volume. The length required to reach outside the back of the Bi filter is $ 50\un{cm}$. This tube also serves as the heat removal conduit. After another $50\un{cm}$ past the Bi a T-section is used to allow a vertical column, also filled with He-II, to be connected to the horizontal tube. This vertical column, with a height $h_{\rm column}  = 1\un{m}$, allows heat to reach the heat exchanger coupled to the flowing He-II cooling line from the separate sub-cooled He-II system. Gravity helps reduce UCNs going up this vertical column and reaching the heat exchanger, which will typically be lossy for UCNs. The key to the heat removal design is to avoid having reductions in cross-section area along the thermal transport path.

The $50\un{cm}$ gap between the Bi and heat exchanger is to give sufficient space for shielding the material used in the heat exchanger (most likely copper) to reduce activation. (The space reserved for biological shielding is $\sim 5\un{m}$.) Fast neutron suppression by an order-of-magnitude can typically be achieved with $15\un{cm}$ thickness of polyethylene or borated-polyethylene.\cite{Fragopoulou2006} Because of the small volume of material in the heat exchanger, more advanced materials can be used also to further reduce the neutron activation flux in this space. It is also possible, instead of the vertical column, to have the column at an angle (e.g. $\sim 45^\circ$ to the horizontal) to create more space for shielding. This would likely have slightly higher UCN loss though. The design of the shielding is beyond the scope of the current work but will be considered in the future as we move towards an engineered design.

The considerations for heat removal are discussed in Sec.~\ref{sec:heatEx}. It is found that $D_{\rm guide} > 18\un{cm}$ is needed to keep the temperature drop from the 40\un{L} converter volume to the heat exchanger to be $< 50\un{mK}$. We will select $D_{\rm guide} = 18\un{cm}$ in our design. We find with UCN trajectory simulations that there is a slight improvement with a larger $D_{\rm guide}$. However, having it too large would make it more difficult to shield (as discussed in Sec.~\ref{sec:shieldingExtractGuide}).

There will be a He-II containment foil that also allows UCN transmission on the horizontal arm of the T-piece. While there are foil-less designs that rely on vertical upward extraction of UCNs from He-II,\cite{Zimmer2007a,Piegsa2014,Leung2016,Zimmer2015,Masuda2012,Ahmed2019} this is not particularly suitable in our design due a combination of the gravity-separated heat exchanger and He-II pressurization (see Fig.~\ref{fig:inverseGeometry3D}). There is also He-II film flow loss associated with foil-less designs. This is typically reduced by a short UCN guide section with a small diameter, however this is not optimal for a current-optimized UCN source and is thus avoided.

A common UCN transmission foil material is aluminum with $U_{\rm Al} = 54\un{neV}$, with a typically observed $\sim 10-20\un{neV}$ higher potential due to oxide layers. Due to reflection and loss in the bulk of the Al foil, we attempt to use gravity to accelerate UCNs before they reach the foil. This is the motivation behind the other two geometries considered: the ``horizontal far-foil'' and the ``vertical-downward'' extraction. In these two geometries, UCN guide bends are placed after the foil to end up with a horizontal guide, which is the most useful geometry for experiments. This also allows a direct comparison between the three geometries considered.

An alternative material explored for the foil is polypropylene (PP), which has a negative neutron optical potential $U_{\rm PP} = -8\un{neV}$. This will have lower UCN reflection and is therefore well-suited for the horizontal near-foil geometry. Experimental measurements of the UCN transmission through foils are typically lower than when including only absorption and upscattering loss in the bulk material,\cite{Pokotilovski2016} which we will call an ``idealized'' foil. Idealized foil properties are sometimes erroneously used in UCN simulations however.

Fortunately, experimental studies of UCN transmission through cryogenic PP foils have been performed in Ref.~\onlinecite{Miranda1988}, where UCN transmission values of $\sim 85\%$ were measured. We have also demonstrated in separate work with free-standing, 5\un{cm} diameter, 30\un{\mu m} thick PP foils that the cryogenic mechanical properties are sufficient for the use as a He-II containment foil. For our larger 18\un{cm} diameter, we can use a support grid.  Based on the experimental studies of Ref.~\onlinecite{Miranda1988}, we arrive at an effective transmission of $\sim 70\%$ for our 30\un{\mu m} thick PP foil compared to idealized PP foil properties. This reduced transmission is caused an additional loss mechanism in the bulk of the foil caused by UCN elastic scattering off material inhomogeneities. We deduced a scattering mean-free-path $\lambda_{\rm scat} = 20\un{\mu m}$ from the experiments of Ref.~\onlinecite{Miranda1988}. In our simulations we further include the results from both the idealized foils and foils with the added loss mechanism (``more realistic loss''). Furthermore, we also include a 90\% transmission for a support grid for our large 18\un{cm} diameter foil such that the effective transmission of the foil assembly is $\sim 60\%$. A detailed discussion of the foil is given in Sec.~\ref{sec:foil}.

In the 40\un{L} He-II converter volume and in the vertical column to the heat exchanger, it will be advantageous to have UCNs undergo ``non-specular'' reflections. It is a well-known problem in UCN-bottle-based neutron lifetime experiments that highly symmetric storage volumes with smooth walls, in particular cylindrical-shaped volumes, allows UCNs to be in long-lived, quasi-stable orbits where they make many specular, glancing-angle reflections. For the cylindrical converter volume in our source, such UCNs will become lost because of the short $\tau_{\rm up}$ of the He-II. In the vertical column, having non-specular reflections on the side walls will allow some UCNs to reflect back into the main horizontal UCN transport channel, thus reducing the probability of interacting with the lossy heat exchanger.

Typically when discussing specularity of a surface, roughness features comparable to the incident wavelength are the ones considered, which for UCNs is $\sim 10-100\un{nm}$.\cite{Atchison2010} However, for our application where we want to induce non-specular reflections, relying on microscopic roughness could prove difficult. Instead, in order to produce our needed non-specular reflections we can consider using macroscopic geometric features on a surface, such as bumps or corrugations, with sizes $\sim 1-10\un{mm}$. This is similar to the technique used in the MAMBO-II neutron lifetime experiment.\cite{Pichlmaier2010a} When we use the term ``non-specular'' reflection, we will refer to a lack of correlation between the incoming and outgoing reflection angles, but where these angles are defined relative to the surface normal given by averaging over an area of $\sim 10\un{cm^2}$. This can be produced both by surface roughness or the geometric features.

UCN trajectory tracking Monte-Carlo simulations are used to evaluate the performance of a UCN extraction system design (described more in Sec.~\ref{sec:UCNtrajectorySim}). In these simulations UCNs are created uniformly in the 40\un{L} He-II converter volume with an energy distribution ${\rm d}P_{\rm UCN}/{\rm d}E_{\rm UCN} \propto \sqrt{E_{\rm UCN}} \, {\rm d}E_{\rm UCN}$ with a cut-off at $U_{\rm ^{58}Ni} - U_{\textrm{He-II}}$ (see Sec.~\ref{sec:strategy}). 

A ``UCN simulation detector'' is used to count, and subsequently terminate, UCNs that reach it in the simulations. We further define that only UCNs with total kinetic energy less than $U_{\rm ^{58} Ni}$ at the UCN simulation detector are counted. These UCN simulation detectors are placed $\sim 30\un{cm}$ after the He-II containment window for the horizontal near-foil geometry or $\sim 30\un{cm}$ after when the UCN guide becomes horizontal for the other two geometries. The choice of this position is used to provide a comparison between the three geometries studied. The largest UCN extraction loss in our source design is up-scattering in He-II (described in Sec.~\ref{sec:HeUpscatterUCN}).

The fraction of UCNs that are counted in the simulation detector relative to the total number of UCNs generated in a simulation is denoted by $\epsilon_{\rm sim}$. Simulations are run with different improvements as summarized previously. The transmission through the foil support grid $\epsilon_{\rm grid} = 90\%$ is multiplied after the simulations to determine the single-passage transmission efficiency out of the source. We also include the transmission loss in long UCN guides.

We are primarily concerned with the useful UCN current that can be delivered to an external experiment. We assume that an experiment can be placed 5\un{m} from the back of the tungsten target. This length is consistent with the biological shielding requirements at the European Spallation Source\cite{Bauer2001} and other facilities.\cite{Filges2009} Therefore, this adds $\sim 4\un{m}$ of UCN guide after the position of the simulation detectors (depending on the geometry). For the UCN transport efficiency in long guides, we will use values based on the experiments in Refs.~\onlinecite{Atchison2010,Serebrov2017}. Loss of UCNs in guide transmission is dominated by non-specular reflections due to microscopic surface roughness because the probability for non-specular reflection is generally much larger than the loss probability. The number of reflections per meter during UCN transport for purely specular reflections is only on the order of $\sim 10$. From a detailed study comparing measurements and atomic force microscope studies of the roughness of UCN reflecting surfaces, Ref.~\onlinecite{Atchison2010} concludes that transmissions of $\sim 75\%$ is possible with guides having a $7\un{cm}$ diameter and a length of 5\un{m}. In Ref.~\onlinecite{Serebrov2017} using measurements with guide assemblies with lengths ``up to several meters'' and having a $8\un{cm}$ diameter, it was concluded that the loss per 1\un{m} of 5.5\% could be obtained. The loss figures from these studies are roughly in agreement. For our 18\un{cm} diameter guides, the transport loss should decrease by approximately a factor of two. The coating on the guides of Ref.~\onlinecite{Serebrov2017} is a $^{58}$NiMo alloy with a optical potential of $320\un{neV}$, which is slightly less than $U_{\rm ^{58}Ni}$. Therefore, we reasonably assign a transport efficiency for our 18\un{cm} diameter, 4\un{m} length UCN guides to be $\epsilon_{\rm guide} \approx 0.8$.

All of the above efficiencies are combined to reach a total single-passage transport efficiency to an external volume $\epsilon_{\rm tot\;single} = \epsilon_{\rm sim} \, \epsilon_{\rm grid}\, \epsilon_{\rm guide}$. This can then be used to calculate the useful UCN current from the source $R_{\rm use} = \epsilon_{\rm tot\;single}\, P_{\rm UCN}$. From this, the extracted UCN density that can be obtained in an external volume is estimated in Sec.~\ref{sec:UCNdensities} using the ``no return'' approximation.

First we describe the expression for the UCN upscattering rate in He-II that is used, and the heat removal properties of He-II that determines diameter of the UCN extraction guide that is needed.

\subsection{UCN upscattering loss with He-II temperature \label{sec:HeUpscatterUCN}}

The primary UCN loss in our source design is caused by upscattering in He-II. Three different processes were included in Ref.~\onlinecite{Golub1979} that contribute to the temperature-dependent UCN upscattering time constant in He-II, $\tau^{-1}_{\rm up}(T)$: one-phonon absorption, two-phonon scattering, and roton-phonon scattering. However, several experiments have shown that only the second mechanism appears to be present.\cite{Yoshiki1992, Piegsa2014,Leung2016} We will use the 95\% C.I limits set on the one-phonon absorption and roton-phonon scattering mechanisms of $<47\%$ and $<29\%$, respectively, that were measured at our temperatures.\cite{Leung2016} Thus, for our upscattering rate estimate, we use:
\begin{eqnarray}
\tau^{-1}_{\rm up}(T) = (0&&.47)(130\un{s^{-1}}) ( {\rm e}^{-(12\un{K})/T}) \nonumber\\
+&& (7.6\times 10^{-3}\un{s^{-1}\,K^{-7}})(T^7) \nonumber\\
&&+ (0.29)(18\un{s^{-1}\,K^{-3/2}}) (T^{3/2} {\rm e}^{-(8.6\un{K})/T}) \,.
\label{eq:tauup}
\end{eqnarray}
Note that the upscattering time constant does not depend on the UCN velocity, but faster moving UCNs will generally spend less time in the He-II so that their upscattering loss is reduced.

We will pressurize our 1.6\un{K} converter He-II bath to $\sim 1\un{bar}$. The upscattering rate does not increase significantly for such a low pressure.\cite{Schmidt-Wellenburg2015} The pressurization moves the state of He-II far from the liquid/vapor line in the phase diagram such that the He-II needs to be locally heated to $> 4.2\un{K}$ before a bubble can be formed. Therefore, we can neglect UCN upscattering caused by helium gas bubbles.\cite{Gudkov2007}

\subsection{Heat removal with UCN extraction guide \label{sec:heatEx}}

Optimizing a UCN source design to allow sufficient heat removal from the UCN converter, or to allow pumping access, while maintaining low UCN transport loss is a common problem, in particular for He-II based sources under high heat loads.\cite{Masuda2014,Ahmed2019,Serebrov2011,Serebrov2015a} The heat exchanger from the flowing He-II line will remove the heat from the He-II converter. As a pessimistic assumption in our UCN transport calculations, we assume that any UCN that reach the heat exchanger will be lost immediately. This is due to the large surface area of the heat exchanger and the typically UCN lossy materials used. To reduce this UCN loss, we can place the heat exchanger on the end of a He-II filled conduit split off from the main UCN extraction path. For the horizontal extraction geometries (see Fig.~\ref{fig:inverseGeometry3D}), this conduit will be a He-II-filled vertical column above the main horizontal UCN extraction guide. Gravity will aid in reducing the UCNs that can reach the heat exchanger.

The heat flux in the static He-II conduit considered in our source ($\sim 0.5\un{W\,cm^{-2}}$) falls in the Gorter-Mellink regime, where the temperature gradient is dominated by turbulence caused by mutual friction between the normal and superfluid components. Following Refs.~\onlinecite{Lebrun2014,Van-Sciver2012}, for a static He-II conduit of length $L$, diameter $D$, with temperatures at the cold end $T_{\rm C}$ and at the warm end $T_{\rm W}$, the steady-state heat flow $\dot{Q}$ is given by:
\begin{equation}
\left( \frac{4\dot{Q}}{\pi D^2} \right)^m L = X(T_{\rm C}) - X(T_{\rm W})\; ,
\label{eq:conductivityConduit}
\end{equation}
where $m \approx (3.4\pm 0.1)$ is found experimentally and $X(T)$ is a parameter that is a function of temperature (physically analogous to a conductivity integral). At the heat exchanger we set $T_{\rm C} = 1.60\un{K}$, and we set a desired $T_{\rm W} = 1.65\un{K}$ (i.e. $\Delta T = T_{\rm W} - T_{\rm C} = 50\un{mK}$) to not significantly affect the UCN upscattering loss. For $L = 2.0\un{m}$ (combining the horizontal and vertical distances to the heat exchanger), we find that $D> 18\un{cm}$ is required. It is worth noting that Eq.~\ref{eq:conductivityConduit} has a strong dependence on $D$ and only a linear dependence on $L$. This means that having a reduction in the cross-section area can produce large unwanted temperature drops. We will set the diameter of the UCN extraction guide $D_{\rm guide} = 18\un{cm}$. Though, as explored with UCN trajectory simulations later, having a slightly larger $D_{\rm guide}$ would actually improve the UCN transport properties as well as reduce the conductive temperature drop.



\subsection{Extraction orientation and He-II containment foil location \label{sec:foil}}

We consider having the UCN extraction guide coming out of the 40~L He-II converter volume horizontally or vertically downwards (see Fig.~\ref{fig:inverseGeometry3D}). A foil with good UCN transmission properties is required for containing the He-II. This foil has to be located after the T-section for the heat exchanger.

A common foil material to use for UCN transmission is aluminum, which has $U_{\rm Al} = 54\un{neV}$. This does not include an oxide layer, which has been found experimentally to effectively add $\sim 10 - 20\un{neV}$. If we define the ``transverse kinetic energy'' as $E_\bot \equiv m v_\bot^2 /2$, where $m$ is the neutron mass and $v_\bot$ is the velocity component perpendicular to the surface of the foil, then only UCNs with $E_\bot > U_{\rm Al}-U_{\textrm{He-II}}$ at the foil can be transmitted (recall $U_{\textrm{He-II}} = 18.5\un{neV}$). If $E_\bot < U_{\rm Al}-U_{\textrm{He-II}}$ then the UCN will be reflected back towards the production region and likely to be lost. When using an Al foil a vertical drop is advantageous to ensure low energy UCNs are accelerated by gravity so they over come the $U_{\rm Al}-U_{\textrm{He-II}}$ potential.

Primarily driven by the UCN transmission properties of the He-II containment foil, the following UCN extraction geometries (see Fig.~\ref{fig:inverseGeometry3D}) are studied: 
\begin{enumerate}
\item \emph{``horizontal near-foil''}: a straight horizontal UCN extraction guide coming out of the He-II converter volume, followed by the foil immediately after the T-section to the heat exchanger. A slight bend is then needed to avoid direct view of radiation and fast neutrons coming from the source.

\item \emph{``horizontal far-foil''}: a straight horizontal guide followed by a downward-pointing 90$^\circ$ elbow, an Al foil, and then another 90$^\circ$ elbow to have the outgoing UCN guide horizontal.

\item \emph{``vertical-downward''}: a vertical downward extraction guide followed by an Al foil after the T-section to the heat exchanger, then a 90$^\circ$ elbow to have the outgoing UCN guide horizontal
\end{enumerate}

The idea behind geometries (2) and (3) is to use gravity to accelerate UCNs so they can more easily overcome the Al foil's potential. Acceleration can also allow UCNs to spend less time in the He-II to reduce upscattering loss. However, acceleration will cause a fraction of UCNs to have too high kinetic energy to be reflected by the guide potential.

During transmission through a material, UCN loss can also occur in the bulk material due to absorption or upscattering. For material much thicker than the UCN wavelength, the loss rate constant is given by $\tau_{\rm abs+up}^{-1} = v \sum n_i (\sigma_{\rm abs} + \sigma_{\rm up})_i$, where $n_i$ is the number density of the $i$'th element in the material (with the summation over the different elements), and $\sigma_{\rm abs}$ and $\sigma_{\rm up}$ are the nuclear absorption and upscattering cross-sections at the UCN velocity $v$. The absorption cross-section scales as $1/v$, therefore $v\sigma_{\rm abs} = v_{\rm th} \sigma_{\rm abs,th}$, where $\sigma_{\rm abs,th}$ is the thermal absorption cross-section at the neutron thermal velocity $v_{\rm th} = 2200\un{m \, s^{-1}}$. For Al, $\sigma_{\rm abs,th} =0.231\un{b}$. In Al, the absorption loss dominates the upscattering loss, so $\tau_{\rm abs+up} = 3\times 10^{-4}\un{s}$. This time constant is independent of the UCN velocity, but slower moving UCNs do suffer a greater loss probability due to a longer time spent in the foil during transmission.

Another possible UCN window material for our application is polypropylene (PP). PP has a negative neutron optical potential $U_{\rm PP} = -8\un{neV}$, therefore almost all UCNs incident on the window can enter the bulk material of the foil. Usually polymer-based UCN windows are not used at room temperature due to the large upscattering cross-section; the incoherent scattering length of hydrogen is 25\un{fm} compared to aluminum's 0.26\un{fm}. However, at low temperatures the up-scattering of UCNs in the bulk is strongly suppressed and insignificant so that for PP $\tau_{\rm abs+up} = 2\times 10^{-4}\un{s}$ can be reached. A discussion of the advantages of PP replacing Al foils for UCN transmission is given in Ref.~\onlinecite{Miranda1988}, where experimental studies were performed. We have also experimentally demonstrated that with PP windows 5\un{cm} in diameter and $30\un{\mu m}$ thick that they are superfluid helium leak tight (at 1.9\un{K}) and can withstand a $\sim 2\un{bar}$ pressure difference (at 77\un{K}). For our required larger 18\un{cm} diameter UCN window, a support grid can be used. A grid with a physical opening area of 90\% can be realized. (E.g. a grid approximately having $1\un{mm}$ wide supports containing openings $< 5\un{cm}$ diameter would achieve this.)

Experimentally observed UCN foil transmission are typically lower than expected from $\tau_{\rm abs+up}$ alone,\cite{Atchison2009,Pokotilovski2016} even when taking into account oxide layers. For instance, for UCNs with $E_{\bot} = 100\un{neV}$ transmitting through a 30\un{\mu m} thick foil with the previous $\tau_{\rm abs+up}$ values, the transmission is expected to be $ \sim 97 - 98 \%$. The most likely mechanism is due to elastic scattering off density inhomogeneities in the bulk (i.e. neutron refraction on the boundaries of crystallites and scattering off defects).\cite{Pokotilovski2016} These elastic reflections will cause some UCNs to go back into the He-II where they can become lost. For cryogenic foils, there is an added problem of frozen contamination build-up that can absorb or elastically scatter UCNs.

To model the non-idealized foil transmission losses, we use the measurements from Ref.~\onlinecite{Miranda1988} that saw a 85\% transmission through two $25\un{\mu m}$ thick PP foils cooled to $\sim 17\un{K}$. This transmission value was observed 20\un{mins} after their foils were cooled. They observed a transmission decrease over time suspected to be caused by frozen contamination build-up. However, the 85\% transmission can be assumed to be dominated by other losses. In their experiment, a stainless steel UCN storage bottle $U_{\rm SS} = 190\un{neV}$ with storage time constant of $\sim 20\un{s}$ was pre-filled in order to have a well-defined UCN spectrum. A valve on the bottle was then opened to allow UCNs to transmit through the foils and then reach a UCN detector. This was compared to the case of no foils to deduce the transmission probability through the foils alone. Due to the difference in geometries between this setup and our source design, there will be different bottling effects if UCNs are elastically scattered in the bulk of the foil. We performed a UCN trajectory simulation study modeling their setup and extracted a UCN elastic scattering mean-free-path $\lambda_{\rm scat} = 20\un{\mu m}$ in the bulk of the PP that reproduces their result. This parameter will be used later in our UCN trajectory simulations to model more realistic loss in the foil.

As observed in the studies of Ref.~\onlinecite{Miranda1988}, the limiting factor on the UCN transmission over time through cryogenic foils will most likely be the build-up of frozen contaminants on its surfaces. We envision exposing our foil to UV light emitted from behind a UCN-reflecting UV-transmitting window to desorb frozen contamination. The heat load from this will be small compared to the 100\un{W} load already on the He-II. We can also consider winding a superconducting coil around the foil to produce a magnetic field gradient that accelerates high-field seeking UCNs (similar to a Stern-Gerlach experiment) through the foil to further reduce losses.\cite{Masuda2014} This latter mode only accelerates 50\% of the UCNs. It is worth noting again here that our estimated extractable UCN density is less sensitive to losses in the transport system due to the ``no return'' approximation.

\subsection{Shielding of the extraction guide and extraction from the front\label{sec:shieldingExtractGuide}}

One aspect of the extraction design not fully explored is the heat load on the He-II that is inside the extraction guide. The source of heating on the He-II converter comes predominantly from fast neutrons (described in Sec.~\ref{sec:summaryMCNP}). We do not require UCN production in the guide, therefore a combination of borated-polyethylene and lead shielding can surround the guide to reduce the heat load there. To reduce fast neutrons by an order of magnitude, around 15\un{cm} thickness of shielding is required. In the UCN extraction geometries discussed so far, such neutron absorbing shielding could affect the CN flux in the 40~L He-II converter volume as well.

To reduce the guide shielding's impact on the CN flux on the 40~L He-II converter, there is the possibility of having the horizontal extraction guide on the opposite side of the tungsten target (i.e. it comes out towards the proton upstream direction). In this geometry, the UCN extraction properties are the same since when UCNs are produced from down-scattered CN, their initial velocity are essentially isotropic in space. Bends in the UCN guide after the He-II containment foil will be needed to bring the UCNs out of the proton beam's path. Such bends are needed in the other geometries as well in order to facilitate shielding of direct sight fast neutrons and gammas. This geometry will make the implementation of the rastered proton beam more difficult as it will need to be controlled so that it does not strike the heat exchanger and UCN guide. This will also reduce the heat distribution of the proton beam on the tungsten target. Another advantage of extracting UCNs from the proton beam upstream side is that the biological shielding can be reduced by $\sim 1\un{m}$. This is because the fast neutron radiation from the tungsten target is asymmetric with respect to the proton beam's momentum. These details will be considered as our design moves towards an engineered design.

\subsection{UCN trajectory simulations \label{sec:UCNtrajectorySim}}

The UCN trajectory tracking Monte-Carlo package developed in Ref.~\onlinecite{Holley2012} is used on the geometries discussed above to estimate the UCN extraction efficiency. UCNs are generated uniformly in the 40\un{L} He-II converter volume with velocity $v < 7.8\un{m\,s^{-1}}$ (equivalent to kinetic energy $E <U_{\rm ^{58}Ni}-U_{\textrm{He-II}} = 316.5\un{neV}$). When UCNs are produced they are expected to fill phase space with almost constant density so that they have isotropic initial velocity and have a spectrum ${\rm d}P_{\rm UCN}/{\rm d}v \propto v^2$ (or ${\rm d}P_{\rm UCN}/{\rm d}E \propto \sqrt{E}$). The initial direction and velocity of UCNs are generated according to these distributions and have their individual trajectories tracked in the presence of gravity. Losses due to upscattering and absorption are included, as well as neutron $\beta$-decay. Upon incidence on a material wall, the transverse energy of the UCN is calculated to determine if the UCN is reflected, transmitted, or lost, following the standard expressions in Refs.~\onlinecite{Golub1991,Ignatovich1990}. If reflection occurs, a Lambertian reflection kernel with a diffuse reflection probability $P_{\rm diffuse}$ is applied. UCNs that exit the final horizontal guide of the different geometries and have kinetic energies less than $U_{\rm ^{58}Ni}$ are counted as being successfully extracted. 

For the surfaces inside the 40~L converter volume and in the guides, the reflection loss factor used is $f = W/U  = 5\times 10^{-4}$, where $W$ is the imaginary part of the neutron optical potential.\cite{Golub1991} A nominal $P_{\rm diffuse} = 3\%$ is used for the surfaces. As described earlier in this section, non-specular reflections might be desirable inside the 40\un{L} He-II converter volume and the vertical column to the heat exchanger. For these cases, $P_{\rm diffuse}$ is increased for some regions in these studies. The UCN simulation detector is placed 30\un{cm} after the He-II containment foil for the horizontal near-foil geometry, and after 30\un{cm} of horizontal guide in the horizontal far-foil and vertical downward geometries. If a UCN reaches the UCN simulation detector, it is terminated and counted if its total kinetic energy is less than $U_{\rm ^{58}Ni}$ at the detector. The probability $\epsilon_{\rm sim}$ is given by the number counted by the detector divided by $N$, the number of UCNs generated which is typically $N= 10,000$.

In the simulations, we separate the contributions to the total loss probability $\Lambda_{\rm sim} = 1 - \epsilon_{\rm sim}$ in to losses (relative to the total number of generated UCNs) from: (1) He-II upscattering $\Lambda_{\rm He\textrm{-}II}$, (2) walls of the guiding system $\Lambda_{\rm wall}$, and (3) due to UCNs that reach the heat exchanger $\Lambda_{\rm ex}$. (i.e. $\Lambda_{\rm sim} \approx \Lambda_{\rm He\textrm{-}II} + \Lambda_{\rm wall} + \Lambda_{\rm ex}$.) There also exists some ``unphysical'' losses in the simulations, which are $0.7\%$ or less, due to, for example, small gaps at where the software attempts to smoothly join surfaces of different shapes. Next we summarize the key findings and optimizations of the UCN extraction geometries studied. We note that we begin with using idealized loss in an Al foil.

For typical horizontal near-foil and far-foil geometries $\Lambda_{\rm He\textrm{-}II}/\Lambda_{\rm sim} \sim 55\%$. For the vertical-downward extraction the He-II loss reduces to $\Lambda_{\rm He\textrm{-}II}/\Lambda_{\rm sim} \sim 35\%$. However, the gain in kinetic energy of $\sim 160\un{neV}$ from the vertical drop causes many more UCNs to be lost at the guides at the bottom such that the total extraction efficiency in the simulations $\epsilon_{\rm sim}(\textrm{downward})/\epsilon_{\rm sim}(\textrm{horizontal}) \sim 25\% - 40\%$ for the downward extraction versus the two horizontal geometries. Therefore, we do not see the downward extraction being a viable choice.

\begin{table}
\caption{Summary of steps taken to reach $\epsilon_{\rm tot\;single} = \epsilon_{\rm sim}\,\epsilon_{\rm grid} \epsilon_{\rm guide}= 26\%$ for the horizontal near-foil UCN extraction geometry for $T=1.6\un{K}$.}
\begin{center}
\begin{ruledtabular}
\begin{tabular}{c c }
Configuration & $\epsilon$ \\\noalign{\smallskip}\hline 
{\footnotesize Baseline (ideal Al foil, $P_{\rm diffuse} = 3\%$ everywhere)}: & 35\% {\footnotesize ($\epsilon_{\rm sim}$)} \\
{\footnotesize Add diffuse reflections in converter volume ($P_{\rm diffuse} = 50\%$):} & 43\% {\footnotesize ($\epsilon_{\rm sim}$)} \\
{\footnotesize Add diffuse reflections in vertical column ($P_{\rm diffuse} = 50\%$):} & 45\% {\footnotesize ($\epsilon_{\rm sim}$)} \\
{\footnotesize Switch from ideal Al foil (54\un{neV}) to ideal PP ($-8\un{neV}$):} & 53\% {\footnotesize ($\epsilon_{\rm sim}$)} \\
{\footnotesize Add more realistic PP elastic scattering ($\lambda_{\rm scat} = 20\un{\mu m}$):} &  36\% {\footnotesize ($\epsilon_{\rm sim}$)} \\ 
{\footnotesize Include PP foil support grid loss ($\epsilon_{\rm grid} = 90\%$):} & 32\% {\footnotesize ($\epsilon_{\rm sim}\epsilon_{\rm grid}$)} \\
{\footnotesize Include 4\un{m} guide loss to external volume ($\epsilon_{\rm guide} = 80\%$):} & {\bf 26\% {\footnotesize ($\epsilon_{\rm tot\;single}$)}} \\
\end{tabular}
\end{ruledtabular}
\end{center}
\label{tab:UCNtransport}
\end{table}

Non-specular reflections in the 40~L converter volume can help break-up otherwise long-lived glancing-reflection orbits in the cylindrical volume, allowing them to escape more quickly and not be lost in the He-II, which is the main loss mechanism for the horizontal geometries. As mentioned at the start of this section, implementing these reflections can be done by having geometric features (e.g. bumps) on the inner walls. However, programming in these features into the UCN transport code would be very cumbersome. Instead, to study the effect of increasing non-specular reflections, we will also use the Lambertian diffuse reflection kernel. Increasing the effective $P_{\rm diffuse}$ in the 40~L converter volume from 3\% to 50\% increases $\epsilon_{\rm sim}$ by $\sim 20-30\%$ for the horizontal near-foil and far-foil geometries. We find that increasing $P_{\rm diffuse}$ beyond 50\% does not produce much further improvement and thus will stick to using $P_{\rm diffuse} = 50\%$ in the 40~L converter volume. 
Another source of loss is the heat exchanger (see Sec.~\ref{sec:heatEx}). The primary way we attempt to reduce this loss is to deliberately induce non-specular reflections on the side walls of the vertical column. This allows more UCNs that enter the vertical column to return to the main horizontal UCN extraction path and away from the heat exchanger. However, some of these UCNs will also travel backwards and forwards in the vertical section and thus become lost due to the long path lengths in the He-II. When $P_{\rm diffuse}$ is increased from 3\% to 50\% in the vertical column, $\Lambda_{\rm sim}$ decreases from 9\% to 2\%, but the corresponding increase in $\Lambda_{\rm He\textrm{-}II}$ is from 52\% to 55\%. The net improvement in $\epsilon_{\rm sim}$ is therefore only $\sim 5\%$, a rather modest gain. We will use $P_{\rm diffuse} = 50\%$ in the vertical column. We also briefly studied placing a UCN reflecting orifice directly in front of heat exchanger. This also gave very modest gains in $\epsilon_{\rm sim}$ and would reduce the heat transport efficiency of the system, thus we do not consider this further.

The final total UCN extraction efficiencies $\epsilon_{\rm sim}$ from using an idealized 30\un{\mu m} Al foil and with the implementations of the non-specular reflections described previously is $\sim 45\%$ for the near-foil geometry and $\sim 30\%$ for the far-foil geometry. The UCN loss at the walls due to the drop in the far-foil geometry is large and does not compensate the reduction in Al foil losses. For the near-foil geometry, if the idealized Al foil is replaced with an idealized PP foil (see Sec.~\ref{sec:foil}) then $\epsilon_{\rm sim} \sim 53\%$ is reached. This improvement primarily comes from the reduced reflection at the foil. Considering there is the further added $\sim 10-20\un{neV}$ potential due to the oxide layer on an Al foil, we consider the PP foil to be a better choice. This conclusion is the same as that reached in the experimental studies of Ref.~\onlinecite{Miranda1988}.

We now add more realistic loss at the foil. Note that these effects should be present in both metal and polymer foils. As explained in Sec.~\ref{sec:foil}, based on an experiment that measured 85\% transmission through two $25\un{\mu m}$ PP foils at 17~K, we can extract an elastic scattering mean-free-path off inhomogeneities in the bulk of the foil to be $\lambda_{\rm scat} \approx 20\un{\mu m}$. This produces $\epsilon_{\rm sim} = 36\%$.

For comparison, if the simulations are run without any foil (i.e. setting the neutron potential to 0\un{neV} without any absorption or scattering mechanism), $\epsilon_{\rm sim} = 53\%$. Therefore, the transmission factor caused by the foil is effectively ($36\%/53\% =$) 68\%. Table~\ref{tab:UCNtransport} summarizes the steps taken to reach this $\epsilon_{\rm sim}$. Table~\ref{tab:UCNtransportStudies} gives the impact on $\epsilon_{\rm sim}$ in the final horizontal near-foil extraction geometry when scanning the diameter of the horizontal UCN guide $D_{\rm guide}$, $P_{\rm diffuse}$ in the 40~L converter volume, height of the vertical column $h_{\rm column}$, and $P_{\rm diffuse}$ in the vertical column's side walls.

\begin{table}
\footnotesize
\caption{The impact on $\epsilon_{\rm sim}$ when scanning some parameters of the UCN extraction system in the final horizontal near-foil geometry. The realistic PP foil loss ($\lambda_{\rm scat} = 20\un{\mu m}$) are used in all the results. The nominal values of the parameters are: diameter of the horizontal UCN extraction guide diameter $D_{\rm guide} = 18\un{cm}$, non-specular reflections in the 40\un{L} He-II converter volume $P_{\rm diffuse} = 50\%$, height of the vertical column to the heat exchanger $h_{\rm column} = 1.0\un{m}$, non-specular reflections in the vertical column $P_{\rm diffuse} = 50\%$. The parameters are at these nominal values unless they are being scanned.  The $1\sigma$ statistical error bar in each $\epsilon_{\rm sim}$ value is around $\pm 0.2 \%$. The value in bold is the final $\epsilon_{\rm sim}$ value used.}
\label{tab:UCNtransportStudies}
\vspace{2mm}
(a) scanning diameter of horizontal UCN extraction guide $D_{\rm guide}$:
\vspace{1.5mm}
\begin{ruledtabular}
\begin{tabular}{c | c c c }
$D_{\rm guide}$ [cm]       &18      & 19     & 20    \\ \hline
$\epsilon_{\rm sim}$  & {\bf 36.4 \%} & 38.4 \% & 41.0 \% \\
\end{tabular}
\end{ruledtabular}
\vspace{4mm}
(b) scanning $P_{\rm diffuse}$ in the 40\un{L} He-II converter volume:
\vspace{1.5mm}
\begin{ruledtabular}
\begin{tabular}{c | c c c c c c c }
$P_{\rm diffuse}$ [\%]       &3      & 10     & 20    & 30   & 40    & 50            & 60 \\ \hline
$\epsilon_{\rm sim}$  &31.7 \% & 34.1 \% & 34.9 \% & 35.7 \% & 36.1 \% & {\bf 36.2 \%}  & 36.6 \% \\
\end{tabular}
\end{ruledtabular}
\vspace{4mm}
(c) scanning vertical column's height $h_{\rm column}$ and $P_{\rm diffuse}$ on its side walls:
\vspace{1.5mm}
\begin{ruledtabular}
\begin{tabular}{c c | c c c c c c }
          &  &\multicolumn{6}{c}{$h_{\rm column}$ [m]}\\
          &  & 0.2 & 0.4 & 0.6 & 0.8 & 1.0 & 1.2 \\ \hline
\multirow{7}{*}{\rotatebox[origin=c]{90}{$P_{\rm diffuse}$ [ \%]}} & 3  &  28.6 \% & 30.0 \% & 31.3 \% & 32.6 \% & 33.6 \% & 34.4 \% \\
&10 & 28.8 \% & 30.3 \% & 32.0 \% & 33.1 \% & 33.7 \% & 34.8 \% \\
&20 & 29.4 \% & 31.2 \% & 32.6 \% & 34.0 \% & 34.8 \% & 35.6 \% \\
&30 & 29.7 \% & 31.6 \% & 33.4 \% & 34.2 \% & 35.3 \% & 35.9 \% \\
&40 & 30.2 \% & 32.7 \% & 33.8 \% & 35.2 \% & 36.0 \% & 36.3 \% \\
&50 & 31.0 \% & 32.9 \% & 34.3 \% & 35.5 \% & {\bf 36.3 \%} & 37.0 \% \\
&60 & 30.9 \% & 33.6 \% & 34.9 \% & 36.2 \% & 36.5 \% & 37.3 \% \\
\end{tabular}
\end{ruledtabular}
\end{table}

We also include the transmission loss due to the foil's support grid. A grid with physical opening of 90\% can be achieved. However, if the grid is made from a good UCN reflector with a high potential (e.g. $^{58}$Ni, natural Ni, Be, or BeO) then UCNs that strike the grid and go back into the He-II still have some probability to make it back to the foil. We take the UCN transmission to be $\epsilon_{\rm grid} = 90\%$, which is a conservative value. We then reach the extraction efficiency out of our source to be $\epsilon_{\rm sim}\,\epsilon_{\rm grid} \approx 31\%$.

As discussed earlier, the UCN transport efficiency for an additional 4\un{m} of UCN guides to reach an external experiment based on measurement values is $\epsilon_{\rm guide} = 80\%$. Including this loss, we reach $\epsilon_{\rm tot\;single} = \epsilon_{\rm sim}\,\epsilon_{\rm grid}\,\epsilon_{\rm guide} = 26\%$. This will be the extraction efficiency we use to determine the useful UCN current from our source design.

\subsection{UCN densities \label{sec:UCNdensities}}

Here, we provide UCN density figures relevant to our design. Due to differences in source geometries, operational modes, and experimental volumes needed to be filled, these densities only provide an approximate comparison between different sources. We first estimate the UCN density that exists inside the source, and then attempt to assess the ``useful'' density in an experimental volume relatively near the source. 

The density produced inside the source sets the maximum UCN density that can be extracted from the source. This density is given by $\rho_{\rm source} = P_{\rm UCN}\,\tau_{\rm source}/V_{\rm source}$, where $\tau_{\rm source}$ is the effective lifetime of UCNs inside the source and $V_{\rm source}$ is the volume of the source. From Eq.~(\ref{eq:tauup}), the upscattering rate in He-II is 0.29\un{s^{-1}} at 1.6\un{K}. This can be scaled by $\Lambda_{\rm He\textrm{-}II}/\Lambda_{\rm tot} \approx 70\%$ to give $\tau_{\rm source} \approx 2.4\un{s}$. Then we arrive at $\rho_{\rm source} \approx 5 \times 10^{4} \un{UCN\,cm^{-3}}$, where we take the total source volume to be 90\un{L} by adding the 40\un{L} converter volume to the volume of the guides and the heat exchanger's vertical column.


The ``useful'' UCN density is what might be achievable inside a UCN bottle located at the nearest experimentally accessible position to the source. The UCN current at a position $\sim 5\un{m}$ away from the tungsten target from the previous section is expected to be $R_{\rm use} = P_{\rm UCN}\,\epsilon_{\rm tot\;single} =  5 \times 10^{8}\un{UCN\,s^{-1}}$ out of the 18~cm diameter UCN guide.

If the bottle's opening diameter $D_{\rm open}$ is matched to this, then we can assume that the rate of UCNs entering the bottle $R_{\rm in} = R_{\rm use}$. Once UCNs are inside the bottle, they can be lost via: (1) UCNs leaving the bottle through the opening again, (2) losses at the walls, or (3) neutron $\beta$-decay. At equilibrium, $R_{\rm in}$ is equal to the total rate of UCN loss in the bottle. If kinetic theory is assumed (see Sec.~\ref{sec:UCNdensities}), which can be established quickly if the bottle walls have non-specular reflection features, then this equilibrium condition becomes:
\begin{equation} \label{eq:equilUCNdensity}
R_{\rm in} = \rho_{\rm bottle}\left( \frac{\pi \bar{v} D_{\rm open}^2}{16} + \frac{\bar{\mu}\bar{v}A_{\rm bottle}}{4} + \frac{V_{\rm bottle}}{\tau_\beta}  \right)
\end{equation}
where $\rho_{\rm bottle}$ is the equilibrium UCN density inside the bottle, $\bar{v}$ is the average UCN velocity, $V_{\rm bottle}$ and $A_{\rm bottle}$ are the volume and wall surface area of the bottle, and $\bar{\mu}$ is the average loss probability per reflection on the walls. Note the dependence on the $V_{\rm bottle}$ and $A_{\rm bottle}$.

In Eq.~\ref{eq:equilUCNdensity}, the assumption that once a UCN leaves the bottle it has zero probability of returning again is made. This we call the ``no return'' (or ``single passage'') approximation. We therefore don't claim gains in $R_{\rm in}$ due to UCNs exiting the bottle, then being stored in the guides or in the source, and then returning back to the external volume. This is a conservative approximation, but a current-optimized source operates somewhat near this condition. 

By making this approximation, our useful UCN density estimates will depend only linearly on the UCN transport loss between the source and the bottle since $R_{\rm in} = R_{\rm use} \propto \epsilon_{\rm tot\,single}$. For UCN filling calculations that model a combined system containing the source, guides and external bottle, UCNs will make several passages from source to bottle before equilibrium is reached. This is where a longer storage time in the source (i.e. moving towards a density-optimized source) is advantageous. If $n$ is the average number of passages between source and bottle UCNs make to approach equilibrium, then the extracted density would scale by $\sim (\epsilon_{\rm tot\,single})^n$. Therefore, buildup of frozen contamination on the guides or foils, which would decrease $\epsilon_{\rm tot\,single}$ over time, will have a larger impact.

To illustrate how the filling from our source scales with the bottle size, Table~\ref{tab:sourceDensity} shows calculations for a spherical bottle. The other values used are: $\tau_\beta = 880\un{s}$, inner coating of the bottle is $^{58}$Ni matched to the guides; $\bar{v} = 7\un{m\, s^{-1}}$, corresponding to $255\un{neV}$ kinetic energy, which is a conservative value; and $\bar{\mu} = 5 \times 10^{-4}$. The decrease of $\rho_{\rm bottle}$ for very large $V_{\rm bottle}$ is small. This is particularly advantageous for experiments with very large volumes since the \emph{total number} of UCNs is given by $\rho_{\rm bottle}\, V_{\rm bottle}$.

There is only a weak-dependence of $\rho_{\rm bottle}$ on $\bar{\mu}$ unless the volume is large. For instance, if $\bar{\mu} = 10 \times 10^{-4}$, then $\rho_{\rm bottle}$ is only reduced by 2\% if $V_{\rm bottle} = 50\un{L}$ while it is reduced by 20\% if $V_{\rm bottle} = 5000\un{L}$. Also shown in the table are the UCN storage time constant for the open bottle $\tau_{\rm bottle}$ (i.e. it includes the time constant for UCNs to escape the open bottle). Therefore, $\tau_{\rm bottle}$ is also the build-up time constant of UCN density during filling. We note that these results will be the same for any source design that operates close to the ``no return'' approximation.

\section{Summary and discussions}

In this paper we have presented our Inverse Geometry spallation-target, current-optimized, He-II UCN source. We proposed to raster-scan the proton beam around a cylindrical tungsten spallation to reduce the time-averaged power density on the target to allow water edge-cooling. Our cylindrical pre-moderator, moderator, and He-II converter volumes are then nested radially inwards at smaller diameters. The overall design goal was to maximize the time-averaged 1\un{meV} CN flux (the primary UCN production energy) in the 40~L He-II converter per He-II heat load and per proton. The boundary conditions we set for this optimization were maximum 100~W He-II heat load, achievable with sub-cooled He-II technology at the 1.6~K temperature range we're considering, and maximum 1~MW proton beam power (at fixed 800\un{MeV} proton energy), consistent with existing neutron spallation sources, in particular at LANSCE.

In Sec.~\ref{sec:baselineDesign} we presented our baseline Inverse Geometry design that yielded a total UCN production rate of $P_{\rm UCN} = 0.2 \times 10^{9}\un{UCN\,s^{-1}}$. This is already 1.7 times higher than the Mark-3-inspired UCN design (described in Sec.~\ref{sec:lujan}). This improvement came predominantly due to a reduced He-II heat load per proton. In Sec.~\ref{sec:modification} we modified our baseline Inverse Geometry design by changing the moderator and He-II canister material from aluminum to beryllium, adding a D$_2$O pre-moderator, and changing the moderator from LH$_2$ to LD$_2$, to obtain $P_{\rm UCN} = 0.7  \times 10^{9}\un{UCN\,s^{-1}}$. This improvement came from a combination of reduced He-II heat load per proton (2.3 times gain) and an increase in CN flux per proton (1.6 times gain).

In Sec.~\ref{sec:optimization} we performed MCNP6.1 simulations with smaller neutron energy bin sizes to a lower CN energy range. With these simulations, we optimized the LD$_2$ thickness, D$_2$O thickness, and tungsten target position. After our optimization, compared to the Mark-3-inspired UCN design we obtained a geometry that provided a factor of 2 gain in the CN flux in the He-II converter per proton and a factor of 8 reduction in the He-II heat load per proton. The CN temperature decreased, with the peak in the differential CN flux dropping from 2.6\un{meV} to 1.7\un{meV}. The He-II heat load was found to be dominated by fast neutrons. It was shown that by moving the tungsten target further downstream (along the proton beam direction) relative to the He-II converter volume, the heating due to fast neutrons dropped more quickly than the CN flux. This allowed higher proton powers (which resulted in an increased $P_{\rm UCN}$) until at some point the proton power reached our 1\un{MW} restriction. This led to $P_{\rm UCN} = 2.1 \times 10^{9}\un{UCN\,s^{-1}}$, the value quoted in the abstract, for our design boundary conditions. These results were summarized in Table~\ref{tab:OptimizeSummary}. Next we studied effects that could negatively impact this UCN production rate. 

We explored the effects on $P_{\rm UCN}$ associated with: (1) the quantum-fluid behavior of He-II that is not incorporated in the MCNP scattering kernel, (2) the warmer He-II temperature than in typical UCN sources, and (3) pressurizing the He-II to $1\un{bar}$, which desirably reduces gas bubble formation. Combining these effects, we estimated that $P_{\rm UCN} = 1.8  \times 10^{9}\un{UCN\,s^{-1}}$ is possible. This is the final $P_{\rm UCN}$ value we arrived at in this work. Our next task was studying how to efficiently extract the produced UCNs from the source and to an experiment 5\un{m} away from the target.

In Sec.~\ref{sec:UCNextraction} we studied the UCN extraction for different extraction geometries with UCN trajectory tracking simulations. We showed that a ``horizontal near-foil'' extraction geometry (see Fig.~\ref{fig:inverseGeometry3D}) performed the best. The design of the UCN extraction system took into account the requirements for the 100\un{W} heat load removal from the 40\un{L} He-II converter volume to the heat exchanger. This requires a He-II filled conduit $>18\un{cm}$ in diameter for a 2\un{m} distance. We settled on a geometry with a 18\un{cm} diameter horizontal He-II filled conduit connected to the converter volume. This 1.3\un{m} long horizontal conduit serves as the primary UCN transport path as well as the first section of the heat removal path. After 1\un{m} along the horizontal conduit, a T-piece is used to split off to He-II filled vertical column (also 18\un{cm} diameter) leading to the heat exchanger. This vertical column serves as the second section of the heat removal path. The height of the vertical column (restricted to a maximum of 1\un{m} for thermal transport reasons) and deliberately inducing non-specular reflections on its side walls were studied to reduce loss of UCNs going up this column and reaching the lossy heat exchanger. Non-specular reflections in the 40\un{L} converter volume were also studied to improve the escape times of the UCNs.

At the end of the horizontal conduit is a He-II containment foil that needs to allow high UCN transmission. We found that a polypropylene foil, which has a $U_{\rm PP} = -8\un{neV}$, offers higher transmission than an aluminum foil $U_{\rm Al} = 54\un{neV}$ (plus another $10-20\un{neV}$ due to an oxide layer). The detector used in our UCN trajectory simulations, which terminates UCNs that reach it, is placed 30\un{cm} downstream from this foil. The probability that a generated UCN reaches this detector is given by $\epsilon_{\rm sim}$. It is found that $\epsilon_{\rm sim} = 53\%$ with no foil in place, and $\epsilon_{\rm sim} = 36\%$ with the PP foil that had more realistic scattering mechanisms included (more later). Therefore, the effective transmission factor of the PP foil is 68\%.

The key parameter we extracted is the total single passage transmission factor for a UCN to reach a position 5\un{m} away from the tungsten target, which is denoted by $\epsilon_{\rm tot\,single} = \epsilon_{\rm sim}\,\epsilon_{\rm grid}\,\epsilon_{\rm guide}$. By using $\epsilon_{\rm sim} = 36\%$ from above, $\epsilon_{\rm grid} = 90\%$ the transmission factor of the support grid for the 18\un{cm} diameter PP foil, and $\epsilon_{\rm guide} =80\%$ deduced from other measurements, we arrived at $\epsilon_{\rm tot\,single} = 26\%$.

We note that the UCN upscattering rate in the He-II $\tau_{\rm up}$, which makes up around half the UCN loss, has been measured at different temperatures by several experiments (e.g. see Refs.~\onlinecite{Ahmed2019,Golub1996,Leung2016,Piegsa2014,Zimmer2007a} and references therein). Therefore, its value is well known. The loss at the guides inside the source is small compared to the He-II loss and thus have little impact. We also paid close attention to the next largest loss mechanism that is caused by scattering at the foil. We implemented not only idealized foil material properties, but further added a loss mechanism due to elastic UCN scattering off inhomogeneities in the bulk of the foil material that is expected to be the dominant cause of reduction in transmission.\cite{Pokotilovski2016} The elastic scattering mean-free-path $\lambda_{\rm scat}$ was determined from an experimental study of UCN transmission through cryogenic foils, which measured a 85\% transmission factor through two $25\un{\mu m}$ PP foils from stainless steel bottled UCNs.\cite{Miranda1988} The implementation of the PP foil with this ``more realistic'' mechanism scaled for our design produced an effective transmission factor due to a $30\un{\mu m}$ PP foil of $\sim 68\%$. This value appears reasonable, given that UCN transmission values measured in Al foils 100\un{um} thick have been observed\cite{Atchison2009} to be $\sim 83\%$. The $\epsilon_{\rm guide}$ value is based on experimentally measured values with $^{58}$NiMo guides with 8\un{cm} diameter in Ref.~\onlinecite{Serebrov2017}. Scaling this linearly for our 18\un{cm} guides, since the guide transmission loss is dominated by the number of wall reflections, we arrived at $\epsilon_{\rm guide} = 80\%$. Therefore, we note that all the primary transmission reduction mechanisms are well accounted for when arriving at $\epsilon_{\rm tot\,single}$.

To sustain high transmission through a cryogenic foil over time, we will need research to develop techniques for periodic removal of frozen contaminants on the foil. We note that this is, or will be, a problem also for several cryogenic UCN sources.\cite{Anghel2009,Becker2015,Atchison2009,Martin2013,Masuda2014,Picker2017}

The useful UCN current delivered to an experiment is defined to be at a position 5\un{m} from the target to give sufficient space for the required radiation shielding. This current is given by $R_{\rm use} = \epsilon_{\rm tot\;single}P_{\rm UCN} = 5  \times 10^{8}\un{UCN\,s^{-1}}$ out of a $18\,{\rm cm}$ diameter guide. This extracted rate is roughly an order of magnitude higher than the best proposed UCN sources so far, and is at least three orders of magnitude higher than existing sources. Such a rate would open a window to new exciting experiments with UCN that were not possible previously. We will continue to explore the possibility of using advanced neutron moderators, such as triphenylmethane,\cite{Young2014} and high albedo CN reflectors, such as diamond nanoparticles,\cite{Cubitt2010} to boost $P_{\rm UCN}$.

In Sec.~\ref{sec:UCNdensities} we provided UCN density figures for our source design. Firstly, the saturation UCN density inside the source was estimated to be approximately $\sim 5 \times 10^{4}\un{cm^{-3}}$, with a survival time of UCNs in the source $\tau_{\rm source} \approx 2\un{s}$. This density is the upper limit of what can be extracted from the source. Secondly, kinetic theory was used to describe the filling of an external volume using $R_{\rm use}$ at a location 5\un{m} away from the tungsten target. The equilibrium accumulated UCN density in an external bottle was found to be $\rho_{\rm bottle} \approx 1.1\times 10^{4}\un{cm^{-3}}$ for volumes $V_{\rm bottle} < 1000\un{L}$ (see Table~\ref{tab:sourceDensity}) under the conservative ``no return'' (or ``single passage'') approximation used to model the bottle filling from our current-optimized source.  

The equilibrium density $\rho_{\rm bottle}$ decreases slowly with $V_{\rm bottle}$. This is characteristic of a current-optimized source, which is particularly advantageous for filling experiments with large volumes. It is also worth noting that density estimates under our ``no return'' approximation, although conservative, are less effected by losses in the UCN transport system.


For context, the He-II source of Refs.~\onlinecite{Serebrov2009b,Serebrov2011,Serebrov2015a,Serebrov2018} provides an interesting comparison to our source design. Our design has approximately 10 times higher UCN current, however, the lifetime in our source is around 10 times shorter due to our warmer He-II temperature. Therefore, the UCN densities deliverable to small experimental volumes end up being approximately the same. However, the decrease in the UCN density for larger volume sizes will be slower in our source. Therefore, our design is particularly advantageous for experiments that can use large experimental volumes as it enables a very large total number of UCNs to be obtained.

Another interesting comparison to provide a broader context of our source design is to consider placing a thin layer\cite{Golub1983} of sD$_2$, 2~L in total, frozen onto the inner side and end walls of the same 40\un{L} converter volume instead of He-II. A rough estimate for the same 1~MW proton beam power from our differential CN flux gives a decrease in $P_{\rm UCN}$ by approximately a factor of 20. The heating of the sD$_2$ is not considered here. While the UCN production cross-section in sD$_2$ averaged over the $<10\un{meV}$ CN energy range is larger than for He-II, our source design is optimized to produce a high flux of low energy CN in the converter volume, which is better suited for He-II. With sD$_2$, the cooling requirements of the crystal under high heat loads will need to be studied, and the UCN extraction efficiency will decrease due to the high losses when UCNs exit or re-enter the sD$_2$ in the converter volume. An optimized sD$_2$ source for high proton beam powers should be the topic of a future paper.

We point out that the current design presented is a so-called ``physics model''. For the MCNPX simulations, it was shown in Ref.~\onlinecite{Muhrer2000} for the case of the Lujan Center target, the fluxes calculated using a so-called ``engineering model'', which is based on the actual engineering design, predicts up to 30\% less flux than the original physics model. The accuracy of the engineering model was later shown to be reasonable.\cite{Muhrer2004} Conventional knowledge suggests that the flux reduction between a physics model and an engineering model can actually be up to 50\%. However, this is still significantly higher than what is achievable with other current and proposed UCN sources world-wide. In our study, we have identified the need for further research into state-of-the-art neutron moderator and reflector materials, and high-performance cryogenic UCN transmission foil systems and other components.

\begin{acknowledgments}

We would like to thank Ruslan Nagimov and Steven Van Sciver for their cryogenic insights. This work was supported by Readiness in Technical Base and Facilities (RTBF), which is funded by the U.S. Department of Energy's Office of National Nuclear Security Administration (NNSA). It has benefited from the use of the Manuel Lujan, Jr. Neutron Scattering Center at Los Alamos National Laboratory, which is funded by the Department of Energy's Office of Basic Energy Sciences. Los Alamos National Laboratory is operated by Triad National Security, LLC for the US Department of Energy's NNSA. In addition, this work has been supported by the National Science Foundation under grants PHY1005233, PHY1615153, and PHY1307426, and Department of Energy's Office of Nuclear Physics under grant DE-FG02-97ER41042, and contract No.~89233218CNA000001 under proposal 2020LANLEEDM.

\end{acknowledgments}

\nocite{*}
\bibliography{LHeSpallationUCNsource.bib}

\begin{thebibliography}{95}%
\makeatletter
\providecommand \@ifxundefined [1]{%
 \@ifx{#1\undefined}
}%
\providecommand \@ifnum [1]{%
 \ifnum #1\expandafter \@firstoftwo
 \else \expandafter \@secondoftwo
 \fi
}%
\providecommand \@ifx [1]{%
 \ifx #1\expandafter \@firstoftwo
 \else \expandafter \@secondoftwo
 \fi
}%
\providecommand \natexlab [1]{#1}%
\providecommand \enquote  [1]{``#1''}%
\providecommand \bibnamefont  [1]{#1}%
\providecommand \bibfnamefont [1]{#1}%
\providecommand \citenamefont [1]{#1}%
\providecommand \href@noop [0]{\@secondoftwo}%
\providecommand \href [0]{\begingroup \@sanitize@url \@href}%
\providecommand \@href[1]{\@@startlink{#1}\@@href}%
\providecommand \@@href[1]{\endgroup#1\@@endlink}%
\providecommand \@sanitize@url [0]{\catcode `\\12\catcode `\$12\catcode
  `\&12\catcode `\#12\catcode `\^12\catcode `\_12\catcode `\%12\relax}%
\providecommand \@@startlink[1]{}%
\providecommand \@@endlink[0]{}%
\providecommand \url  [0]{\begingroup\@sanitize@url \@url }%
\providecommand \@url [1]{\endgroup\@href {#1}{\urlprefix }}%
\providecommand \urlprefix  [0]{URL }%
\providecommand \Eprint [0]{\href }%
\providecommand \doibase [0]{http://dx.doi.org/}%
\providecommand \selectlanguage [0]{\@gobble}%
\providecommand \bibinfo  [0]{\@secondoftwo}%
\providecommand \bibfield  [0]{\@secondoftwo}%
\providecommand \translation [1]{[#1]}%
\providecommand \BibitemOpen [0]{}%
\providecommand \bibitemStop [0]{}%
\providecommand \bibitemNoStop [0]{.\EOS\space}%
\providecommand \EOS [0]{\spacefactor3000\relax}%
\providecommand \BibitemShut  [1]{\csname bibitem#1\endcsname}%
\let\auto@bib@innerbib\@empty
\bibitem [{\citenamefont {Nico}\ and\ \citenamefont {Snow}(2005)}]{Nico2005}%
  \BibitemOpen
  \bibfield  {author} {\bibinfo {author} {\bibfnamefont {J.~S.}\ \bibnamefont
  {Nico}}\ and\ \bibinfo {author} {\bibfnamefont {W.~M.}\ \bibnamefont
  {Snow}},\ }\href {\doibase 10.1146/annurev.nucl.55.090704.151611} {\bibfield
  {journal} {\bibinfo  {journal} {Annual Review of Nuclear and Particle
  Science}\ }\textbf {\bibinfo {volume} {55}},\ \bibinfo {pages} {27} (\bibinfo
  {year} {2005})}\BibitemShut {NoStop}%
\bibitem [{\citenamefont {Abele}(2008)}]{Abele2008a}%
  \BibitemOpen
  \bibfield  {author} {\bibinfo {author} {\bibfnamefont {H.}~\bibnamefont
  {Abele}},\ }\href@noop {} {\bibfield  {journal} {\bibinfo  {journal}
  {Progress in Particle and Nuclear Physics}\ }\textbf {\bibinfo {volume}
  {60}},\ \bibinfo {pages} {1} (\bibinfo {year} {2008})}\BibitemShut {NoStop}%
\bibitem [{\citenamefont {Dubbers}\ and\ \citenamefont
  {Schmidt}(2011)}]{Dubbers2011}%
  \BibitemOpen
  \bibfield  {author} {\bibinfo {author} {\bibfnamefont {D.}~\bibnamefont
  {Dubbers}}\ and\ \bibinfo {author} {\bibfnamefont {M.~G.}\ \bibnamefont
  {Schmidt}},\ }\href@noop {} {\bibfield  {journal} {\bibinfo  {journal}
  {Reviews of Modern Physics}\ }\textbf {\bibinfo {volume} {83}},\ \bibinfo
  {pages} {1111} (\bibinfo {year} {2011})}\BibitemShut {NoStop}%
\bibitem [{\citenamefont {Golub}(1996)}]{Golub1996}%
  \BibitemOpen
  \bibfield  {author} {\bibinfo {author} {\bibfnamefont {R.}~\bibnamefont
  {Golub}},\ }\href {\doibase 10.1103/RevModPhys.68.329} {\bibfield  {journal}
  {\bibinfo  {journal} {Reviews of Modern Physics}\ }\textbf {\bibinfo {volume}
  {68}},\ \bibinfo {pages} {329} (\bibinfo {year} {1996})}\BibitemShut
  {NoStop}%
\bibitem [{\citenamefont {Wietfeldt}\ and\ \citenamefont
  {Greene}(2011)}]{Wietfeldt2011}%
  \BibitemOpen
  \bibfield  {author} {\bibinfo {author} {\bibfnamefont {F.~E.}\ \bibnamefont
  {Wietfeldt}}\ and\ \bibinfo {author} {\bibfnamefont {G.~L.}\ \bibnamefont
  {Greene}},\ }\href@noop {} {\bibfield  {journal} {\bibinfo  {journal}
  {Reviews of Modern Physics}\ }\textbf {\bibinfo {volume} {83}},\ \bibinfo
  {pages} {1173} (\bibinfo {year} {2011})}\BibitemShut {NoStop}%
\bibitem [{\citenamefont {Nico}(2009)}]{Nico2009}%
  \BibitemOpen
  \bibfield  {author} {\bibinfo {author} {\bibfnamefont {J.~S.}\ \bibnamefont
  {Nico}},\ }\href@noop {} {\bibfield  {journal} {\bibinfo  {journal} {Journal
  of Physics G: Nuclear and Particle Physics}\ }\textbf {\bibinfo {volume}
  {36}},\ \bibinfo {pages} {104001} (\bibinfo {year} {2009})}\BibitemShut
  {NoStop}%
\bibitem [{\citenamefont {Young}\ \emph
  {et~al.}(2014{\natexlab{a}})\citenamefont {Young}, \citenamefont {Clayton},
  \citenamefont {Filippone}, \citenamefont {Geltenbort}, \citenamefont {Ito},
  \citenamefont {Liu}, \citenamefont {Makela}, \citenamefont {Morris},
  \citenamefont {Plaster}, \citenamefont {Saunders}, \citenamefont {Seestrom},\
  and\ \citenamefont {Vogelaar}}]{Young2014b}%
  \BibitemOpen
  \bibfield  {author} {\bibinfo {author} {\bibfnamefont {A.~R.}\ \bibnamefont
  {Young}}, \bibinfo {author} {\bibfnamefont {S.}~\bibnamefont {Clayton}},
  \bibinfo {author} {\bibfnamefont {B.~W.}\ \bibnamefont {Filippone}}, \bibinfo
  {author} {\bibfnamefont {P.}~\bibnamefont {Geltenbort}}, \bibinfo {author}
  {\bibfnamefont {T.~M.}\ \bibnamefont {Ito}}, \bibinfo {author} {\bibfnamefont
  {C.-Y.}\ \bibnamefont {Liu}}, \bibinfo {author} {\bibfnamefont
  {M.}~\bibnamefont {Makela}}, \bibinfo {author} {\bibfnamefont {C.~L.}\
  \bibnamefont {Morris}}, \bibinfo {author} {\bibfnamefont {B.}~\bibnamefont
  {Plaster}}, \bibinfo {author} {\bibfnamefont {A.}~\bibnamefont {Saunders}},
  \bibinfo {author} {\bibfnamefont {S.~J.}\ \bibnamefont {Seestrom}}, \ and\
  \bibinfo {author} {\bibfnamefont {R.~B.}\ \bibnamefont {Vogelaar}},\ }\href
  {http://stacks.iop.org/0954-3899/41/i=11/a=114007} {\bibfield  {journal}
  {\bibinfo  {journal} {Journal of Physics G}\ }\textbf {\bibinfo {volume}
  {41}},\ \bibinfo {pages} {114007} (\bibinfo {year}
  {2014}{\natexlab{a}})}\BibitemShut {NoStop}%
\bibitem [{\citenamefont {Lamoreaux}\ and\ \citenamefont
  {Golub}(2009)}]{Lamoreaux2009}%
  \BibitemOpen
  \bibfield  {author} {\bibinfo {author} {\bibfnamefont {S.~K.}\ \bibnamefont
  {Lamoreaux}}\ and\ \bibinfo {author} {\bibfnamefont {R.}~\bibnamefont
  {Golub}},\ }\href@noop {} {\bibfield  {journal} {\bibinfo  {journal} {Journal
  of Physics G}\ }\textbf {\bibinfo {volume} {36}},\ \bibinfo {pages} {104002}
  (\bibinfo {year} {2009})}\BibitemShut {NoStop}%
\bibitem [{\citenamefont {Chupp}\ \emph {et~al.}(2019)\citenamefont {Chupp},
  \citenamefont {Fierlinger}, \citenamefont {Ramsey-Musolf},\ and\
  \citenamefont {Singh}}]{Chupp2019}%
  \BibitemOpen
  \bibfield  {author} {\bibinfo {author} {\bibfnamefont {T.~E.}\ \bibnamefont
  {Chupp}}, \bibinfo {author} {\bibfnamefont {P.}~\bibnamefont {Fierlinger}},
  \bibinfo {author} {\bibfnamefont {M.~J.}\ \bibnamefont {Ramsey-Musolf}}, \
  and\ \bibinfo {author} {\bibfnamefont {J.~T.}\ \bibnamefont {Singh}},\ }\href
  {\doibase 10.1103/RevModPhys.91.015001} {\bibfield  {journal} {\bibinfo
  {journal} {Reviews of Modern Physics}\ }\textbf {\bibinfo {volume} {91}},\
  \bibinfo {pages} {015001} (\bibinfo {year} {2019})}\BibitemShut {NoStop}%
\bibitem [{\citenamefont {Nesvizhevsky}\ \emph {et~al.}(2003)\citenamefont
  {Nesvizhevsky}, \citenamefont {B\"orner}, \citenamefont {Gagarski},
  \citenamefont {Petoukhov}, \citenamefont {Petrov}, \citenamefont {Abele},
  \citenamefont {Bae\ss{}ler}, \citenamefont {Divkovic}, \citenamefont
  {Rue\ss{}}, \citenamefont {St\"oferle}, \citenamefont {Westphal},
  \citenamefont {Strelkov}, \citenamefont {Protasov},\ and\ \citenamefont
  {Voronin}}]{Nesvizhevsky2003}%
  \BibitemOpen
  \bibfield  {author} {\bibinfo {author} {\bibfnamefont {V.~V.}\ \bibnamefont
  {Nesvizhevsky}}, \bibinfo {author} {\bibfnamefont {H.~G.}\ \bibnamefont
  {B\"orner}}, \bibinfo {author} {\bibfnamefont {A.~M.}\ \bibnamefont
  {Gagarski}}, \bibinfo {author} {\bibfnamefont {A.~K.}\ \bibnamefont
  {Petoukhov}}, \bibinfo {author} {\bibfnamefont {G.~A.}\ \bibnamefont
  {Petrov}}, \bibinfo {author} {\bibfnamefont {H.}~\bibnamefont {Abele}},
  \bibinfo {author} {\bibfnamefont {S.}~\bibnamefont {Bae\ss{}ler}}, \bibinfo
  {author} {\bibfnamefont {G.}~\bibnamefont {Divkovic}}, \bibinfo {author}
  {\bibfnamefont {F.~J.}\ \bibnamefont {Rue\ss{}}}, \bibinfo {author}
  {\bibfnamefont {T.}~\bibnamefont {St\"oferle}}, \bibinfo {author}
  {\bibfnamefont {A.}~\bibnamefont {Westphal}}, \bibinfo {author}
  {\bibfnamefont {A.~V.}\ \bibnamefont {Strelkov}}, \bibinfo {author}
  {\bibfnamefont {K.~V.}\ \bibnamefont {Protasov}}, \ and\ \bibinfo {author}
  {\bibfnamefont {A.~Y.}\ \bibnamefont {Voronin}},\ }\href {\doibase
  10.1103/PhysRevD.67.102002} {\bibfield  {journal} {\bibinfo  {journal}
  {Physical Review D}\ }\textbf {\bibinfo {volume} {67}},\ \bibinfo {pages}
  {102002} (\bibinfo {year} {2003})}\BibitemShut {NoStop}%
\bibitem [{\citenamefont {Bae\ss{}ler}\ \emph {et~al.}(2007)\citenamefont
  {Bae\ss{}ler}, \citenamefont {Nesvizhevsky}, \citenamefont {Protasov},\ and\
  \citenamefont {Voronin}}]{Baessler2007}%
  \BibitemOpen
  \bibfield  {author} {\bibinfo {author} {\bibfnamefont {S.}~\bibnamefont
  {Bae\ss{}ler}}, \bibinfo {author} {\bibfnamefont {V.~V.}\ \bibnamefont
  {Nesvizhevsky}}, \bibinfo {author} {\bibfnamefont {K.~V.}\ \bibnamefont
  {Protasov}}, \ and\ \bibinfo {author} {\bibfnamefont {A.~Y.}\ \bibnamefont
  {Voronin}},\ }\href {\doibase 10.1103/PhysRevD.75.075006} {\bibfield
  {journal} {\bibinfo  {journal} {Physical Review D}\ }\textbf {\bibinfo
  {volume} {75}},\ \bibinfo {pages} {075006} (\bibinfo {year}
  {2007})}\BibitemShut {NoStop}%
\bibitem [{\citenamefont {Jenke}\ \emph {et~al.}(2011)\citenamefont {Jenke},
  \citenamefont {Geltenbort}, \citenamefont {Lemmel},\ and\ \citenamefont
  {Abele}}]{Jenke2011}%
  \BibitemOpen
  \bibfield  {author} {\bibinfo {author} {\bibfnamefont {T.}~\bibnamefont
  {Jenke}}, \bibinfo {author} {\bibfnamefont {P.}~\bibnamefont {Geltenbort}},
  \bibinfo {author} {\bibfnamefont {H.}~\bibnamefont {Lemmel}}, \ and\ \bibinfo
  {author} {\bibfnamefont {H.}~\bibnamefont {Abele}},\ }\href
  {https://doi.org/10.1038/nphys1970} {\bibfield  {journal} {\bibinfo
  {journal} {Nature Physics}\ }\textbf {\bibinfo {volume} {7}},\ \bibinfo
  {pages} {468 EP } (\bibinfo {year} {2011})}\BibitemShut {NoStop}%
\bibitem [{\citenamefont {Jenke}\ \emph {et~al.}(2014)\citenamefont {Jenke},
  \citenamefont {Cronenberg}, \citenamefont {Burgd\"orfer}, \citenamefont
  {Chizhova}, \citenamefont {Geltenbort}, \citenamefont {Ivanov}, \citenamefont
  {Lauer}, \citenamefont {Lins}, \citenamefont {Rotter}, \citenamefont {Saul},
  \citenamefont {Schmidt},\ and\ \citenamefont {Abele}}]{Jenke2014}%
  \BibitemOpen
  \bibfield  {author} {\bibinfo {author} {\bibfnamefont {T.}~\bibnamefont
  {Jenke}}, \bibinfo {author} {\bibfnamefont {G.}~\bibnamefont {Cronenberg}},
  \bibinfo {author} {\bibfnamefont {J.}~\bibnamefont {Burgd\"orfer}}, \bibinfo
  {author} {\bibfnamefont {L.~A.}\ \bibnamefont {Chizhova}}, \bibinfo {author}
  {\bibfnamefont {P.}~\bibnamefont {Geltenbort}}, \bibinfo {author}
  {\bibfnamefont {A.~N.}\ \bibnamefont {Ivanov}}, \bibinfo {author}
  {\bibfnamefont {T.}~\bibnamefont {Lauer}}, \bibinfo {author} {\bibfnamefont
  {T.}~\bibnamefont {Lins}}, \bibinfo {author} {\bibfnamefont {S.}~\bibnamefont
  {Rotter}}, \bibinfo {author} {\bibfnamefont {H.}~\bibnamefont {Saul}},
  \bibinfo {author} {\bibfnamefont {U.}~\bibnamefont {Schmidt}}, \ and\
  \bibinfo {author} {\bibfnamefont {H.}~\bibnamefont {Abele}},\ }\href
  {\doibase 10.1103/PhysRevLett.112.151105} {\bibfield  {journal} {\bibinfo
  {journal} {Physical Review Letters}\ }\textbf {\bibinfo {volume} {112}},\
  \bibinfo {pages} {151105} (\bibinfo {year} {2014})}\BibitemShut {NoStop}%
\bibitem [{\citenamefont {Altarev}\ \emph {et~al.}(2009)\citenamefont
  {Altarev}, \citenamefont {Baker}, \citenamefont {Ban}, \citenamefont {Bison},
  \citenamefont {Bodek}, \citenamefont {Daum}, \citenamefont {Fierlinger},
  \citenamefont {Geltenbort}, \citenamefont {Green}, \citenamefont {van~der
  Grinten}, \citenamefont {Gutsmiedl}, \citenamefont {Harris}, \citenamefont
  {Heil}, \citenamefont {Henneck}, \citenamefont {Horras}, \citenamefont
  {Iaydjiev}, \citenamefont {Ivanov}, \citenamefont {Khomutov}, \citenamefont
  {Kirch}, \citenamefont {Kistryn}, \citenamefont {Knecht}, \citenamefont
  {Knowles}, \citenamefont {Kozela}, \citenamefont {Kuchler}, \citenamefont
  {Ku\ifmmode~\acute{z}\else \'{z}\fi{}niak}, \citenamefont {Lauer},
  \citenamefont {Lauss}, \citenamefont {Lefort}, \citenamefont
  {Mtchedlishvili}, \citenamefont {Naviliat-Cuncic}, \citenamefont {Pazgalev},
  \citenamefont {Pendlebury}, \citenamefont {Petzoldt}, \citenamefont {Pierre},
  \citenamefont {Pignol}, \citenamefont {Qu\'em\'ener}, \citenamefont
  {Rebetez}, \citenamefont {Rebreyend}, \citenamefont {Roccia}, \citenamefont
  {Rogel}, \citenamefont {Severijns}, \citenamefont {Shiers}, \citenamefont
  {Sobolev}, \citenamefont {Weis}, \citenamefont {Zejma},\ and\ \citenamefont
  {Zsigmond}}]{Altarev2009}%
  \BibitemOpen
  \bibfield  {author} {\bibinfo {author} {\bibfnamefont {I.}~\bibnamefont
  {Altarev}}, \bibinfo {author} {\bibfnamefont {C.~A.}\ \bibnamefont {Baker}},
  \bibinfo {author} {\bibfnamefont {G.}~\bibnamefont {Ban}}, \bibinfo {author}
  {\bibfnamefont {G.}~\bibnamefont {Bison}}, \bibinfo {author} {\bibfnamefont
  {K.}~\bibnamefont {Bodek}}, \bibinfo {author} {\bibfnamefont
  {M.}~\bibnamefont {Daum}}, \bibinfo {author} {\bibfnamefont {P.}~\bibnamefont
  {Fierlinger}}, \bibinfo {author} {\bibfnamefont {P.}~\bibnamefont
  {Geltenbort}}, \bibinfo {author} {\bibfnamefont {K.}~\bibnamefont {Green}},
  \bibinfo {author} {\bibfnamefont {M.~G.~D.}\ \bibnamefont {van~der Grinten}},
  \bibinfo {author} {\bibfnamefont {E.}~\bibnamefont {Gutsmiedl}}, \bibinfo
  {author} {\bibfnamefont {P.~G.}\ \bibnamefont {Harris}}, \bibinfo {author}
  {\bibfnamefont {W.}~\bibnamefont {Heil}}, \bibinfo {author} {\bibfnamefont
  {R.}~\bibnamefont {Henneck}}, \bibinfo {author} {\bibfnamefont
  {M.}~\bibnamefont {Horras}}, \bibinfo {author} {\bibfnamefont
  {P.}~\bibnamefont {Iaydjiev}}, \bibinfo {author} {\bibfnamefont {S.~N.}\
  \bibnamefont {Ivanov}}, \bibinfo {author} {\bibfnamefont {N.}~\bibnamefont
  {Khomutov}}, \bibinfo {author} {\bibfnamefont {K.}~\bibnamefont {Kirch}},
  \bibinfo {author} {\bibfnamefont {S.}~\bibnamefont {Kistryn}}, \bibinfo
  {author} {\bibfnamefont {A.}~\bibnamefont {Knecht}}, \bibinfo {author}
  {\bibfnamefont {P.}~\bibnamefont {Knowles}}, \bibinfo {author} {\bibfnamefont
  {A.}~\bibnamefont {Kozela}}, \bibinfo {author} {\bibfnamefont
  {F.}~\bibnamefont {Kuchler}}, \bibinfo {author} {\bibfnamefont
  {M.}~\bibnamefont {Ku\ifmmode~\acute{z}\else \'{z}\fi{}niak}}, \bibinfo
  {author} {\bibfnamefont {T.}~\bibnamefont {Lauer}}, \bibinfo {author}
  {\bibfnamefont {B.}~\bibnamefont {Lauss}}, \bibinfo {author} {\bibfnamefont
  {T.}~\bibnamefont {Lefort}}, \bibinfo {author} {\bibfnamefont
  {A.}~\bibnamefont {Mtchedlishvili}}, \bibinfo {author} {\bibfnamefont
  {O.}~\bibnamefont {Naviliat-Cuncic}}, \bibinfo {author} {\bibfnamefont
  {A.}~\bibnamefont {Pazgalev}}, \bibinfo {author} {\bibfnamefont {J.~M.}\
  \bibnamefont {Pendlebury}}, \bibinfo {author} {\bibfnamefont
  {G.}~\bibnamefont {Petzoldt}}, \bibinfo {author} {\bibfnamefont
  {E.}~\bibnamefont {Pierre}}, \bibinfo {author} {\bibfnamefont
  {G.}~\bibnamefont {Pignol}}, \bibinfo {author} {\bibfnamefont
  {G.}~\bibnamefont {Qu\'em\'ener}}, \bibinfo {author} {\bibfnamefont
  {M.}~\bibnamefont {Rebetez}}, \bibinfo {author} {\bibfnamefont
  {D.}~\bibnamefont {Rebreyend}}, \bibinfo {author} {\bibfnamefont
  {S.}~\bibnamefont {Roccia}}, \bibinfo {author} {\bibfnamefont
  {G.}~\bibnamefont {Rogel}}, \bibinfo {author} {\bibfnamefont
  {N.}~\bibnamefont {Severijns}}, \bibinfo {author} {\bibfnamefont
  {D.}~\bibnamefont {Shiers}}, \bibinfo {author} {\bibfnamefont
  {Y.}~\bibnamefont {Sobolev}}, \bibinfo {author} {\bibfnamefont
  {A.}~\bibnamefont {Weis}}, \bibinfo {author} {\bibfnamefont {J.}~\bibnamefont
  {Zejma}}, \ and\ \bibinfo {author} {\bibfnamefont {G.}~\bibnamefont
  {Zsigmond}},\ }\href {\doibase 10.1103/PhysRevLett.103.081602} {\bibfield
  {journal} {\bibinfo  {journal} {Physical Review Letters}\ }\textbf {\bibinfo
  {volume} {103}},\ \bibinfo {pages} {081602} (\bibinfo {year}
  {2009})}\BibitemShut {NoStop}%
\bibitem [{\citenamefont {Steyerl}\ \emph {et~al.}(1986)\citenamefont
  {Steyerl}, \citenamefont {Nagel}, \citenamefont {Schreiber}, \citenamefont
  {Steinhauser}, \citenamefont {G{\"a}hler}, \citenamefont {Gl{\"a}ser},
  \citenamefont {Ageron}, \citenamefont {Astruc}, \citenamefont {Drexel},
  \citenamefont {Gervais},\ and\ \citenamefont {Mampe}}]{Steyerl1986}%
  \BibitemOpen
  \bibfield  {author} {\bibinfo {author} {\bibfnamefont {A.}~\bibnamefont
  {Steyerl}}, \bibinfo {author} {\bibfnamefont {H.}~\bibnamefont {Nagel}},
  \bibinfo {author} {\bibfnamefont {F.-X.}\ \bibnamefont {Schreiber}}, \bibinfo
  {author} {\bibfnamefont {K.-A.}\ \bibnamefont {Steinhauser}}, \bibinfo
  {author} {\bibfnamefont {R.}~\bibnamefont {G{\"a}hler}}, \bibinfo {author}
  {\bibfnamefont {W.}~\bibnamefont {Gl{\"a}ser}}, \bibinfo {author}
  {\bibfnamefont {P.}~\bibnamefont {Ageron}}, \bibinfo {author} {\bibfnamefont
  {J.}~\bibnamefont {Astruc}}, \bibinfo {author} {\bibfnamefont
  {W.}~\bibnamefont {Drexel}}, \bibinfo {author} {\bibfnamefont
  {G.}~\bibnamefont {Gervais}}, \ and\ \bibinfo {author} {\bibfnamefont
  {W.}~\bibnamefont {Mampe}},\ }\href {\doibase
  http://dx.doi.org/10.1016/0375-9601(86)90587-6} {\bibfield  {journal}
  {\bibinfo  {journal} {Physics Letters A}\ }\textbf {\bibinfo {volume}
  {116}},\ \bibinfo {pages} {347 } (\bibinfo {year} {1986})}\BibitemShut
  {NoStop}%
\bibitem [{\citenamefont {Golub}\ and\ \citenamefont
  {Pendlebury}(1977)}]{Golub1977}%
  \BibitemOpen
  \bibfield  {author} {\bibinfo {author} {\bibfnamefont {R.}~\bibnamefont
  {Golub}}\ and\ \bibinfo {author} {\bibfnamefont {J.~M.}\ \bibnamefont
  {Pendlebury}},\ }\href@noop {} {\bibfield  {journal} {\bibinfo  {journal}
  {Physics Letters A}\ }\textbf {\bibinfo {volume} {62}},\ \bibinfo {pages}
  {337} (\bibinfo {year} {1977})}\BibitemShut {NoStop}%
\bibitem [{\citenamefont {Morris}\ \emph {et~al.}(2002)\citenamefont {Morris},
  \citenamefont {Anaya}, \citenamefont {Bowles}, \citenamefont {Filippone},
  \citenamefont {Geltenbort}, \citenamefont {Hill}, \citenamefont {Hino},
  \citenamefont {Hoedl}, \citenamefont {Hogan}, \citenamefont {Ito},
  \citenamefont {Kawai}, \citenamefont {Kirch}, \citenamefont {Lamoreaux},
  \citenamefont {Liu}, \citenamefont {Makela}, \citenamefont {Marek},
  \citenamefont {Martin}, \citenamefont {Mortensen}, \citenamefont
  {Pichlmaier}, \citenamefont {Saunders}, \citenamefont {Seestrom},
  \citenamefont {Smith}, \citenamefont {Teasdale}, \citenamefont {Tipton},
  \citenamefont {Utsuro}, \citenamefont {Young},\ and\ \citenamefont
  {Yuan}}]{Morris2002}%
  \BibitemOpen
  \bibfield  {author} {\bibinfo {author} {\bibfnamefont {C.~L.}\ \bibnamefont
  {Morris}}, \bibinfo {author} {\bibfnamefont {J.~M.}\ \bibnamefont {Anaya}},
  \bibinfo {author} {\bibfnamefont {T.~J.}\ \bibnamefont {Bowles}}, \bibinfo
  {author} {\bibfnamefont {B.~W.}\ \bibnamefont {Filippone}}, \bibinfo {author}
  {\bibfnamefont {P.}~\bibnamefont {Geltenbort}}, \bibinfo {author}
  {\bibfnamefont {R.~E.}\ \bibnamefont {Hill}}, \bibinfo {author}
  {\bibfnamefont {M.}~\bibnamefont {Hino}}, \bibinfo {author} {\bibfnamefont
  {S.}~\bibnamefont {Hoedl}}, \bibinfo {author} {\bibfnamefont {G.~E.}\
  \bibnamefont {Hogan}}, \bibinfo {author} {\bibfnamefont {T.~M.}\ \bibnamefont
  {Ito}}, \bibinfo {author} {\bibfnamefont {T.}~\bibnamefont {Kawai}}, \bibinfo
  {author} {\bibfnamefont {K.}~\bibnamefont {Kirch}}, \bibinfo {author}
  {\bibfnamefont {S.~K.}\ \bibnamefont {Lamoreaux}}, \bibinfo {author}
  {\bibfnamefont {C.-Y.}\ \bibnamefont {Liu}}, \bibinfo {author} {\bibfnamefont
  {M.}~\bibnamefont {Makela}}, \bibinfo {author} {\bibfnamefont {L.~J.}\
  \bibnamefont {Marek}}, \bibinfo {author} {\bibfnamefont {J.~W.}\ \bibnamefont
  {Martin}}, \bibinfo {author} {\bibfnamefont {R.~N.}\ \bibnamefont
  {Mortensen}}, \bibinfo {author} {\bibfnamefont {A.}~\bibnamefont
  {Pichlmaier}}, \bibinfo {author} {\bibfnamefont {A.}~\bibnamefont
  {Saunders}}, \bibinfo {author} {\bibfnamefont {S.~J.}\ \bibnamefont
  {Seestrom}}, \bibinfo {author} {\bibfnamefont {D.}~\bibnamefont {Smith}},
  \bibinfo {author} {\bibfnamefont {W.}~\bibnamefont {Teasdale}}, \bibinfo
  {author} {\bibfnamefont {B.}~\bibnamefont {Tipton}}, \bibinfo {author}
  {\bibfnamefont {M.}~\bibnamefont {Utsuro}}, \bibinfo {author} {\bibfnamefont
  {A.~R.}\ \bibnamefont {Young}}, \ and\ \bibinfo {author} {\bibfnamefont
  {J.}~\bibnamefont {Yuan}},\ }\href {\doibase 10.1103/PhysRevLett.89.272501}
  {\bibfield  {journal} {\bibinfo  {journal} {Physical Review Letters}\
  }\textbf {\bibinfo {volume} {89}},\ \bibinfo {pages} {272501} (\bibinfo
  {year} {2002})}\BibitemShut {NoStop}%
\bibitem [{\citenamefont {Saunders}\ \emph {et~al.}(2013)\citenamefont
  {Saunders}, \citenamefont {Makela}, \citenamefont {Bagdasarova},
  \citenamefont {Back}, \citenamefont {Boissevain}, \citenamefont {Broussard},
  \citenamefont {Bowles}, \citenamefont {Carr}, \citenamefont {Currie},
  \citenamefont {Filippone}, \citenamefont {Garc{\'\i}a}, \citenamefont
  {Geltenbort}, \citenamefont {Hickerson}, \citenamefont {Hill}, \citenamefont
  {Hoagland}, \citenamefont {Hoedl}, \citenamefont {Holley}, \citenamefont
  {Hogan}, \citenamefont {Ito}, \citenamefont {Lamoreaux}, \citenamefont {Liu},
  \citenamefont {Liu}, \citenamefont {Mammei}, \citenamefont {Martin},
  \citenamefont {Melconian}, \citenamefont {Mendenhall}, \citenamefont
  {Morris}, \citenamefont {Mortensen}, \citenamefont {Pattie}, \citenamefont
  {Pitt}, \citenamefont {Plaster}, \citenamefont {Ramsey}, \citenamefont
  {Rios}, \citenamefont {Sallaska}, \citenamefont {Seestrom}, \citenamefont
  {Sharapov}, \citenamefont {Sjue}, \citenamefont {Sondheim}, \citenamefont
  {Teasdale}, \citenamefont {Young}, \citenamefont {VornDick}, \citenamefont
  {Vogelaar}, \citenamefont {Wang},\ and\ \citenamefont {Xu}}]{Saunders2013}%
  \BibitemOpen
  \bibfield  {author} {\bibinfo {author} {\bibfnamefont {A.}~\bibnamefont
  {Saunders}}, \bibinfo {author} {\bibfnamefont {M.}~\bibnamefont {Makela}},
  \bibinfo {author} {\bibfnamefont {Y.}~\bibnamefont {Bagdasarova}}, \bibinfo
  {author} {\bibfnamefont {H.~O.}\ \bibnamefont {Back}}, \bibinfo {author}
  {\bibfnamefont {J.}~\bibnamefont {Boissevain}}, \bibinfo {author}
  {\bibfnamefont {L.~J.}\ \bibnamefont {Broussard}}, \bibinfo {author}
  {\bibfnamefont {T.~J.}\ \bibnamefont {Bowles}}, \bibinfo {author}
  {\bibfnamefont {R.}~\bibnamefont {Carr}}, \bibinfo {author} {\bibfnamefont
  {S.~A.}\ \bibnamefont {Currie}}, \bibinfo {author} {\bibfnamefont
  {B.}~\bibnamefont {Filippone}}, \bibinfo {author} {\bibfnamefont
  {A.}~\bibnamefont {Garc{\'\i}a}}, \bibinfo {author} {\bibfnamefont
  {P.}~\bibnamefont {Geltenbort}}, \bibinfo {author} {\bibfnamefont {K.~P.}\
  \bibnamefont {Hickerson}}, \bibinfo {author} {\bibfnamefont {R.~E.}\
  \bibnamefont {Hill}}, \bibinfo {author} {\bibfnamefont {J.}~\bibnamefont
  {Hoagland}}, \bibinfo {author} {\bibfnamefont {S.}~\bibnamefont {Hoedl}},
  \bibinfo {author} {\bibfnamefont {A.~T.}\ \bibnamefont {Holley}}, \bibinfo
  {author} {\bibfnamefont {G.}~\bibnamefont {Hogan}}, \bibinfo {author}
  {\bibfnamefont {T.~M.}\ \bibnamefont {Ito}}, \bibinfo {author} {\bibfnamefont
  {S.}~\bibnamefont {Lamoreaux}}, \bibinfo {author} {\bibfnamefont {C.-Y.}\
  \bibnamefont {Liu}}, \bibinfo {author} {\bibfnamefont {J.}~\bibnamefont
  {Liu}}, \bibinfo {author} {\bibfnamefont {R.~R.}\ \bibnamefont {Mammei}},
  \bibinfo {author} {\bibfnamefont {J.}~\bibnamefont {Martin}}, \bibinfo
  {author} {\bibfnamefont {D.}~\bibnamefont {Melconian}}, \bibinfo {author}
  {\bibfnamefont {M.~P.}\ \bibnamefont {Mendenhall}}, \bibinfo {author}
  {\bibfnamefont {C.~L.}\ \bibnamefont {Morris}}, \bibinfo {author}
  {\bibfnamefont {R.~N.}\ \bibnamefont {Mortensen}}, \bibinfo {author}
  {\bibfnamefont {R.~W.}\ \bibnamefont {Pattie}}, \bibinfo {author}
  {\bibfnamefont {M.}~\bibnamefont {Pitt}}, \bibinfo {author} {\bibfnamefont
  {B.}~\bibnamefont {Plaster}}, \bibinfo {author} {\bibfnamefont
  {J.}~\bibnamefont {Ramsey}}, \bibinfo {author} {\bibfnamefont
  {R.}~\bibnamefont {Rios}}, \bibinfo {author} {\bibfnamefont {A.}~\bibnamefont
  {Sallaska}}, \bibinfo {author} {\bibfnamefont {S.~J.}\ \bibnamefont
  {Seestrom}}, \bibinfo {author} {\bibfnamefont {E.~I.}\ \bibnamefont
  {Sharapov}}, \bibinfo {author} {\bibfnamefont {S.}~\bibnamefont {Sjue}},
  \bibinfo {author} {\bibfnamefont {W.~E.}\ \bibnamefont {Sondheim}}, \bibinfo
  {author} {\bibfnamefont {W.}~\bibnamefont {Teasdale}}, \bibinfo {author}
  {\bibfnamefont {A.~R.}\ \bibnamefont {Young}}, \bibinfo {author}
  {\bibfnamefont {B.}~\bibnamefont {VornDick}}, \bibinfo {author}
  {\bibfnamefont {R.~B.}\ \bibnamefont {Vogelaar}}, \bibinfo {author}
  {\bibfnamefont {Z.}~\bibnamefont {Wang}}, \ and\ \bibinfo {author}
  {\bibfnamefont {Y.}~\bibnamefont {Xu}},\ }\href {\doibase
  http://dx.doi.org/10.1063/1.4770063} {\bibfield  {journal} {\bibinfo
  {journal} {Review of Scientific Instruments}\ }\textbf {\bibinfo {volume}
  {84}},\ \bibinfo {eid} {013304} (\bibinfo {year} {2013})}\BibitemShut
  {NoStop}%
\bibitem [{\citenamefont {Ito}\ \emph {et~al.}(2018)\citenamefont {Ito},
  \citenamefont {Adamek}, \citenamefont {Callahan}, \citenamefont {Choi},
  \citenamefont {Clayton}, \citenamefont {Cude-Woods}, \citenamefont {Currie},
  \citenamefont {Ding}, \citenamefont {Fellers}, \citenamefont {Geltenbort},
  \citenamefont {Lamoreaux}, \citenamefont {Liu}, \citenamefont {MacDonald},
  \citenamefont {Makela}, \citenamefont {Morris}, \citenamefont {Pattie},
  \citenamefont {Ramsey}, \citenamefont {Salvat}, \citenamefont {Saunders},
  \citenamefont {Sharapov}, \citenamefont {Sjue}, \citenamefont {Sprow},
  \citenamefont {Tang}, \citenamefont {Weaver}, \citenamefont {Wei},\ and\
  \citenamefont {Young}}]{Ito2018}%
  \BibitemOpen
  \bibfield  {author} {\bibinfo {author} {\bibfnamefont {T.~M.}\ \bibnamefont
  {Ito}}, \bibinfo {author} {\bibfnamefont {E.~R.}\ \bibnamefont {Adamek}},
  \bibinfo {author} {\bibfnamefont {N.~B.}\ \bibnamefont {Callahan}}, \bibinfo
  {author} {\bibfnamefont {J.~H.}\ \bibnamefont {Choi}}, \bibinfo {author}
  {\bibfnamefont {S.~M.}\ \bibnamefont {Clayton}}, \bibinfo {author}
  {\bibfnamefont {C.}~\bibnamefont {Cude-Woods}}, \bibinfo {author}
  {\bibfnamefont {S.}~\bibnamefont {Currie}}, \bibinfo {author} {\bibfnamefont
  {X.}~\bibnamefont {Ding}}, \bibinfo {author} {\bibfnamefont {D.~E.}\
  \bibnamefont {Fellers}}, \bibinfo {author} {\bibfnamefont {P.}~\bibnamefont
  {Geltenbort}}, \bibinfo {author} {\bibfnamefont {S.~K.}\ \bibnamefont
  {Lamoreaux}}, \bibinfo {author} {\bibfnamefont {C.-Y.}\ \bibnamefont {Liu}},
  \bibinfo {author} {\bibfnamefont {S.}~\bibnamefont {MacDonald}}, \bibinfo
  {author} {\bibfnamefont {M.}~\bibnamefont {Makela}}, \bibinfo {author}
  {\bibfnamefont {C.~L.}\ \bibnamefont {Morris}}, \bibinfo {author}
  {\bibfnamefont {R.~W.}\ \bibnamefont {Pattie}}, \bibinfo {author}
  {\bibfnamefont {J.~C.}\ \bibnamefont {Ramsey}}, \bibinfo {author}
  {\bibfnamefont {D.~J.}\ \bibnamefont {Salvat}}, \bibinfo {author}
  {\bibfnamefont {A.}~\bibnamefont {Saunders}}, \bibinfo {author}
  {\bibfnamefont {E.~I.}\ \bibnamefont {Sharapov}}, \bibinfo {author}
  {\bibfnamefont {S.}~\bibnamefont {Sjue}}, \bibinfo {author} {\bibfnamefont
  {A.~P.}\ \bibnamefont {Sprow}}, \bibinfo {author} {\bibfnamefont
  {Z.}~\bibnamefont {Tang}}, \bibinfo {author} {\bibfnamefont {H.~L.}\
  \bibnamefont {Weaver}}, \bibinfo {author} {\bibfnamefont {W.}~\bibnamefont
  {Wei}}, \ and\ \bibinfo {author} {\bibfnamefont {A.~R.}\ \bibnamefont
  {Young}},\ }\href {\doibase 10.1103/PhysRevC.97.012501} {\bibfield  {journal}
  {\bibinfo  {journal} {Physical Review C}\ }\textbf {\bibinfo {volume} {97}},\
  \bibinfo {pages} {012501} (\bibinfo {year} {2018})}\BibitemShut {NoStop}%
\bibitem [{\citenamefont {Anghel}\ \emph {et~al.}(2009)\citenamefont {Anghel},
  \citenamefont {Atchison}, \citenamefont {Blau}, \citenamefont {van~den
  Brandt}, \citenamefont {Daum}, \citenamefont {Doelling}, \citenamefont
  {Dubs}, \citenamefont {Duperrex}, \citenamefont {Fuchs}, \citenamefont
  {George}, \citenamefont {G{\"u}ltl}, \citenamefont {Hautle}, \citenamefont
  {Heidenreich}, \citenamefont {Heinrich}, \citenamefont {Henneck},
  \citenamefont {Heule}, \citenamefont {Hofmann}, \citenamefont {Joray},
  \citenamefont {Kasprzak}, \citenamefont {Kirch}, \citenamefont {Knecht},
  \citenamefont {Konter}, \citenamefont {Korhonen}, \citenamefont {Kuzniak},
  \citenamefont {Lauss}, \citenamefont {Mezger}, \citenamefont
  {Mtchedlishvili}, \citenamefont {Petzoldt}, \citenamefont {Pichlmaier},
  \citenamefont {Reggiani}, \citenamefont {Reiser}, \citenamefont {Rohrer},
  \citenamefont {Seidel}, \citenamefont {Spitzer}, \citenamefont {Thomsen},
  \citenamefont {Wagner}, \citenamefont {Wohlmuther}, \citenamefont {Zsigmond},
  \citenamefont {Zuellig}, \citenamefont {Bodek}, \citenamefont {Kistryn},
  \citenamefont {Zejma}, \citenamefont {Geltenbort}, \citenamefont {Plonka},\
  and\ \citenamefont {Grigoriev}}]{Anghel2009}%
  \BibitemOpen
  \bibfield  {author} {\bibinfo {author} {\bibfnamefont {A.}~\bibnamefont
  {Anghel}}, \bibinfo {author} {\bibfnamefont {F.}~\bibnamefont {Atchison}},
  \bibinfo {author} {\bibfnamefont {B.}~\bibnamefont {Blau}}, \bibinfo {author}
  {\bibfnamefont {B.}~\bibnamefont {van~den Brandt}}, \bibinfo {author}
  {\bibfnamefont {M.}~\bibnamefont {Daum}}, \bibinfo {author} {\bibfnamefont
  {R.}~\bibnamefont {Doelling}}, \bibinfo {author} {\bibfnamefont
  {M.}~\bibnamefont {Dubs}}, \bibinfo {author} {\bibfnamefont {P.-A.}\
  \bibnamefont {Duperrex}}, \bibinfo {author} {\bibfnamefont {A.}~\bibnamefont
  {Fuchs}}, \bibinfo {author} {\bibfnamefont {D.}~\bibnamefont {George}},
  \bibinfo {author} {\bibfnamefont {L.}~\bibnamefont {G{\"u}ltl}}, \bibinfo
  {author} {\bibfnamefont {P.}~\bibnamefont {Hautle}}, \bibinfo {author}
  {\bibfnamefont {G.}~\bibnamefont {Heidenreich}}, \bibinfo {author}
  {\bibfnamefont {F.}~\bibnamefont {Heinrich}}, \bibinfo {author}
  {\bibfnamefont {R.}~\bibnamefont {Henneck}}, \bibinfo {author} {\bibfnamefont
  {S.}~\bibnamefont {Heule}}, \bibinfo {author} {\bibfnamefont
  {T.}~\bibnamefont {Hofmann}}, \bibinfo {author} {\bibfnamefont
  {S.}~\bibnamefont {Joray}}, \bibinfo {author} {\bibfnamefont
  {M.}~\bibnamefont {Kasprzak}}, \bibinfo {author} {\bibfnamefont
  {K.}~\bibnamefont {Kirch}}, \bibinfo {author} {\bibfnamefont
  {A.}~\bibnamefont {Knecht}}, \bibinfo {author} {\bibfnamefont
  {J.}~\bibnamefont {Konter}}, \bibinfo {author} {\bibfnamefont
  {T.}~\bibnamefont {Korhonen}}, \bibinfo {author} {\bibfnamefont
  {M.}~\bibnamefont {Kuzniak}}, \bibinfo {author} {\bibfnamefont
  {B.}~\bibnamefont {Lauss}}, \bibinfo {author} {\bibfnamefont
  {A.}~\bibnamefont {Mezger}}, \bibinfo {author} {\bibfnamefont
  {A.}~\bibnamefont {Mtchedlishvili}}, \bibinfo {author} {\bibfnamefont
  {G.}~\bibnamefont {Petzoldt}}, \bibinfo {author} {\bibfnamefont
  {A.}~\bibnamefont {Pichlmaier}}, \bibinfo {author} {\bibfnamefont
  {D.}~\bibnamefont {Reggiani}}, \bibinfo {author} {\bibfnamefont
  {R.}~\bibnamefont {Reiser}}, \bibinfo {author} {\bibfnamefont
  {U.}~\bibnamefont {Rohrer}}, \bibinfo {author} {\bibfnamefont
  {M.}~\bibnamefont {Seidel}}, \bibinfo {author} {\bibfnamefont
  {H.}~\bibnamefont {Spitzer}}, \bibinfo {author} {\bibfnamefont
  {K.}~\bibnamefont {Thomsen}}, \bibinfo {author} {\bibfnamefont
  {W.}~\bibnamefont {Wagner}}, \bibinfo {author} {\bibfnamefont
  {M.}~\bibnamefont {Wohlmuther}}, \bibinfo {author} {\bibfnamefont
  {G.}~\bibnamefont {Zsigmond}}, \bibinfo {author} {\bibfnamefont
  {J.}~\bibnamefont {Zuellig}}, \bibinfo {author} {\bibfnamefont
  {K.}~\bibnamefont {Bodek}}, \bibinfo {author} {\bibfnamefont
  {S.}~\bibnamefont {Kistryn}}, \bibinfo {author} {\bibfnamefont
  {J.}~\bibnamefont {Zejma}}, \bibinfo {author} {\bibfnamefont
  {P.}~\bibnamefont {Geltenbort}}, \bibinfo {author} {\bibfnamefont
  {C.}~\bibnamefont {Plonka}}, \ and\ \bibinfo {author} {\bibfnamefont
  {S.}~\bibnamefont {Grigoriev}},\ }\href {\doibase
  http://dx.doi.org/10.1016/j.nima.2009.07.077} {\bibfield  {journal} {\bibinfo
   {journal} {Nuclear Instruments and Methods in Physics Research Section A}\
  }\textbf {\bibinfo {volume} {611}},\ \bibinfo {pages} {272 } (\bibinfo {year}
  {2009})}\BibitemShut {NoStop}%
\bibitem [{\citenamefont {Becker}\ \emph {et~al.}(2015)\citenamefont {Becker},
  \citenamefont {Bison}, \citenamefont {Blau}, \citenamefont {Chowdhuri},
  \citenamefont {Eikenberg}, \citenamefont {Fertl}, \citenamefont {Kirch},
  \citenamefont {Lauss}, \citenamefont {Perret}, \citenamefont {Reggiani},
  \citenamefont {Ries}, \citenamefont {Schmidt-Wellenburg}, \citenamefont
  {Talanov}, \citenamefont {Wohlmuther},\ and\ \citenamefont
  {Zsigmond}}]{Becker2015}%
  \BibitemOpen
  \bibfield  {author} {\bibinfo {author} {\bibfnamefont {H.}~\bibnamefont
  {Becker}}, \bibinfo {author} {\bibfnamefont {G.}~\bibnamefont {Bison}},
  \bibinfo {author} {\bibfnamefont {B.}~\bibnamefont {Blau}}, \bibinfo {author}
  {\bibfnamefont {Z.}~\bibnamefont {Chowdhuri}}, \bibinfo {author}
  {\bibfnamefont {J.}~\bibnamefont {Eikenberg}}, \bibinfo {author}
  {\bibfnamefont {M.}~\bibnamefont {Fertl}}, \bibinfo {author} {\bibfnamefont
  {K.}~\bibnamefont {Kirch}}, \bibinfo {author} {\bibfnamefont
  {B.}~\bibnamefont {Lauss}}, \bibinfo {author} {\bibfnamefont
  {G.}~\bibnamefont {Perret}}, \bibinfo {author} {\bibfnamefont
  {D.}~\bibnamefont {Reggiani}}, \bibinfo {author} {\bibfnamefont
  {D.}~\bibnamefont {Ries}}, \bibinfo {author} {\bibfnamefont {P.}~\bibnamefont
  {Schmidt-Wellenburg}}, \bibinfo {author} {\bibfnamefont {V.}~\bibnamefont
  {Talanov}}, \bibinfo {author} {\bibfnamefont {M.}~\bibnamefont {Wohlmuther}},
  \ and\ \bibinfo {author} {\bibfnamefont {G.}~\bibnamefont {Zsigmond}},\
  }\href@noop {} {\bibfield  {journal} {\bibinfo  {journal} {Nuclear
  Instruments and Methods in Physics Research Section A}\ }\textbf {\bibinfo
  {volume} {777}},\ \bibinfo {pages} {20} (\bibinfo {year} {2015})}\BibitemShut
  {NoStop}%
\bibitem [{\citenamefont {Frei}\ \emph {et~al.}(2007)\citenamefont {Frei},
  \citenamefont {Sobolev}, \citenamefont {Altarev}, \citenamefont {Eberhardt},
  \citenamefont {Gschrey}, \citenamefont {Gutsmiedl}, \citenamefont {Hackl},
  \citenamefont {Hampel}, \citenamefont {Hartmann}, \citenamefont {Heil},
  \citenamefont {Kratz}, \citenamefont {Lauer}, \citenamefont
  {Li{\'{z}}on~Aguilar}, \citenamefont {M{\"u}ller}, \citenamefont {Paul},
  \citenamefont {Pokotilovski}, \citenamefont {Schmid}, \citenamefont
  {Tassini}, \citenamefont {Tortorella}, \citenamefont {Trautmann},
  \citenamefont {Trinks},\ and\ \citenamefont {Wiehl}}]{Frei2007}%
  \BibitemOpen
  \bibfield  {author} {\bibinfo {author} {\bibfnamefont {A.}~\bibnamefont
  {Frei}}, \bibinfo {author} {\bibfnamefont {Y.}~\bibnamefont {Sobolev}},
  \bibinfo {author} {\bibfnamefont {I.}~\bibnamefont {Altarev}}, \bibinfo
  {author} {\bibfnamefont {K.}~\bibnamefont {Eberhardt}}, \bibinfo {author}
  {\bibfnamefont {A.}~\bibnamefont {Gschrey}}, \bibinfo {author} {\bibfnamefont
  {E.}~\bibnamefont {Gutsmiedl}}, \bibinfo {author} {\bibfnamefont
  {R.}~\bibnamefont {Hackl}}, \bibinfo {author} {\bibfnamefont
  {G.}~\bibnamefont {Hampel}}, \bibinfo {author} {\bibfnamefont {F.~J.}\
  \bibnamefont {Hartmann}}, \bibinfo {author} {\bibfnamefont {W.}~\bibnamefont
  {Heil}}, \bibinfo {author} {\bibfnamefont {J.~V.}\ \bibnamefont {Kratz}},
  \bibinfo {author} {\bibfnamefont {T.}~\bibnamefont {Lauer}}, \bibinfo
  {author} {\bibfnamefont {A.}~\bibnamefont {Li{\'{z}}on~Aguilar}}, \bibinfo
  {author} {\bibfnamefont {A.~R.}\ \bibnamefont {M{\"u}ller}}, \bibinfo
  {author} {\bibfnamefont {S.}~\bibnamefont {Paul}}, \bibinfo {author}
  {\bibfnamefont {Y.}~\bibnamefont {Pokotilovski}}, \bibinfo {author}
  {\bibfnamefont {W.}~\bibnamefont {Schmid}}, \bibinfo {author} {\bibfnamefont
  {L.}~\bibnamefont {Tassini}}, \bibinfo {author} {\bibfnamefont
  {D.}~\bibnamefont {Tortorella}}, \bibinfo {author} {\bibfnamefont
  {N.}~\bibnamefont {Trautmann}}, \bibinfo {author} {\bibfnamefont
  {U.}~\bibnamefont {Trinks}}, \ and\ \bibinfo {author} {\bibfnamefont
  {N.}~\bibnamefont {Wiehl}},\ }\href {\doibase 10.1140/epja/i2007-10494-2}
  {\bibfield  {journal} {\bibinfo  {journal} {The European Physical Journal A}\
  }\textbf {\bibinfo {volume} {34}},\ \bibinfo {pages} {119} (\bibinfo {year}
  {2007})}\BibitemShut {NoStop}%
\bibitem [{\citenamefont {Kahlenberg}\ \emph {et~al.}(2017)\citenamefont
  {Kahlenberg}, \citenamefont {Ries}, \citenamefont {Ross}, \citenamefont
  {Siemensen}, \citenamefont {Beck}, \citenamefont {Geppert}, \citenamefont
  {Heil}, \citenamefont {Hild}, \citenamefont {Karch}, \citenamefont {Karpuk},
  \citenamefont {Kories}, \citenamefont {Kretschmer}, \citenamefont {Lauss},
  \citenamefont {Reich}, \citenamefont {Sobolev},\ and\ \citenamefont
  {Trautmann}}]{Kahlenberg2017}%
  \BibitemOpen
  \bibfield  {author} {\bibinfo {author} {\bibfnamefont {J.}~\bibnamefont
  {Kahlenberg}}, \bibinfo {author} {\bibfnamefont {D.}~\bibnamefont {Ries}},
  \bibinfo {author} {\bibfnamefont {K.~U.}\ \bibnamefont {Ross}}, \bibinfo
  {author} {\bibfnamefont {C.}~\bibnamefont {Siemensen}}, \bibinfo {author}
  {\bibfnamefont {M.}~\bibnamefont {Beck}}, \bibinfo {author} {\bibfnamefont
  {C.}~\bibnamefont {Geppert}}, \bibinfo {author} {\bibfnamefont
  {W.}~\bibnamefont {Heil}}, \bibinfo {author} {\bibfnamefont {N.}~\bibnamefont
  {Hild}}, \bibinfo {author} {\bibfnamefont {J.}~\bibnamefont {Karch}},
  \bibinfo {author} {\bibfnamefont {S.}~\bibnamefont {Karpuk}}, \bibinfo
  {author} {\bibfnamefont {F.}~\bibnamefont {Kories}}, \bibinfo {author}
  {\bibfnamefont {M.}~\bibnamefont {Kretschmer}}, \bibinfo {author}
  {\bibfnamefont {B.}~\bibnamefont {Lauss}}, \bibinfo {author} {\bibfnamefont
  {T.}~\bibnamefont {Reich}}, \bibinfo {author} {\bibfnamefont
  {Y.}~\bibnamefont {Sobolev}}, \ and\ \bibinfo {author} {\bibfnamefont
  {N.}~\bibnamefont {Trautmann}},\ }\href@noop {} {\bibfield  {journal}
  {\bibinfo  {journal} {The European Physical Journal A}\ }\textbf {\bibinfo
  {volume} {53}},\ \bibinfo {pages} {226} (\bibinfo {year} {2017})}\BibitemShut
  {NoStop}%
\bibitem [{\citenamefont {Korobkina}\ \emph {et~al.}(2014)\citenamefont
  {Korobkina}, \citenamefont {Medlin}, \citenamefont {Wehring}, \citenamefont
  {Hawari}, \citenamefont {Huffman}, \citenamefont {Young}, \citenamefont
  {Beaumont},\ and\ \citenamefont {Palmquist}}]{Korobkina2014}%
  \BibitemOpen
  \bibfield  {author} {\bibinfo {author} {\bibfnamefont {E.}~\bibnamefont
  {Korobkina}}, \bibinfo {author} {\bibfnamefont {G.}~\bibnamefont {Medlin}},
  \bibinfo {author} {\bibfnamefont {B.}~\bibnamefont {Wehring}}, \bibinfo
  {author} {\bibfnamefont {A.}~\bibnamefont {Hawari}}, \bibinfo {author}
  {\bibfnamefont {P.}~\bibnamefont {Huffman}}, \bibinfo {author} {\bibfnamefont
  {A.}~\bibnamefont {Young}}, \bibinfo {author} {\bibfnamefont
  {B.}~\bibnamefont {Beaumont}}, \ and\ \bibinfo {author} {\bibfnamefont
  {G.}~\bibnamefont {Palmquist}},\ }\href {\doibase
  http://dx.doi.org/10.1016/j.nima.2014.08.016} {\bibfield  {journal} {\bibinfo
   {journal} {Nuclear Instruments and Methods in Physics Research Section A}\
  }\textbf {\bibinfo {volume} {767}},\ \bibinfo {pages} {169 } (\bibinfo {year}
  {2014})}\BibitemShut {NoStop}%
\bibitem [{\citenamefont {Trinks}\ \emph {et~al.}(2000)\citenamefont {Trinks},
  \citenamefont {Hartmann}, \citenamefont {Paul},\ and\ \citenamefont
  {Schott}}]{Trinks2000a}%
  \BibitemOpen
  \bibfield  {author} {\bibinfo {author} {\bibfnamefont {U.}~\bibnamefont
  {Trinks}}, \bibinfo {author} {\bibfnamefont {F.~J.}\ \bibnamefont
  {Hartmann}}, \bibinfo {author} {\bibfnamefont {S.}~\bibnamefont {Paul}}, \
  and\ \bibinfo {author} {\bibfnamefont {W.}~\bibnamefont {Schott}},\ }\href
  {\doibase http://dx.doi.org/10.1016/S0168-9002(99)01059-1} {\bibfield
  {journal} {\bibinfo  {journal} {Nuclear Instruments and Methods in Physics
  Research Section A}\ }\textbf {\bibinfo {volume} {440}},\ \bibinfo {pages}
  {666} (\bibinfo {year} {2000})}\BibitemShut {NoStop}%
\bibitem [{\citenamefont {Frei}(2012)}]{Frei2012}%
  \BibitemOpen
  \bibfield  {author} {\bibinfo {author} {\bibfnamefont {A.}~\bibnamefont
  {Frei}},\ }\href@noop {} {\bibfield  {journal} {\bibinfo  {journal} {Private
  communication}\ } (\bibinfo {year} {2012})}\BibitemShut {NoStop}%
\bibitem [{\citenamefont {Sommers}, \citenamefont {Dash},\ and\ \citenamefont
  {Goldstein}(1955)}]{Sommers1955}%
  \BibitemOpen
  \bibfield  {author} {\bibinfo {author} {\bibfnamefont {H.~S.}\ \bibnamefont
  {Sommers}}, \bibinfo {author} {\bibfnamefont {J.~G.}\ \bibnamefont {Dash}}, \
  and\ \bibinfo {author} {\bibfnamefont {L.}~\bibnamefont {Goldstein}},\
  }\href@noop {} {\bibfield  {journal} {\bibinfo  {journal} {Physical Review}\
  }\textbf {\bibinfo {volume} {97}},\ \bibinfo {pages} {855} (\bibinfo {year}
  {1955})}\BibitemShut {NoStop}%
\bibitem [{\citenamefont {Masuda}\ \emph {et~al.}(2012)\citenamefont {Masuda},
  \citenamefont {Hatanaka}, \citenamefont {Jeong}, \citenamefont {Kawasaki},
  \citenamefont {Matsumiya}, \citenamefont {Matsuta}, \citenamefont {Mihara},\
  and\ \citenamefont {Watanabe}}]{Masuda2012}%
  \BibitemOpen
  \bibfield  {author} {\bibinfo {author} {\bibfnamefont {Y.}~\bibnamefont
  {Masuda}}, \bibinfo {author} {\bibfnamefont {K.}~\bibnamefont {Hatanaka}},
  \bibinfo {author} {\bibfnamefont {S.-C.}\ \bibnamefont {Jeong}}, \bibinfo
  {author} {\bibfnamefont {S.}~\bibnamefont {Kawasaki}}, \bibinfo {author}
  {\bibfnamefont {R.}~\bibnamefont {Matsumiya}}, \bibinfo {author}
  {\bibfnamefont {K.}~\bibnamefont {Matsuta}}, \bibinfo {author} {\bibfnamefont
  {M.}~\bibnamefont {Mihara}}, \ and\ \bibinfo {author} {\bibfnamefont
  {Y.}~\bibnamefont {Watanabe}},\ }\href@noop {} {\bibfield  {journal}
  {\bibinfo  {journal} {Physical Review Letters}\ }\textbf {\bibinfo {volume}
  {108}},\ \bibinfo {pages} {134801} (\bibinfo {year} {2012})}\BibitemShut
  {NoStop}%
\bibitem [{\citenamefont {Martin}\ and\ \citenamefont {{Japan-Canada nEDM
  Collaboration}}(2013)}]{Martin2013}%
  \BibitemOpen
  \bibfield  {author} {\bibinfo {author} {\bibfnamefont {J.~W.}\ \bibnamefont
  {Martin}}\ and\ \bibinfo {author} {\bibnamefont {{Japan-Canada nEDM
  Collaboration}}},\ }\href {\doibase http://dx.doi.org/10.1063/1.4826737}
  {\bibfield  {journal} {\bibinfo  {journal} {AIP Conference Proceedings}\
  }\textbf {\bibinfo {volume} {1560}},\ \bibinfo {pages} {134} (\bibinfo {year}
  {2013})}\BibitemShut {NoStop}%
\bibitem [{\citenamefont {Masuda}\ \emph {et~al.}(2014)\citenamefont {Masuda},
  \citenamefont {Hatanaka}, \citenamefont {Jeong}, \citenamefont {Kawasaki},
  \citenamefont {Matsumiya}, \citenamefont {Matsuta}, \citenamefont {Mihara},\
  and\ \citenamefont {Watanabe}}]{Masuda2014}%
  \BibitemOpen
  \bibfield  {author} {\bibinfo {author} {\bibfnamefont {Y.}~\bibnamefont
  {Masuda}}, \bibinfo {author} {\bibfnamefont {K.}~\bibnamefont {Hatanaka}},
  \bibinfo {author} {\bibfnamefont {S.-C.}\ \bibnamefont {Jeong}}, \bibinfo
  {author} {\bibfnamefont {S.}~\bibnamefont {Kawasaki}}, \bibinfo {author}
  {\bibfnamefont {R.}~\bibnamefont {Matsumiya}}, \bibinfo {author}
  {\bibfnamefont {K.}~\bibnamefont {Matsuta}}, \bibinfo {author} {\bibfnamefont
  {M.}~\bibnamefont {Mihara}}, \ and\ \bibinfo {author} {\bibfnamefont
  {Y.}~\bibnamefont {Watanabe}},\ }\href {\doibase
  http://dx.doi.org/10.1016/j.phpro.2013.12.020} {\bibfield  {journal}
  {\bibinfo  {journal} {Physics Procedia}\ }\textbf {\bibinfo {volume} {51}},\
  \bibinfo {pages} {89} (\bibinfo {year} {2014})}\BibitemShut {NoStop}%
\bibitem [{\citenamefont {Picker}(2017)}]{Picker2017}%
  \BibitemOpen
  \bibfield  {author} {\bibinfo {author} {\bibfnamefont {R.}~\bibnamefont
  {Picker}},\ }\href@noop {} {\bibfield  {journal} {\bibinfo  {journal} {The
  Physical Society of Japan Conference Proceedings}\ }\textbf {\bibinfo
  {volume} {13}} (\bibinfo {year} {2017})}\BibitemShut {NoStop}%
\bibitem [{\citenamefont {Ahmed}\ \emph {et~al.}(2019)\citenamefont {Ahmed},
  \citenamefont {Altiere}, \citenamefont {Andalib}, \citenamefont {Bell},
  \citenamefont {Bidinosti}, \citenamefont {Cudmore}, \citenamefont {Das},
  \citenamefont {Davis}, \citenamefont {Franke}, \citenamefont {Gericke},
  \citenamefont {Giampa}, \citenamefont {Gnyp}, \citenamefont {Hansen-Romu},
  \citenamefont {Hatanaka}, \citenamefont {Hayamizu}, \citenamefont {Jamieson},
  \citenamefont {Jones}, \citenamefont {Kawasaki}, \citenamefont {Kikawa},
  \citenamefont {Kitaguchi}, \citenamefont {Klassen}, \citenamefont {Konaka},
  \citenamefont {Korkmaz}, \citenamefont {Kuchler}, \citenamefont {Lang},
  \citenamefont {Lee}, \citenamefont {Lindner}, \citenamefont {Madison},
  \citenamefont {Makida}, \citenamefont {Mammei}, \citenamefont {Mammei},
  \citenamefont {Martin}, \citenamefont {Matsumiya}, \citenamefont {Miller},
  \citenamefont {Mishima}, \citenamefont {Momose}, \citenamefont {Okamura},
  \citenamefont {Page}, \citenamefont {Picker}, \citenamefont {Pierre},
  \citenamefont {Ramsay}, \citenamefont {Rebenitsch}, \citenamefont {Rehm},
  \citenamefont {Schreyer}, \citenamefont {Shimizu}, \citenamefont {Sidhu},
  \citenamefont {Sikora}, \citenamefont {Smith}, \citenamefont {Tanihata},
  \citenamefont {Thorsteinson}, \citenamefont {Vanbergen}, \citenamefont {van
  Oers},\ and\ \citenamefont {Watanabe}}]{Ahmed2019}%
  \BibitemOpen
  \bibfield  {author} {\bibinfo {author} {\bibfnamefont {S.}~\bibnamefont
  {Ahmed}}, \bibinfo {author} {\bibfnamefont {E.}~\bibnamefont {Altiere}},
  \bibinfo {author} {\bibfnamefont {T.}~\bibnamefont {Andalib}}, \bibinfo
  {author} {\bibfnamefont {B.}~\bibnamefont {Bell}}, \bibinfo {author}
  {\bibfnamefont {C.~P.}\ \bibnamefont {Bidinosti}}, \bibinfo {author}
  {\bibfnamefont {E.}~\bibnamefont {Cudmore}}, \bibinfo {author} {\bibfnamefont
  {M.}~\bibnamefont {Das}}, \bibinfo {author} {\bibfnamefont {C.~A.}\
  \bibnamefont {Davis}}, \bibinfo {author} {\bibfnamefont {B.}~\bibnamefont
  {Franke}}, \bibinfo {author} {\bibfnamefont {M.}~\bibnamefont {Gericke}},
  \bibinfo {author} {\bibfnamefont {P.}~\bibnamefont {Giampa}}, \bibinfo
  {author} {\bibfnamefont {P.}~\bibnamefont {Gnyp}}, \bibinfo {author}
  {\bibfnamefont {S.}~\bibnamefont {Hansen-Romu}}, \bibinfo {author}
  {\bibfnamefont {K.}~\bibnamefont {Hatanaka}}, \bibinfo {author}
  {\bibfnamefont {T.}~\bibnamefont {Hayamizu}}, \bibinfo {author}
  {\bibfnamefont {B.}~\bibnamefont {Jamieson}}, \bibinfo {author}
  {\bibfnamefont {D.}~\bibnamefont {Jones}}, \bibinfo {author} {\bibfnamefont
  {S.}~\bibnamefont {Kawasaki}}, \bibinfo {author} {\bibfnamefont
  {T.}~\bibnamefont {Kikawa}}, \bibinfo {author} {\bibfnamefont
  {M.}~\bibnamefont {Kitaguchi}}, \bibinfo {author} {\bibfnamefont
  {W.}~\bibnamefont {Klassen}}, \bibinfo {author} {\bibfnamefont
  {A.}~\bibnamefont {Konaka}}, \bibinfo {author} {\bibfnamefont
  {E.}~\bibnamefont {Korkmaz}}, \bibinfo {author} {\bibfnamefont
  {F.}~\bibnamefont {Kuchler}}, \bibinfo {author} {\bibfnamefont
  {M.}~\bibnamefont {Lang}}, \bibinfo {author} {\bibfnamefont {L.}~\bibnamefont
  {Lee}}, \bibinfo {author} {\bibfnamefont {T.}~\bibnamefont {Lindner}},
  \bibinfo {author} {\bibfnamefont {K.~W.}\ \bibnamefont {Madison}}, \bibinfo
  {author} {\bibfnamefont {Y.}~\bibnamefont {Makida}}, \bibinfo {author}
  {\bibfnamefont {J.}~\bibnamefont {Mammei}}, \bibinfo {author} {\bibfnamefont
  {R.}~\bibnamefont {Mammei}}, \bibinfo {author} {\bibfnamefont {J.~W.}\
  \bibnamefont {Martin}}, \bibinfo {author} {\bibfnamefont {R.}~\bibnamefont
  {Matsumiya}}, \bibinfo {author} {\bibfnamefont {E.}~\bibnamefont {Miller}},
  \bibinfo {author} {\bibfnamefont {K.}~\bibnamefont {Mishima}}, \bibinfo
  {author} {\bibfnamefont {T.}~\bibnamefont {Momose}}, \bibinfo {author}
  {\bibfnamefont {T.}~\bibnamefont {Okamura}}, \bibinfo {author} {\bibfnamefont
  {S.}~\bibnamefont {Page}}, \bibinfo {author} {\bibfnamefont {R.}~\bibnamefont
  {Picker}}, \bibinfo {author} {\bibfnamefont {E.}~\bibnamefont {Pierre}},
  \bibinfo {author} {\bibfnamefont {W.~D.}\ \bibnamefont {Ramsay}}, \bibinfo
  {author} {\bibfnamefont {L.}~\bibnamefont {Rebenitsch}}, \bibinfo {author}
  {\bibfnamefont {F.}~\bibnamefont {Rehm}}, \bibinfo {author} {\bibfnamefont
  {W.}~\bibnamefont {Schreyer}}, \bibinfo {author} {\bibfnamefont {H.~M.}\
  \bibnamefont {Shimizu}}, \bibinfo {author} {\bibfnamefont {S.}~\bibnamefont
  {Sidhu}}, \bibinfo {author} {\bibfnamefont {A.}~\bibnamefont {Sikora}},
  \bibinfo {author} {\bibfnamefont {J.}~\bibnamefont {Smith}}, \bibinfo
  {author} {\bibfnamefont {I.}~\bibnamefont {Tanihata}}, \bibinfo {author}
  {\bibfnamefont {B.}~\bibnamefont {Thorsteinson}}, \bibinfo {author}
  {\bibfnamefont {S.}~\bibnamefont {Vanbergen}}, \bibinfo {author}
  {\bibfnamefont {W.~T.~H.}\ \bibnamefont {van Oers}}, \ and\ \bibinfo {author}
  {\bibfnamefont {Y.~X.}\ \bibnamefont {Watanabe}} (\bibinfo {collaboration}
  {TUCAN Collaboration}),\ }\href {\doibase 10.1103/PhysRevC.99.025503}
  {\bibfield  {journal} {\bibinfo  {journal} {Physical Review C}\ }\textbf
  {\bibinfo {volume} {99}},\ \bibinfo {pages} {025503} (\bibinfo {year}
  {2019})}\BibitemShut {NoStop}%
\bibitem [{\citenamefont {Serebrov}\ \emph {et~al.}(2009)\citenamefont
  {Serebrov}, \citenamefont {Mityuklyaev}, \citenamefont {Zakharov},
  \citenamefont {Erykalov}, \citenamefont {Onegin}, \citenamefont {Fomin},
  \citenamefont {Ilatovskiy}, \citenamefont {Orlov}, \citenamefont {Konoplev},
  \citenamefont {Krivshitch}, \citenamefont {Samsonov}, \citenamefont {Ezhov},
  \citenamefont {Fedorov}, \citenamefont {Keshyshev}, \citenamefont
  {Boldarev},\ and\ \citenamefont {Marchenko}}]{Serebrov2009b}%
  \BibitemOpen
  \bibfield  {author} {\bibinfo {author} {\bibfnamefont {A.}~\bibnamefont
  {Serebrov}}, \bibinfo {author} {\bibfnamefont {V.}~\bibnamefont
  {Mityuklyaev}}, \bibinfo {author} {\bibfnamefont {A.}~\bibnamefont
  {Zakharov}}, \bibinfo {author} {\bibfnamefont {A.}~\bibnamefont {Erykalov}},
  \bibinfo {author} {\bibfnamefont {M.}~\bibnamefont {Onegin}}, \bibinfo
  {author} {\bibfnamefont {A.}~\bibnamefont {Fomin}}, \bibinfo {author}
  {\bibfnamefont {V.}~\bibnamefont {Ilatovskiy}}, \bibinfo {author}
  {\bibfnamefont {S.}~\bibnamefont {Orlov}}, \bibinfo {author} {\bibfnamefont
  {K.}~\bibnamefont {Konoplev}}, \bibinfo {author} {\bibfnamefont
  {A.}~\bibnamefont {Krivshitch}}, \bibinfo {author} {\bibfnamefont
  {V.}~\bibnamefont {Samsonov}}, \bibinfo {author} {\bibfnamefont
  {V.}~\bibnamefont {Ezhov}}, \bibinfo {author} {\bibfnamefont
  {V.}~\bibnamefont {Fedorov}}, \bibinfo {author} {\bibfnamefont
  {K.}~\bibnamefont {Keshyshev}}, \bibinfo {author} {\bibfnamefont
  {S.}~\bibnamefont {Boldarev}}, \ and\ \bibinfo {author} {\bibfnamefont
  {V.}~\bibnamefont {Marchenko}},\ }\href {\doibase
  http://dx.doi.org/10.1016/j.nima.2009.07.078} {\bibfield  {journal} {\bibinfo
   {journal} {Nuclear Instruments and Methods in Physics Research Section A}\
  }\textbf {\bibinfo {volume} {611}},\ \bibinfo {pages} {276} (\bibinfo {year}
  {2009})}\BibitemShut {NoStop}%
\bibitem [{\citenamefont {Serebrov}\ \emph {et~al.}(2010)\citenamefont
  {Serebrov}, \citenamefont {Mityukhlyaev}, \citenamefont {Zakharov},
  \citenamefont {Erykalov}, \citenamefont {Onegin}, \citenamefont {Fomin},
  \citenamefont {Ilatovskiy}, \citenamefont {Orlov}, \citenamefont {Konoplev},
  \citenamefont {Krivshich}, \citenamefont {Samsonov}, \citenamefont {Ezhov},
  \citenamefont {Fedorov}, \citenamefont {Keshishev}, \citenamefont
  {Boldarev},\ and\ \citenamefont {Marchenko}}]{Serebrov2010}%
  \BibitemOpen
  \bibfield  {author} {\bibinfo {author} {\bibfnamefont {A.~P.}\ \bibnamefont
  {Serebrov}}, \bibinfo {author} {\bibfnamefont {V.~A.}\ \bibnamefont
  {Mityukhlyaev}}, \bibinfo {author} {\bibfnamefont {A.~A.}\ \bibnamefont
  {Zakharov}}, \bibinfo {author} {\bibfnamefont {A.~N.}\ \bibnamefont
  {Erykalov}}, \bibinfo {author} {\bibfnamefont {M.~S.}\ \bibnamefont
  {Onegin}}, \bibinfo {author} {\bibfnamefont {A.~K.}\ \bibnamefont {Fomin}},
  \bibinfo {author} {\bibfnamefont {V.~A.}\ \bibnamefont {Ilatovskiy}},
  \bibinfo {author} {\bibfnamefont {S.~P.}\ \bibnamefont {Orlov}}, \bibinfo
  {author} {\bibfnamefont {K.~A.}\ \bibnamefont {Konoplev}}, \bibinfo {author}
  {\bibfnamefont {A.~G.}\ \bibnamefont {Krivshich}}, \bibinfo {author}
  {\bibfnamefont {V.~M.}\ \bibnamefont {Samsonov}}, \bibinfo {author}
  {\bibfnamefont {V.~F.}\ \bibnamefont {Ezhov}}, \bibinfo {author}
  {\bibfnamefont {V.~V.}\ \bibnamefont {Fedorov}}, \bibinfo {author}
  {\bibfnamefont {K.~O.}\ \bibnamefont {Keshishev}}, \bibinfo {author}
  {\bibfnamefont {S.~T.}\ \bibnamefont {Boldarev}}, \ and\ \bibinfo {author}
  {\bibfnamefont {V.~I.}\ \bibnamefont {Marchenko}},\ }\href {\doibase
  10.1134/S106378341005032X} {\bibfield  {journal} {\bibinfo  {journal}
  {Physics of the Solid State}\ }\textbf {\bibinfo {volume} {52}},\ \bibinfo
  {pages} {1034} (\bibinfo {year} {2010})}\BibitemShut {NoStop}%
\bibitem [{\citenamefont {Serebrov}\ \emph {et~al.}(2011)\citenamefont
  {Serebrov}, \citenamefont {Boldarev}, \citenamefont {Erykalov}, \citenamefont
  {Ezhov}, \citenamefont {Fedorov}, \citenamefont {Fomin}, \citenamefont
  {Ilatovskiy}, \citenamefont {Keshyshev}, \citenamefont {Konoplev},
  \citenamefont {Krivshitch}, \citenamefont {Marchenko}, \citenamefont
  {Mityuklyaev}, \citenamefont {Onegin}, \citenamefont {Orlov}, \citenamefont
  {Samsonov},\ and\ \citenamefont {Zakharov}}]{Serebrov2011}%
  \BibitemOpen
  \bibfield  {author} {\bibinfo {author} {\bibfnamefont {A.}~\bibnamefont
  {Serebrov}}, \bibinfo {author} {\bibfnamefont {S.}~\bibnamefont {Boldarev}},
  \bibinfo {author} {\bibfnamefont {A.}~\bibnamefont {Erykalov}}, \bibinfo
  {author} {\bibfnamefont {V.}~\bibnamefont {Ezhov}}, \bibinfo {author}
  {\bibfnamefont {V.}~\bibnamefont {Fedorov}}, \bibinfo {author} {\bibfnamefont
  {A.}~\bibnamefont {Fomin}}, \bibinfo {author} {\bibfnamefont
  {V.}~\bibnamefont {Ilatovskiy}}, \bibinfo {author} {\bibfnamefont
  {K.}~\bibnamefont {Keshyshev}}, \bibinfo {author} {\bibfnamefont
  {K.}~\bibnamefont {Konoplev}}, \bibinfo {author} {\bibfnamefont
  {A.}~\bibnamefont {Krivshitch}}, \bibinfo {author} {\bibfnamefont
  {V.}~\bibnamefont {Marchenko}}, \bibinfo {author} {\bibfnamefont
  {V.}~\bibnamefont {Mityuklyaev}}, \bibinfo {author} {\bibfnamefont
  {M.}~\bibnamefont {Onegin}}, \bibinfo {author} {\bibfnamefont
  {S.}~\bibnamefont {Orlov}}, \bibinfo {author} {\bibfnamefont
  {V.}~\bibnamefont {Samsonov}}, \ and\ \bibinfo {author} {\bibfnamefont
  {A.}~\bibnamefont {Zakharov}},\ }\href {\doibase
  http://dx.doi.org/10.1016/j.phpro.2011.06.044} {\bibfield  {journal}
  {\bibinfo  {journal} {Physics Procedia}\ }\textbf {\bibinfo {volume} {17}},\
  \bibinfo {pages} {251} (\bibinfo {year} {2011})}\BibitemShut {NoStop}%
\bibitem [{\citenamefont {Serebrov}\ and\ \citenamefont
  {Fomin}(2015)}]{Serebrov2015a}%
  \BibitemOpen
  \bibfield  {author} {\bibinfo {author} {\bibfnamefont {A.~P.}\ \bibnamefont
  {Serebrov}}\ and\ \bibinfo {author} {\bibfnamefont {A.~K.}\ \bibnamefont
  {Fomin}},\ }\href {\doibase 10.1134/S106378421508023X} {\bibfield  {journal}
  {\bibinfo  {journal} {Technical Physics}\ }\textbf {\bibinfo {volume} {60}},\
  \bibinfo {pages} {1238} (\bibinfo {year} {2015})}\BibitemShut {NoStop}%
\bibitem [{\citenamefont {Serebrov}(2018)}]{Serebrov2018}%
  \BibitemOpen
  \bibfield  {author} {\bibinfo {author} {\bibfnamefont {A.~P.}\ \bibnamefont
  {Serebrov}},\ }\href {\doibase 10.1134/S1063778818020175} {\bibfield
  {journal} {\bibinfo  {journal} {Physics of Atomic Nuclei}\ }\textbf {\bibinfo
  {volume} {81}},\ \bibinfo {pages} {214} (\bibinfo {year} {2018})}\BibitemShut
  {NoStop}%
\bibitem [{\citenamefont {Zimmer}\ \emph {et~al.}(2007)\citenamefont {Zimmer},
  \citenamefont {Baumann}, \citenamefont {Fertl}, \citenamefont {Franke},
  \citenamefont {Mironov}, \citenamefont {Plonka}, \citenamefont {Rich},
  \citenamefont {Schmidt-Wellenburg}, \citenamefont {Wirth},\ and\
  \citenamefont {van~den Brandt}}]{Zimmer2007a}%
  \BibitemOpen
  \bibfield  {author} {\bibinfo {author} {\bibfnamefont {O.}~\bibnamefont
  {Zimmer}}, \bibinfo {author} {\bibfnamefont {K.}~\bibnamefont {Baumann}},
  \bibinfo {author} {\bibfnamefont {M.}~\bibnamefont {Fertl}}, \bibinfo
  {author} {\bibfnamefont {B.}~\bibnamefont {Franke}}, \bibinfo {author}
  {\bibfnamefont {S.}~\bibnamefont {Mironov}}, \bibinfo {author} {\bibfnamefont
  {C.}~\bibnamefont {Plonka}}, \bibinfo {author} {\bibfnamefont
  {D.}~\bibnamefont {Rich}}, \bibinfo {author} {\bibfnamefont {P.}~\bibnamefont
  {Schmidt-Wellenburg}}, \bibinfo {author} {\bibfnamefont {H.~F.}\ \bibnamefont
  {Wirth}}, \ and\ \bibinfo {author} {\bibfnamefont {B.}~\bibnamefont {van~den
  Brandt}},\ }\href@noop {} {\bibfield  {journal} {\bibinfo  {journal}
  {Physical Review Letters}\ }\textbf {\bibinfo {volume} {99}},\ \bibinfo
  {pages} {104801} (\bibinfo {year} {2007})}\BibitemShut {NoStop}%
\bibitem [{\citenamefont {Piegsa}\ \emph {et~al.}(2014)\citenamefont {Piegsa},
  \citenamefont {Fertl}, \citenamefont {Ivanov}, \citenamefont {Kreuz},
  \citenamefont {Leung}, \citenamefont {Schmidt-Wellenburg}, \citenamefont
  {Soldner},\ and\ \citenamefont {Zimmer}}]{Piegsa2014}%
  \BibitemOpen
  \bibfield  {author} {\bibinfo {author} {\bibfnamefont {F.~M.}\ \bibnamefont
  {Piegsa}}, \bibinfo {author} {\bibfnamefont {M.}~\bibnamefont {Fertl}},
  \bibinfo {author} {\bibfnamefont {S.~N.}\ \bibnamefont {Ivanov}}, \bibinfo
  {author} {\bibfnamefont {M.}~\bibnamefont {Kreuz}}, \bibinfo {author}
  {\bibfnamefont {K.~K.~H.}\ \bibnamefont {Leung}}, \bibinfo {author}
  {\bibfnamefont {P.}~\bibnamefont {Schmidt-Wellenburg}}, \bibinfo {author}
  {\bibfnamefont {T.}~\bibnamefont {Soldner}}, \ and\ \bibinfo {author}
  {\bibfnamefont {O.}~\bibnamefont {Zimmer}},\ }\href {\doibase
  10.1103/PhysRevC.90.015501} {\bibfield  {journal} {\bibinfo  {journal}
  {Physical Review C}\ }\textbf {\bibinfo {volume} {90}},\ \bibinfo {pages}
  {015501} (\bibinfo {year} {2014})}\BibitemShut {NoStop}%
\bibitem [{\citenamefont {Leung}\ \emph {et~al.}(2016)\citenamefont {Leung},
  \citenamefont {Ivanov}, \citenamefont {Piegsa}, \citenamefont {Simson},\ and\
  \citenamefont {Zimmer}}]{Leung2016}%
  \BibitemOpen
  \bibfield  {author} {\bibinfo {author} {\bibfnamefont {K.~K.~H.}\
  \bibnamefont {Leung}}, \bibinfo {author} {\bibfnamefont {S.}~\bibnamefont
  {Ivanov}}, \bibinfo {author} {\bibfnamefont {F.~M.}\ \bibnamefont {Piegsa}},
  \bibinfo {author} {\bibfnamefont {M.}~\bibnamefont {Simson}}, \ and\ \bibinfo
  {author} {\bibfnamefont {O.}~\bibnamefont {Zimmer}},\ }\href {\doibase
  10.1103/PhysRevC.93.025501} {\bibfield  {journal} {\bibinfo  {journal}
  {Physical Review C}\ }\textbf {\bibinfo {volume} {93}},\ \bibinfo {pages}
  {025501} (\bibinfo {year} {2016})}\BibitemShut {NoStop}%
\bibitem [{\citenamefont {Zimmer}(2018)}]{Zimmer2018}%
  \BibitemOpen
  \bibfield  {author} {\bibinfo {author} {\bibfnamefont {O.}~\bibnamefont
  {Zimmer}},\ }\href@noop {} {\enquote {\bibinfo {title} {Private
  communication},}\ } (\bibinfo {year} {2018})\BibitemShut {NoStop}%
\bibitem [{\citenamefont {Zimmer}\ and\ \citenamefont
  {Golub}(2015)}]{Zimmer2015}%
  \BibitemOpen
  \bibfield  {author} {\bibinfo {author} {\bibfnamefont {O.}~\bibnamefont
  {Zimmer}}\ and\ \bibinfo {author} {\bibfnamefont {R.}~\bibnamefont {Golub}},\
  }\href {\doibase 10.1103/PhysRevC.92.015501} {\bibfield  {journal} {\bibinfo
  {journal} {Physical Review C}\ }\textbf {\bibinfo {volume} {92}},\ \bibinfo
  {pages} {015501} (\bibinfo {year} {2015})}\BibitemShut {NoStop}%
\bibitem [{\citenamefont {Young}\ \emph
  {et~al.}(2014{\natexlab{b}})\citenamefont {Young}, \citenamefont {Huegle},
  \citenamefont {Makela}, \citenamefont {Morris}, \citenamefont {Muhrer},\ and\
  \citenamefont {Saunders}}]{Young2014}%
  \BibitemOpen
  \bibfield  {author} {\bibinfo {author} {\bibfnamefont {A.}~\bibnamefont
  {Young}}, \bibinfo {author} {\bibfnamefont {T.}~\bibnamefont {Huegle}},
  \bibinfo {author} {\bibfnamefont {M.}~\bibnamefont {Makela}}, \bibinfo
  {author} {\bibfnamefont {C.}~\bibnamefont {Morris}}, \bibinfo {author}
  {\bibfnamefont {G.}~\bibnamefont {Muhrer}}, \ and\ \bibinfo {author}
  {\bibfnamefont {A.}~\bibnamefont {Saunders}},\ }\href {\doibase
  http://dx.doi.org/10.1016/j.phpro.2013.12.021} {\bibfield  {journal}
  {\bibinfo  {journal} {Physics Procedia}\ }\textbf {\bibinfo {volume} {51}},\
  \bibinfo {pages} {93 } (\bibinfo {year} {2014}{\natexlab{b}})}\BibitemShut
  {NoStop}%
\bibitem [{\citenamefont {Mocko}\ and\ \citenamefont
  {Muhrer}(2013)}]{Mocko2013}%
  \BibitemOpen
  \bibfield  {author} {\bibinfo {author} {\bibfnamefont {M.}~\bibnamefont
  {Mocko}}\ and\ \bibinfo {author} {\bibfnamefont {G.}~\bibnamefont {Muhrer}},\
  }\href {\doibase http://dx.doi.org/10.1016/j.nima.2012.11.103} {\bibfield
  {journal} {\bibinfo  {journal} {Nuclear Instruments and Methods in Physics
  Research Section A}\ }\textbf {\bibinfo {volume} {704}},\ \bibinfo {pages}
  {27 } (\bibinfo {year} {2013})}\BibitemShut {NoStop}%
\bibitem [{\citenamefont {Carpenter}\ \emph {et~al.}(2005)\citenamefont
  {Carpenter}, \citenamefont {Micklich}, \citenamefont {Richardson},
  \citenamefont {Littrell}, \citenamefont {Loong}, \citenamefont {Teller},
  \citenamefont {Agamalian}, \citenamefont {Arai}, \citenamefont {Benmore},
  \citenamefont {Bennington}, \citenamefont {Brod}, \citenamefont {Carlisle},
  \citenamefont {Crawford}, \citenamefont {Daum}, \citenamefont {Felcher},
  \citenamefont {Gahler},\ and\ \citenamefont {Geltenbort}}]{Carpenter2005}%
  \BibitemOpen
  \bibfield  {author} {\bibinfo {author} {\bibfnamefont {J.}~\bibnamefont
  {Carpenter}}, \bibinfo {author} {\bibfnamefont {B.}~\bibnamefont {Micklich}},
  \bibinfo {author} {\bibfnamefont {J.}~\bibnamefont {Richardson}}, \bibinfo
  {author} {\bibfnamefont {K.}~\bibnamefont {Littrell}}, \bibinfo {author}
  {\bibfnamefont {C.-K.}\ \bibnamefont {Loong}}, \bibinfo {author}
  {\bibfnamefont {R.}~\bibnamefont {Teller}}, \bibinfo {author} {\bibfnamefont
  {M.}~\bibnamefont {Agamalian}}, \bibinfo {author} {\bibfnamefont
  {M.}~\bibnamefont {Arai}}, \bibinfo {author} {\bibfnamefont {C.}~\bibnamefont
  {Benmore}}, \bibinfo {author} {\bibfnamefont {S.}~\bibnamefont {Bennington}},
  \bibinfo {author} {\bibfnamefont {P.}~\bibnamefont {Brod}}, \bibinfo {author}
  {\bibfnamefont {C.}~\bibnamefont {Carlisle}}, \bibinfo {author}
  {\bibfnamefont {K.}~\bibnamefont {Crawford}}, \bibinfo {author}
  {\bibfnamefont {M.}~\bibnamefont {Daum}}, \bibinfo {author} {\bibfnamefont
  {G.}~\bibnamefont {Felcher}}, \bibinfo {author} {\bibfnamefont
  {R.}~\bibnamefont {Gahler}}, \ and\ \bibinfo {author} {\bibfnamefont
  {P.}~\bibnamefont {Geltenbort}},\ }in\ \href {www.ipns.gov} {\emph {\bibinfo
  {booktitle} {Workshop on Applications of a Very Cold Neutron Source
  (08/21/2005 - 08/24/2005)}}}\ (\bibinfo {year} {2005})\BibitemShut {NoStop}%
\bibitem [{\citenamefont {Micklich}(2007)}]{Micklich2007}%
  \BibitemOpen
  \bibfield  {author} {\bibinfo {author} {\bibfnamefont {B.}~\bibnamefont
  {Micklich}},\ }\href {\doibase arxiv:1908.03550} {\bibfield  {journal}
  {\bibinfo  {journal} {Conference proceedings from the 3rd High-Power Targetry
  Workshop September 10-14, 2007, Bad Zurzach, Switzerland}\ } (\bibinfo {year}
  {2007}),\ arxiv:1908.03550}\BibitemShut {NoStop}%
\bibitem [{\citenamefont {LeBrun}(1994)}]{LeBrun1994}%
  \BibitemOpen
  \bibfield  {author} {\bibinfo {author} {\bibfnamefont {P.}~\bibnamefont
  {LeBrun}},\ }in\ \href@noop {} {\emph {\bibinfo {booktitle} {15th
  International Cryogenic Engineering Conference {ICEC 15} June 7-10, Genova,
  Italy}}}\ (\bibinfo {year} {1994})\BibitemShut {NoStop}%
\bibitem [{\citenamefont {Lebrun}\ and\ \citenamefont
  {Tavian}(2014)}]{Lebrun2014}%
  \BibitemOpen
  \bibfield  {author} {\bibinfo {author} {\bibfnamefont {P.}~\bibnamefont
  {Lebrun}}\ and\ \bibinfo {author} {\bibfnamefont {L.}~\bibnamefont
  {Tavian}},\ }\href {http://cds.cern.ch/record/1974065} {\bibfield  {journal}
  {\bibinfo  {journal} {CAS-CERN Accelerator School: Superconductivity for
  Accelerators, Erice, Italy, 24 April - 4 May 2013, edited by R. Bailey}\ ,\
  \bibinfo {pages} {24 p}} (\bibinfo {year} {2014})}\BibitemShut {NoStop}%
\bibitem [{\citenamefont {Gudkov}(2007)}]{Gudkov2007}%
  \BibitemOpen
  \bibfield  {author} {\bibinfo {author} {\bibfnamefont {V.}~\bibnamefont
  {Gudkov}},\ }\href {\doibase http://dx.doi.org/10.1016/j.nima.2007.06.043}
  {\bibfield  {journal} {\bibinfo  {journal} {Nuclear Instruments and Methods
  in Physics Research Section A}\ }\textbf {\bibinfo {volume} {580}},\ \bibinfo
  {pages} {1390 } (\bibinfo {year} {2007})}\BibitemShut {NoStop}%
\bibitem [{\citenamefont {Van~Sciver}(2012)}]{Van-Sciver2012}%
  \BibitemOpen
  \bibfield  {author} {\bibinfo {author} {\bibfnamefont {S.}~\bibnamefont
  {Van~Sciver}},\ }\href@noop {} {\emph {\bibinfo {title} {Helium
  Cryogenics}}},\ International Cryogenics Monograph Series\ (\bibinfo
  {publisher} {Springer New York},\ \bibinfo {year} {2012})\BibitemShut
  {NoStop}%
\bibitem [{\citenamefont {Pobell}(1996)}]{Pobell1996}%
  \BibitemOpen
  \bibfield  {author} {\bibinfo {author} {\bibfnamefont {F.}~\bibnamefont
  {Pobell}},\ }\href@noop {} {\emph {\bibinfo {title} {Matter and Methods at
  Low Temperatures}}}\ (\bibinfo  {publisher} {Springer-Verlag},\ \bibinfo
  {year} {1996})\BibitemShut {NoStop}%
\bibitem [{\citenamefont {Zimmer}(2014)}]{Zimmer2014}%
  \BibitemOpen
  \bibfield  {author} {\bibinfo {author} {\bibfnamefont {O.}~\bibnamefont
  {Zimmer}},\ }\href {\doibase https://doi.org/10.1016/j.phpro.2013.12.019}
  {\bibfield  {journal} {\bibinfo  {journal} {Physics Procedia}\ }\textbf
  {\bibinfo {volume} {51}},\ \bibinfo {pages} {85 } (\bibinfo {year}
  {2014})}\BibitemShut {NoStop}%
\bibitem [{\citenamefont {Lisowski}\ and\ \citenamefont
  {Schoenberg}(2006)}]{Lisowski2006}%
  \BibitemOpen
  \bibfield  {author} {\bibinfo {author} {\bibfnamefont {P.~W.}\ \bibnamefont
  {Lisowski}}\ and\ \bibinfo {author} {\bibfnamefont {K.~F.}\ \bibnamefont
  {Schoenberg}},\ }\href {\doibase
  http://dx.doi.org/10.1016/j.nima.2006.02.178} {\bibfield  {journal} {\bibinfo
   {journal} {Nuclear Instruments and Methods in Physics Research Section A}\
  }\textbf {\bibinfo {volume} {562}},\ \bibinfo {pages} {910 } (\bibinfo {year}
  {2006})}\BibitemShut {NoStop}%
\bibitem [{\citenamefont {Clendenin}\ \emph {et~al.}(1996)\citenamefont
  {Clendenin}, \citenamefont {Rinolfi}, \citenamefont {Takata},\ and\
  \citenamefont {Warner}}]{Clendenin1996}%
  \BibitemOpen
  \bibfield  {author} {\bibinfo {author} {\bibfnamefont {J.}~\bibnamefont
  {Clendenin}}, \bibinfo {author} {\bibfnamefont {L.}~\bibnamefont {Rinolfi}},
  \bibinfo {author} {\bibfnamefont {K.}~\bibnamefont {Takata}}, \ and\ \bibinfo
  {author} {\bibfnamefont {D.}~\bibnamefont {Warner}},\ }\href@noop {}
  {\bibfield  {journal} {\bibinfo  {journal} {Proceedings of {XVIII}
  International Linac Conference}\ ,\ \bibinfo {pages} {52}} (\bibinfo {year}
  {1996})}\BibitemShut {NoStop}%
\bibitem [{\citenamefont {Sinnis}(2019)}]{Sinnis2019}%
  \BibitemOpen
  \bibfield  {author} {\bibinfo {author} {\bibfnamefont {C.}~\bibnamefont
  {Sinnis}},\ }\href@noop {} {\enquote {\bibinfo {title} {A strategy for
  {LANSCE} futures (draft)},}\ }\bibinfo {howpublished} {Los Alamos Report
  LAUR-18-20953} (\bibinfo {year} {2019})\BibitemShut {NoStop}%
\bibitem [{\citenamefont {Bauer}(2001)}]{Bauer2001}%
  \BibitemOpen
  \bibfield  {author} {\bibinfo {author} {\bibfnamefont {G.}~\bibnamefont
  {Bauer}},\ }\href {\doibase https://doi.org/10.1016/S0168-9002(01)00167-X}
  {\bibfield  {journal} {\bibinfo  {journal} {Nuclear Instruments and Methods
  in Physics Research Section A}\ }\textbf {\bibinfo {volume} {463}},\ \bibinfo
  {pages} {505 } (\bibinfo {year} {2001})},\ \bibinfo {note} {accelerator
  driven systems}\BibitemShut {NoStop}%
\bibitem [{\citenamefont {Howell}\ \emph {et~al.}(2017)\citenamefont {Howell},
  \citenamefont {DeGraff}, \citenamefont {Galambos},\ and\ \citenamefont
  {Kim}}]{Howell2017}%
  \BibitemOpen
  \bibfield  {author} {\bibinfo {author} {\bibfnamefont {M.}~\bibnamefont
  {Howell}}, \bibinfo {author} {\bibfnamefont {B.}~\bibnamefont {DeGraff}},
  \bibinfo {author} {\bibfnamefont {J.}~\bibnamefont {Galambos}}, \ and\
  \bibinfo {author} {\bibfnamefont {S.-H.}\ \bibnamefont {Kim}},\ }\href
  {http://stacks.iop.org/1757-899X/278/i=1/a=012185} {\bibfield  {journal}
  {\bibinfo  {journal} {IOP Conference Series: Materials Science and
  Engineering}\ }\textbf {\bibinfo {volume} {278}},\ \bibinfo {pages} {012185}
  (\bibinfo {year} {2017})}\BibitemShut {NoStop}%
\bibitem [{\citenamefont {Korobkina}\ \emph {et~al.}(2002)\citenamefont
  {Korobkina}, \citenamefont {Golub}, \citenamefont {Wehring},\ and\
  \citenamefont {Young}}]{Korobkina2002}%
  \BibitemOpen
  \bibfield  {author} {\bibinfo {author} {\bibfnamefont {E.}~\bibnamefont
  {Korobkina}}, \bibinfo {author} {\bibfnamefont {R.}~\bibnamefont {Golub}},
  \bibinfo {author} {\bibfnamefont {B.}~\bibnamefont {Wehring}}, \ and\
  \bibinfo {author} {\bibfnamefont {A.}~\bibnamefont {Young}},\ }\href
  {\doibase https://doi.org/10.1016/S0375-9601(02)01052-6} {\bibfield
  {journal} {\bibinfo  {journal} {Physics Letters A}\ }\textbf {\bibinfo
  {volume} {301}},\ \bibinfo {pages} {462 } (\bibinfo {year}
  {2002})}\BibitemShut {NoStop}%
\bibitem [{\citenamefont {Andersen}\ \emph {et~al.}(1994)\citenamefont
  {Andersen}, \citenamefont {Stirling}, \citenamefont {Scherm}, \citenamefont
  {Stunault},\ and\ \citenamefont {F{\aa}k}}]{Andersen1994a}%
  \BibitemOpen
  \bibfield  {author} {\bibinfo {author} {\bibfnamefont {K.~H.}\ \bibnamefont
  {Andersen}}, \bibinfo {author} {\bibfnamefont {W.~G.}\ \bibnamefont
  {Stirling}}, \bibinfo {author} {\bibfnamefont {R.}~\bibnamefont {Scherm}},
  \bibinfo {author} {\bibfnamefont {A.}~\bibnamefont {Stunault}}, \ and\
  \bibinfo {author} {\bibfnamefont {B.}~\bibnamefont {F{\aa}k}},\ }\href@noop
  {} {\bibfield  {journal} {\bibinfo  {journal} {Journal of Physics: Condensed
  Matter}\ }\textbf {\bibinfo {volume} {6}},\ \bibinfo {pages} {821} (\bibinfo
  {year} {1994})}\BibitemShut {NoStop}%
\bibitem [{\citenamefont {Andersen}(2015)}]{Andersen2015}%
  \BibitemOpen
  \bibfield  {author} {\bibinfo {author} {\bibfnamefont {K.}~\bibnamefont
  {Andersen}},\ }\href@noop {} {}\bibinfo {howpublished} {Private
  communication} (\bibinfo {year} {2015})\BibitemShut {NoStop}%
\bibitem [{\citenamefont {Schmidt-Wellenburg}\ \emph
  {et~al.}(2015)\citenamefont {Schmidt-Wellenburg}, \citenamefont {Bossy},
  \citenamefont {Farhi}, \citenamefont {Fertl}, \citenamefont {Leung},
  \citenamefont {Rahli}, \citenamefont {Soldner},\ and\ \citenamefont
  {Zimmer}}]{Schmidt-Wellenburg2015}%
  \BibitemOpen
  \bibfield  {author} {\bibinfo {author} {\bibfnamefont {P.}~\bibnamefont
  {Schmidt-Wellenburg}}, \bibinfo {author} {\bibfnamefont {J.}~\bibnamefont
  {Bossy}}, \bibinfo {author} {\bibfnamefont {E.}~\bibnamefont {Farhi}},
  \bibinfo {author} {\bibfnamefont {M.}~\bibnamefont {Fertl}}, \bibinfo
  {author} {\bibfnamefont {K.~K.~H.}\ \bibnamefont {Leung}}, \bibinfo {author}
  {\bibfnamefont {A.}~\bibnamefont {Rahli}}, \bibinfo {author} {\bibfnamefont
  {T.}~\bibnamefont {Soldner}}, \ and\ \bibinfo {author} {\bibfnamefont
  {O.}~\bibnamefont {Zimmer}},\ }\href {\doibase 10.1103/PhysRevC.92.024004}
  {\bibfield  {journal} {\bibinfo  {journal} {Physical Review C}\ }\textbf
  {\bibinfo {volume} {92}},\ \bibinfo {pages} {024004} (\bibinfo {year}
  {2015})}\BibitemShut {NoStop}%
\bibitem [{\citenamefont {Mezei}\ and\ \citenamefont
  {Stirling}(1983)}]{Mezei1983}%
  \BibitemOpen
  \bibfield  {author} {\bibinfo {author} {\bibfnamefont {F.}~\bibnamefont
  {Mezei}}\ and\ \bibinfo {author} {\bibfnamefont {W.~G.}\ \bibnamefont
  {Stirling}},\ }in\ \href
  {http://inis.iaea.org/search/search.aspx?orig_q=RN:19030137} {\emph {\bibinfo
  {booktitle} {The 75th jubilee conference on helium-4}}},\ \bibinfo {editor}
  {edited by\ \bibinfo {editor} {\bibfnamefont {J.~G.~M.}\ \bibnamefont
  {Armitage}}}\ (\bibinfo  {publisher} {World Scientific Pub Co},\ \bibinfo
  {address} {United States},\ \bibinfo {year} {1983})\BibitemShut {NoStop}%
\bibitem [{\citenamefont {{D.B. Pelowitz, Ed.}}(2011)}]{MCNPX}%
  \BibitemOpen
  \bibfield  {author} {\bibinfo {author} {\bibnamefont {{D.B. Pelowitz,
  Ed.}}},\ }\href@noop {} {\enquote {\bibinfo {title} {{MCNPX} users manual
  version 2.7.0 {LA-CP-11-00438}},}\ } (\bibinfo {year} {2011})\BibitemShut
  {NoStop}%
\bibitem [{\citenamefont {Berger}\ \emph {et~al.}(2005)\citenamefont {Berger},
  \citenamefont {Coursey}, \citenamefont {Zucker},\ and\ \citenamefont
  {Chang}}]{STAR}%
  \BibitemOpen
  \bibfield  {author} {\bibinfo {author} {\bibfnamefont {M.}~\bibnamefont
  {Berger}}, \bibinfo {author} {\bibfnamefont {J.}~\bibnamefont {Coursey}},
  \bibinfo {author} {\bibfnamefont {M.}~\bibnamefont {Zucker}}, \ and\ \bibinfo
  {author} {\bibfnamefont {J.}~\bibnamefont {Chang}},\ }\href
  {http://physics.nist.gov/Star} {\bibfield  {journal} {\bibinfo  {journal}
  {National Institute of Standards and Technology, Gaithersburg, MD.}\ }
  (\bibinfo {year} {2005})}\BibitemShut {NoStop}%
\bibitem [{\citenamefont {Bultman}(1998)}]{Bultman1998}%
  \BibitemOpen
  \bibfield  {author} {\bibinfo {author} {\bibfnamefont {N.~K.}\ \bibnamefont
  {Bultman}},\ }\href@noop {} {\bibfield  {journal} {\bibinfo  {journal}
  {Proceedings of ICANS XIV, June 14-19, Utica, IL, USA}\ ,\ \bibinfo {pages}
  {345}} (\bibinfo {year} {1998})}\BibitemShut {NoStop}%
\bibitem [{\citenamefont {Shea}\ \emph {et~al.}(2013)\citenamefont {Shea},
  \citenamefont {Pitcher}, \citenamefont {Thomsen}, \citenamefont {Andersen},
  \citenamefont {Bentley}, \citenamefont {Henry}, \citenamefont {Sabbagh},
  \citenamefont {Takibayev},\ and\ \citenamefont {M{\o}ller}}]{Shea2013}%
  \BibitemOpen
  \bibfield  {author} {\bibinfo {author} {\bibfnamefont {T.~J.}\ \bibnamefont
  {Shea}}, \bibinfo {author} {\bibfnamefont {E.~J.}\ \bibnamefont {Pitcher}},
  \bibinfo {author} {\bibfnamefont {H.~D.}\ \bibnamefont {Thomsen}}, \bibinfo
  {author} {\bibfnamefont {K.}~\bibnamefont {Andersen}}, \bibinfo {author}
  {\bibfnamefont {P.}~\bibnamefont {Bentley}}, \bibinfo {author} {\bibfnamefont
  {P.}~\bibnamefont {Henry}}, \bibinfo {author} {\bibfnamefont
  {P.}~\bibnamefont {Sabbagh}}, \bibinfo {author} {\bibfnamefont
  {A.}~\bibnamefont {Takibayev}}, \ and\ \bibinfo {author} {\bibfnamefont
  {S.~P.}\ \bibnamefont {M{\o}ller}},\ }\href@noop {} {\bibfield  {journal}
  {\bibinfo  {journal} {Proceedings of PAC2013 (Pasadena, CA)}\ ,\ \bibinfo
  {pages} {MOPMA04}} (\bibinfo {year} {2013})}\BibitemShut {NoStop}%
\bibitem [{\citenamefont {Shea}\ \emph {et~al.}(2014)\citenamefont {Shea},
  \citenamefont {Coney}, \citenamefont {Linander}, \citenamefont {Jansson},
  \citenamefont {Pitcher}, \citenamefont {Nordt}, \citenamefont {Thomas},\ and\
  \citenamefont {Thomsen}}]{Shea2014}%
  \BibitemOpen
  \bibfield  {author} {\bibinfo {author} {\bibfnamefont {T.~J.}\ \bibnamefont
  {Shea}}, \bibinfo {author} {\bibfnamefont {L.}~\bibnamefont {Coney}},
  \bibinfo {author} {\bibfnamefont {R.}~\bibnamefont {Linander}}, \bibinfo
  {author} {\bibfnamefont {A.}~\bibnamefont {Jansson}}, \bibinfo {author}
  {\bibfnamefont {E.~J.}\ \bibnamefont {Pitcher}}, \bibinfo {author}
  {\bibfnamefont {A.}~\bibnamefont {Nordt}}, \bibinfo {author} {\bibfnamefont
  {C.}~\bibnamefont {Thomas}}, \ and\ \bibinfo {author} {\bibfnamefont {H.~D.}\
  \bibnamefont {Thomsen}},\ }\href@noop {} {\bibfield  {journal} {\bibinfo
  {journal} {International Collaboration of Advanced Neutron Sources ICANS XXI
  (Mito, Japan)}\ } (\bibinfo {year} {2014})}\BibitemShut {NoStop}%
\bibitem [{\citenamefont {Thomsen}\ and\ \citenamefont
  {M{\o}ller}(2014)}]{Thomsen2014}%
  \BibitemOpen
  \bibfield  {author} {\bibinfo {author} {\bibfnamefont {H.}~\bibnamefont
  {Thomsen}}\ and\ \bibinfo {author} {\bibfnamefont {S.}~\bibnamefont
  {M{\o}ller}},\ }\href@noop {} {\bibfield  {journal} {\bibinfo  {journal}
  {IPAC14}\ } (\bibinfo {year} {2014})}\BibitemShut {NoStop}%
\bibitem [{\citenamefont {Shea}(2019)}]{Shea2019}%
  \BibitemOpen
  \bibfield  {author} {\bibinfo {author} {\bibfnamefont {T.~J.}\ \bibnamefont
  {Shea}},\ }\href@noop {} {\enquote {\bibinfo {title} {Private
  communication},}\ } (\bibinfo {year} {2019})\BibitemShut {NoStop}%
\bibitem [{\citenamefont {Russell}\ \emph {et~al.}(1997)\citenamefont
  {Russell}, \citenamefont {Ferguson}, \citenamefont {Pitcher},\ and\
  \citenamefont {Court}}]{Russell1997}%
  \BibitemOpen
  \bibfield  {author} {\bibinfo {author} {\bibfnamefont {G.~J.}\ \bibnamefont
  {Russell}}, \bibinfo {author} {\bibfnamefont {P.~D.}\ \bibnamefont
  {Ferguson}}, \bibinfo {author} {\bibfnamefont {E.~J.}\ \bibnamefont
  {Pitcher}}, \ and\ \bibinfo {author} {\bibfnamefont {J.~D.}\ \bibnamefont
  {Court}},\ }\href {\doibase http://dx.doi.org/10.1063/1.52599} {\bibfield
  {journal} {\bibinfo  {journal} {AIP Conference Proceedings}\ }\textbf
  {\bibinfo {volume} {392}},\ \bibinfo {pages} {361} (\bibinfo {year}
  {1997})}\BibitemShut {NoStop}%
\bibitem [{\citenamefont {Ino}\ \emph {et~al.}(2004)\citenamefont {Ino},
  \citenamefont {Ooi}, \citenamefont {Kiyanagi}, \citenamefont {Kasugai},
  \citenamefont {Maekawa}, \citenamefont {Takada}, \citenamefont {Muhrer},
  \citenamefont {Pitcher},\ and\ \citenamefont {Russell}}]{Ino2004}%
  \BibitemOpen
  \bibfield  {author} {\bibinfo {author} {\bibfnamefont {T.}~\bibnamefont
  {Ino}}, \bibinfo {author} {\bibfnamefont {M.}~\bibnamefont {Ooi}}, \bibinfo
  {author} {\bibfnamefont {Y.}~\bibnamefont {Kiyanagi}}, \bibinfo {author}
  {\bibfnamefont {Y.}~\bibnamefont {Kasugai}}, \bibinfo {author} {\bibfnamefont
  {F.}~\bibnamefont {Maekawa}}, \bibinfo {author} {\bibfnamefont
  {H.}~\bibnamefont {Takada}}, \bibinfo {author} {\bibfnamefont
  {G.}~\bibnamefont {Muhrer}}, \bibinfo {author} {\bibfnamefont {E.~J.}\
  \bibnamefont {Pitcher}}, \ and\ \bibinfo {author} {\bibfnamefont {G.~J.}\
  \bibnamefont {Russell}},\ }\href {\doibase
  http://dx.doi.org/10.1016/j.nima.2004.02.003} {\bibfield  {journal} {\bibinfo
   {journal} {Nuclear Instruments and Methods in Physics Research Section A}\
  }\textbf {\bibinfo {volume} {525}},\ \bibinfo {pages} {496} (\bibinfo {year}
  {2004})}\BibitemShut {NoStop}%
\bibitem [{\citenamefont {Muhrer}\ \emph {et~al.}(2004)\citenamefont {Muhrer},
  \citenamefont {Pitcher}, \citenamefont {Russell}, \citenamefont {Ino},
  \citenamefont {Ooi},\ and\ \citenamefont {Kiyanagi}}]{Muhrer2004}%
  \BibitemOpen
  \bibfield  {author} {\bibinfo {author} {\bibfnamefont {G.}~\bibnamefont
  {Muhrer}}, \bibinfo {author} {\bibfnamefont {E.}~\bibnamefont {Pitcher}},
  \bibinfo {author} {\bibfnamefont {G.}~\bibnamefont {Russell}}, \bibinfo
  {author} {\bibfnamefont {T.}~\bibnamefont {Ino}}, \bibinfo {author}
  {\bibfnamefont {M.}~\bibnamefont {Ooi}}, \ and\ \bibinfo {author}
  {\bibfnamefont {Y.}~\bibnamefont {Kiyanagi}},\ }\href {\doibase
  http://dx.doi.org/10.1016/j.nima.2004.03.191} {\bibfield  {journal} {\bibinfo
   {journal} {Nuclear Instruments and Methods in Physics Research Section A}\
  }\textbf {\bibinfo {volume} {527}},\ \bibinfo {pages} {531 } (\bibinfo {year}
  {2004})}\BibitemShut {NoStop}%
\bibitem [{\citenamefont {Muhrer}(2012)}]{Muhrer2012}%
  \BibitemOpen
  \bibfield  {author} {\bibinfo {author} {\bibfnamefont {G.}~\bibnamefont
  {Muhrer}},\ }\href {\doibase http://dx.doi.org/10.1016/j.nima.2011.10.018}
  {\bibfield  {journal} {\bibinfo  {journal} {Nuclear Instruments and Methods
  in Physics Research Section A}\ }\textbf {\bibinfo {volume} {664}},\ \bibinfo
  {pages} {38 } (\bibinfo {year} {2012})}\BibitemShut {NoStop}%
\bibitem [{\citenamefont {Kaplan}(1963)}]{Kaplan1963}%
  \BibitemOpen
  \bibfield  {author} {\bibinfo {author} {\bibfnamefont {I.}~\bibnamefont
  {Kaplan}},\ }\href@noop {} {\emph {\bibinfo {title} {Nuclear Physics (second
  ed.)}}}\ (\bibinfo  {publisher} {Addison-Wesley Publishing Company},\
  \bibinfo {year} {1963})\BibitemShut {NoStop}%
\bibitem [{\citenamefont {Goorley}\ \emph {et~al.}(2012)\citenamefont
  {Goorley}, \citenamefont {James}, \citenamefont {Booth}, \citenamefont
  {Brown}, \citenamefont {Bull}, \citenamefont {Cox}, \citenamefont {Durkee},
  \citenamefont {Elson}, \citenamefont {Fensin}, \citenamefont {Forster},
  \citenamefont {Hendricks}, \citenamefont {Hughes}, \citenamefont {Johns},
  \citenamefont {Kiedrowski}, \citenamefont {Martz}, \citenamefont {Mashnik},
  \citenamefont {McKinney}, \citenamefont {Pelowitz}, \citenamefont {Prael},
  \citenamefont {Sweezy}, \citenamefont {Waters}, \citenamefont {Wilcox},\ and\
  \citenamefont {Zukaitis}}]{Goorley2012}%
  \BibitemOpen
  \bibfield  {author} {\bibinfo {author} {\bibfnamefont {T.}~\bibnamefont
  {Goorley}}, \bibinfo {author} {\bibfnamefont {M.}~\bibnamefont {James}},
  \bibinfo {author} {\bibfnamefont {T.}~\bibnamefont {Booth}}, \bibinfo
  {author} {\bibfnamefont {F.}~\bibnamefont {Brown}}, \bibinfo {author}
  {\bibfnamefont {J.}~\bibnamefont {Bull}}, \bibinfo {author} {\bibfnamefont
  {L.~J.}\ \bibnamefont {Cox}}, \bibinfo {author} {\bibfnamefont
  {J.}~\bibnamefont {Durkee}}, \bibinfo {author} {\bibfnamefont
  {J.}~\bibnamefont {Elson}}, \bibinfo {author} {\bibfnamefont
  {M.}~\bibnamefont {Fensin}}, \bibinfo {author} {\bibfnamefont {R.~A.}\
  \bibnamefont {Forster}}, \bibinfo {author} {\bibfnamefont {J.}~\bibnamefont
  {Hendricks}}, \bibinfo {author} {\bibfnamefont {H.~G.}\ \bibnamefont
  {Hughes}}, \bibinfo {author} {\bibfnamefont {R.}~\bibnamefont {Johns}},
  \bibinfo {author} {\bibfnamefont {B.}~\bibnamefont {Kiedrowski}}, \bibinfo
  {author} {\bibfnamefont {R.}~\bibnamefont {Martz}}, \bibinfo {author}
  {\bibfnamefont {S.}~\bibnamefont {Mashnik}}, \bibinfo {author} {\bibfnamefont
  {G.}~\bibnamefont {McKinney}}, \bibinfo {author} {\bibfnamefont
  {D.}~\bibnamefont {Pelowitz}}, \bibinfo {author} {\bibfnamefont
  {R.}~\bibnamefont {Prael}}, \bibinfo {author} {\bibfnamefont
  {J.}~\bibnamefont {Sweezy}}, \bibinfo {author} {\bibfnamefont
  {L.}~\bibnamefont {Waters}}, \bibinfo {author} {\bibfnamefont
  {T.}~\bibnamefont {Wilcox}}, \ and\ \bibinfo {author} {\bibfnamefont
  {T.}~\bibnamefont {Zukaitis}},\ }\href {\doibase 10.13182/NT11-135}
  {\bibfield  {journal} {\bibinfo  {journal} {Nuclear Technology}\ }\textbf
  {\bibinfo {volume} {180}},\ \bibinfo {pages} {298} (\bibinfo {year}
  {2012})}\BibitemShut {NoStop}%
\bibitem [{\citenamefont {MacFarlane}\ and\ \citenamefont
  {Kahler}(2010)}]{MacFarlane2010}%
  \BibitemOpen
  \bibfield  {author} {\bibinfo {author} {\bibfnamefont {R.}~\bibnamefont
  {MacFarlane}}\ and\ \bibinfo {author} {\bibfnamefont {A.}~\bibnamefont
  {Kahler}},\ }\href {\doibase https://doi.org/10.1016/j.nds.2010.11.001}
  {\bibfield  {journal} {\bibinfo  {journal} {Nuclear Data Sheets}\ }\textbf
  {\bibinfo {volume} {111}},\ \bibinfo {pages} {2739} (\bibinfo {year}
  {2010})}\BibitemShut {NoStop}%
\bibitem [{\citenamefont {Cho}\ and\ \citenamefont {Chang}(2002)}]{Cho2002}%
  \BibitemOpen
  \bibfield  {author} {\bibinfo {author} {\bibfnamefont {Y.-S.}\ \bibnamefont
  {Cho}}\ and\ \bibinfo {author} {\bibfnamefont {J.}~\bibnamefont {Chang}},\
  }\href {\doibase 10.1080/00223131.2002.10875069} {\bibfield  {journal}
  {\bibinfo  {journal} {Journal of Nuclear Science and Technology}\ }\textbf
  {\bibinfo {volume} {39}},\ \bibinfo {pages} {176} (\bibinfo {year}
  {2002})}\BibitemShut {NoStop}%
\bibitem [{\citenamefont {Cubitt}\ \emph {et~al.}(2010)\citenamefont {Cubitt},
  \citenamefont {Lychagin}, \citenamefont {Muzychka}, \citenamefont {Nekhaev},
  \citenamefont {Nesvizhevsky}, \citenamefont {Pignol}, \citenamefont
  {Protasov},\ and\ \citenamefont {Strelkov}}]{Cubitt2010}%
  \BibitemOpen
  \bibfield  {author} {\bibinfo {author} {\bibfnamefont {R.}~\bibnamefont
  {Cubitt}}, \bibinfo {author} {\bibfnamefont {E.}~\bibnamefont {Lychagin}},
  \bibinfo {author} {\bibfnamefont {A.}~\bibnamefont {Muzychka}}, \bibinfo
  {author} {\bibfnamefont {G.}~\bibnamefont {Nekhaev}}, \bibinfo {author}
  {\bibfnamefont {V.}~\bibnamefont {Nesvizhevsky}}, \bibinfo {author}
  {\bibfnamefont {G.}~\bibnamefont {Pignol}}, \bibinfo {author} {\bibfnamefont
  {K.}~\bibnamefont {Protasov}}, \ and\ \bibinfo {author} {\bibfnamefont
  {A.}~\bibnamefont {Strelkov}},\ }\href {\doibase
  https://doi.org/10.1016/j.nima.2010.07.049} {\bibfield  {journal} {\bibinfo
  {journal} {Nuclear Instruments and Methods in Physics Research Section A}\
  }\textbf {\bibinfo {volume} {622}},\ \bibinfo {pages} {182} (\bibinfo {year}
  {2010})}\BibitemShut {NoStop}%
\bibitem [{\citenamefont {Turchin}(1965)}]{Turchin1965}%
  \BibitemOpen
  \bibfield  {author} {\bibinfo {author} {\bibfnamefont {V.~F.}\ \bibnamefont
  {Turchin}},\ }\href@noop {} {\emph {\bibinfo {title} {Slow Neutrons}}}\
  (\bibinfo  {publisher} {Israel Program for Scientific Translations},\
  \bibinfo {year} {1965})\BibitemShut {NoStop}%
\bibitem [{\citenamefont {Berk}(1993)}]{Berk1993}%
  \BibitemOpen
  \bibfield  {author} {\bibinfo {author} {\bibfnamefont {N.~F.}\ \bibnamefont
  {Berk}},\ }\href {\doibase doi:10.6028/jres.098.002} {\bibfield  {journal}
  {\bibinfo  {journal} {Journal of Research of the National Institute of
  Standards and Technology}\ }\textbf {\bibinfo {volume} {98}},\ \bibinfo
  {pages} {15} (\bibinfo {year} {1993})}\BibitemShut {NoStop}%
\bibitem [{\citenamefont {Fragopoulou}\ \emph {et~al.}(2006)\citenamefont
  {Fragopoulou}, \citenamefont {Manolopoulou}, \citenamefont {Stoulos},
  \citenamefont {Brandt}, \citenamefont {Westmeier}, \citenamefont
  {Krivopustov}, \citenamefont {Sosnin}, \citenamefont {Golovatyuk},\ and\
  \citenamefont {Zamani}}]{Fragopoulou2006}%
  \BibitemOpen
  \bibfield  {author} {\bibinfo {author} {\bibfnamefont {M.}~\bibnamefont
  {Fragopoulou}}, \bibinfo {author} {\bibfnamefont {M.}~\bibnamefont
  {Manolopoulou}}, \bibinfo {author} {\bibfnamefont {S.}~\bibnamefont
  {Stoulos}}, \bibinfo {author} {\bibfnamefont {R.}~\bibnamefont {Brandt}},
  \bibinfo {author} {\bibfnamefont {W.}~\bibnamefont {Westmeier}}, \bibinfo
  {author} {\bibfnamefont {M.}~\bibnamefont {Krivopustov}}, \bibinfo {author}
  {\bibfnamefont {A.}~\bibnamefont {Sosnin}}, \bibinfo {author} {\bibfnamefont
  {S.}~\bibnamefont {Golovatyuk}}, \ and\ \bibinfo {author} {\bibfnamefont
  {M.}~\bibnamefont {Zamani}},\ }\href {\doibase 10.1088/1742-6596/41/1/058}
  {\bibfield  {journal} {\bibinfo  {journal} {Journal of Physics: Conference
  Series}\ }\textbf {\bibinfo {volume} {41}},\ \bibinfo {pages} {514} (\bibinfo
  {year} {2006})}\BibitemShut {NoStop}%
\bibitem [{\citenamefont {Pokotilovski}(2016)}]{Pokotilovski2016}%
  \BibitemOpen
  \bibfield  {author} {\bibinfo {author} {\bibfnamefont {Y.~N.}\ \bibnamefont
  {Pokotilovski}},\ }\href {\doibase 10.1051/epjap/2016150073} {\bibfield
  {journal} {\bibinfo  {journal} {The European Physical Journal Applied
  Physics}\ }\textbf {\bibinfo {volume} {73}},\ \bibinfo {pages} {20302}
  (\bibinfo {year} {2016})}\BibitemShut {NoStop}%
\bibitem [{\citenamefont {Miranda}(1988)}]{Miranda1988}%
  \BibitemOpen
  \bibfield  {author} {\bibinfo {author} {\bibfnamefont {P.~C.}\ \bibnamefont
  {Miranda}},\ }\href {http://stacks.iop.org/0022-3727/21/i=9/a=003} {\bibfield
   {journal} {\bibinfo  {journal} {Journal of Physics D: Applied Physics}\
  }\textbf {\bibinfo {volume} {21}},\ \bibinfo {pages} {1326} (\bibinfo {year}
  {1988})}\BibitemShut {NoStop}%
\bibitem [{\citenamefont {Atchison}\ \emph {et~al.}(2010)\citenamefont
  {Atchison}, \citenamefont {Daum}, \citenamefont {Henneck}, \citenamefont
  {Heule}, \citenamefont {Horisberger}, \citenamefont {Kasprzak}, \citenamefont
  {Kirch}, \citenamefont {Knecht}, \citenamefont {Ku{\.z}niak}, \citenamefont
  {Lauss}, \citenamefont {Mtchedlishvili}, \citenamefont {Meier}, \citenamefont
  {Petzoldt}, \citenamefont {Plonka-Spehr}, \citenamefont {Schelldorfer},
  \citenamefont {Straumann},\ and\ \citenamefont {Zsigmond}}]{Atchison2010}%
  \BibitemOpen
  \bibfield  {author} {\bibinfo {author} {\bibfnamefont {F.}~\bibnamefont
  {Atchison}}, \bibinfo {author} {\bibfnamefont {M.}~\bibnamefont {Daum}},
  \bibinfo {author} {\bibfnamefont {R.}~\bibnamefont {Henneck}}, \bibinfo
  {author} {\bibfnamefont {S.}~\bibnamefont {Heule}}, \bibinfo {author}
  {\bibfnamefont {M.}~\bibnamefont {Horisberger}}, \bibinfo {author}
  {\bibfnamefont {M.}~\bibnamefont {Kasprzak}}, \bibinfo {author}
  {\bibfnamefont {K.}~\bibnamefont {Kirch}}, \bibinfo {author} {\bibfnamefont
  {A.}~\bibnamefont {Knecht}}, \bibinfo {author} {\bibfnamefont
  {M.}~\bibnamefont {Ku{\.z}niak}}, \bibinfo {author} {\bibfnamefont
  {B.}~\bibnamefont {Lauss}}, \bibinfo {author} {\bibfnamefont
  {A.}~\bibnamefont {Mtchedlishvili}}, \bibinfo {author} {\bibfnamefont
  {M.}~\bibnamefont {Meier}}, \bibinfo {author} {\bibfnamefont
  {G.}~\bibnamefont {Petzoldt}}, \bibinfo {author} {\bibfnamefont
  {C.}~\bibnamefont {Plonka-Spehr}}, \bibinfo {author} {\bibfnamefont
  {R.}~\bibnamefont {Schelldorfer}}, \bibinfo {author} {\bibfnamefont
  {U.}~\bibnamefont {Straumann}}, \ and\ \bibinfo {author} {\bibfnamefont
  {G.}~\bibnamefont {Zsigmond}},\ }\href {\doibase 10.1140/epja/i2010-10926-x}
  {\bibfield  {journal} {\bibinfo  {journal} {The European Physical Journal A}\
  }\textbf {\bibinfo {volume} {44}},\ \bibinfo {pages} {23} (\bibinfo {year}
  {2010})}\BibitemShut {NoStop}%
\bibitem [{\citenamefont {Pichlmaier}\ \emph {et~al.}(2010)\citenamefont
  {Pichlmaier}, \citenamefont {Varlamov}, \citenamefont {Schreckenbach},\ and\
  \citenamefont {Geltenbort}}]{Pichlmaier2010a}%
  \BibitemOpen
  \bibfield  {author} {\bibinfo {author} {\bibfnamefont {A.}~\bibnamefont
  {Pichlmaier}}, \bibinfo {author} {\bibfnamefont {V.}~\bibnamefont
  {Varlamov}}, \bibinfo {author} {\bibfnamefont {K.}~\bibnamefont
  {Schreckenbach}}, \ and\ \bibinfo {author} {\bibfnamefont {P.}~\bibnamefont
  {Geltenbort}},\ }\href {\doibase
  http://dx.doi.org/10.1016/j.physletb.2010.08.032} {\bibfield  {journal}
  {\bibinfo  {journal} {Physics Letters B}\ }\textbf {\bibinfo {volume}
  {693}},\ \bibinfo {pages} {221} (\bibinfo {year} {2010})}\BibitemShut
  {NoStop}%
\bibitem [{\citenamefont {Filges}\ and\ \citenamefont
  {Goldenbaum}(2009)}]{Filges2009}%
  \BibitemOpen
  \bibfield  {author} {\bibinfo {author} {\bibfnamefont {D.}~\bibnamefont
  {Filges}}\ and\ \bibinfo {author} {\bibfnamefont {F.}~\bibnamefont
  {Goldenbaum}},\ }\href@noop {} {\emph {\bibinfo {title} {Handbook of
  spallation research : theory, experiments and applications}}}\ (\bibinfo
  {publisher} {Weinheim, Wiley-VCH},\ \bibinfo {year} {2009})\BibitemShut
  {NoStop}%
\bibitem [{\citenamefont {Serebrov}\ \emph {et~al.}(2017)\citenamefont
  {Serebrov}, \citenamefont {Vasil'ev}, \citenamefont {Lasakov}, \citenamefont
  {Siber}, \citenamefont {Murashkin}, \citenamefont {Egorov}, \citenamefont
  {Fomin}, \citenamefont {Sbitnev}, \citenamefont {Geltenbort},\ and\
  \citenamefont {Zimmer}}]{Serebrov2017}%
  \BibitemOpen
  \bibfield  {author} {\bibinfo {author} {\bibfnamefont {A.~P.}\ \bibnamefont
  {Serebrov}}, \bibinfo {author} {\bibfnamefont {A.~V.}\ \bibnamefont
  {Vasil'ev}}, \bibinfo {author} {\bibfnamefont {M.~S.}\ \bibnamefont
  {Lasakov}}, \bibinfo {author} {\bibfnamefont {E.~V.}\ \bibnamefont {Siber}},
  \bibinfo {author} {\bibfnamefont {A.~N.}\ \bibnamefont {Murashkin}}, \bibinfo
  {author} {\bibfnamefont {A.~I.}\ \bibnamefont {Egorov}}, \bibinfo {author}
  {\bibfnamefont {A.~K.}\ \bibnamefont {Fomin}}, \bibinfo {author}
  {\bibfnamefont {S.~V.}\ \bibnamefont {Sbitnev}}, \bibinfo {author}
  {\bibfnamefont {P.}~\bibnamefont {Geltenbort}}, \ and\ \bibinfo {author}
  {\bibfnamefont {O.}~\bibnamefont {Zimmer}},\ }\href {\doibase
  10.1134/S1063784217010212} {\bibfield  {journal} {\bibinfo  {journal}
  {Technical Physics}\ }\textbf {\bibinfo {volume} {62}},\ \bibinfo {pages}
  {164} (\bibinfo {year} {2017})}\BibitemShut {NoStop}%
\bibitem [{\citenamefont {Golub}(1979)}]{Golub1979}%
  \BibitemOpen
  \bibfield  {author} {\bibinfo {author} {\bibfnamefont {R.}~\bibnamefont
  {Golub}},\ }\href@noop {} {\bibfield  {journal} {\bibinfo  {journal} {Physics
  Letters A}\ }\textbf {\bibinfo {volume} {72}},\ \bibinfo {pages} {387}
  (\bibinfo {year} {1979})}\BibitemShut {NoStop}%
\bibitem [{\citenamefont {Yoshiki}\ \emph {et~al.}(1992)\citenamefont
  {Yoshiki}, \citenamefont {Sakai}, \citenamefont {Ogura}, \citenamefont
  {Kawai}, \citenamefont {Masuda}, \citenamefont {Nakajima}, \citenamefont
  {Takayama}, \citenamefont {Tanaka},\ and\ \citenamefont
  {Yamaguchi}}]{Yoshiki1992}%
  \BibitemOpen
  \bibfield  {author} {\bibinfo {author} {\bibfnamefont {H.}~\bibnamefont
  {Yoshiki}}, \bibinfo {author} {\bibfnamefont {K.}~\bibnamefont {Sakai}},
  \bibinfo {author} {\bibfnamefont {M.}~\bibnamefont {Ogura}}, \bibinfo
  {author} {\bibfnamefont {T.}~\bibnamefont {Kawai}}, \bibinfo {author}
  {\bibfnamefont {Y.}~\bibnamefont {Masuda}}, \bibinfo {author} {\bibfnamefont
  {T.}~\bibnamefont {Nakajima}}, \bibinfo {author} {\bibfnamefont
  {T.}~\bibnamefont {Takayama}}, \bibinfo {author} {\bibfnamefont
  {S.}~\bibnamefont {Tanaka}}, \ and\ \bibinfo {author} {\bibfnamefont
  {A.}~\bibnamefont {Yamaguchi}},\ }\href@noop {} {\bibfield  {journal}
  {\bibinfo  {journal} {Physical Review Letters}\ }\textbf {\bibinfo {volume}
  {68}} (\bibinfo {year} {1992})}\BibitemShut {NoStop}%
\bibitem [{\citenamefont {Atchison}\ \emph {et~al.}(2009)\citenamefont
  {Atchison}, \citenamefont {Blau}, \citenamefont {Bollhalder}, \citenamefont
  {Daum}, \citenamefont {Fierlinger}, \citenamefont {Geltenbort}, \citenamefont
  {Hampel}, \citenamefont {Kasprzak}, \citenamefont {Kirch}, \citenamefont
  {K{\~A}¶chli}, \citenamefont {Kuczewski}, \citenamefont {Leber},
  \citenamefont {Locher}, \citenamefont {Meier}, \citenamefont {Ochse},
  \citenamefont {Pichlmaier}, \citenamefont {Plonka}, \citenamefont {Reiser},
  \citenamefont {Ulrich}, \citenamefont {Wang}, \citenamefont {Wiehl},
  \citenamefont {Zimmer},\ and\ \citenamefont {Zsigmond}}]{Atchison2009}%
  \BibitemOpen
  \bibfield  {author} {\bibinfo {author} {\bibfnamefont {F.}~\bibnamefont
  {Atchison}}, \bibinfo {author} {\bibfnamefont {B.}~\bibnamefont {Blau}},
  \bibinfo {author} {\bibfnamefont {A.}~\bibnamefont {Bollhalder}}, \bibinfo
  {author} {\bibfnamefont {M.}~\bibnamefont {Daum}}, \bibinfo {author}
  {\bibfnamefont {P.}~\bibnamefont {Fierlinger}}, \bibinfo {author}
  {\bibfnamefont {P.}~\bibnamefont {Geltenbort}}, \bibinfo {author}
  {\bibfnamefont {G.}~\bibnamefont {Hampel}}, \bibinfo {author} {\bibfnamefont
  {M.}~\bibnamefont {Kasprzak}}, \bibinfo {author} {\bibfnamefont
  {K.}~\bibnamefont {Kirch}}, \bibinfo {author} {\bibfnamefont
  {S.}~\bibnamefont {K{\~A}¶chli}}, \bibinfo {author} {\bibfnamefont
  {B.}~\bibnamefont {Kuczewski}}, \bibinfo {author} {\bibfnamefont
  {H.}~\bibnamefont {Leber}}, \bibinfo {author} {\bibfnamefont
  {M.}~\bibnamefont {Locher}}, \bibinfo {author} {\bibfnamefont
  {M.}~\bibnamefont {Meier}}, \bibinfo {author} {\bibfnamefont
  {S.}~\bibnamefont {Ochse}}, \bibinfo {author} {\bibfnamefont
  {A.}~\bibnamefont {Pichlmaier}}, \bibinfo {author} {\bibfnamefont
  {C.}~\bibnamefont {Plonka}}, \bibinfo {author} {\bibfnamefont
  {R.}~\bibnamefont {Reiser}}, \bibinfo {author} {\bibfnamefont
  {J.}~\bibnamefont {Ulrich}}, \bibinfo {author} {\bibfnamefont
  {X.}~\bibnamefont {Wang}}, \bibinfo {author} {\bibfnamefont {N.}~\bibnamefont
  {Wiehl}}, \bibinfo {author} {\bibfnamefont {O.}~\bibnamefont {Zimmer}}, \
  and\ \bibinfo {author} {\bibfnamefont {G.}~\bibnamefont {Zsigmond}},\ }\href
  {\doibase https://doi.org/10.1016/j.nima.2009.06.047} {\bibfield  {journal}
  {\bibinfo  {journal} {Nuclear Instruments and Methods in Physics Research
  Section A}\ }\textbf {\bibinfo {volume} {608}},\ \bibinfo {pages} {144 }
  (\bibinfo {year} {2009})}\BibitemShut {NoStop}%
\bibitem [{\citenamefont {Holley}(2012)}]{Holley2012}%
  \BibitemOpen
  \bibfield  {author} {\bibinfo {author} {\bibfnamefont {A.~T.}\ \bibnamefont
  {Holley}},\ }\emph {\bibinfo {title} {Ultracold neutron polarimetry in a
  measurement of the $beta$ asymmetry}},\ \href
  {http://www.lib.ncsu.edu/resolver/1840.16/7583} {Ph.D. thesis},\ \bibinfo
  {school} {North Carolina State University} (\bibinfo {year}
  {2012})\BibitemShut {NoStop}%
\bibitem [{\citenamefont {Golub}, \citenamefont {Richardson},\ and\
  \citenamefont {Lamoreaux}(1991)}]{Golub1991}%
  \BibitemOpen
  \bibfield  {author} {\bibinfo {author} {\bibfnamefont {R.}~\bibnamefont
  {Golub}}, \bibinfo {author} {\bibfnamefont {D.}~\bibnamefont {Richardson}}, \
  and\ \bibinfo {author} {\bibfnamefont {S.~K.}\ \bibnamefont {Lamoreaux}},\
  }\href@noop {} {\emph {\bibinfo {title} {Ultra-Cold Neutrons}}}\ (\bibinfo
  {publisher} {IOP Publishing Ltd.},\ \bibinfo {year} {1991})\BibitemShut
  {NoStop}%
\bibitem [{\citenamefont {Ignatovich}(1990)}]{Ignatovich1990}%
  \BibitemOpen
  \bibfield  {author} {\bibinfo {author} {\bibfnamefont {V.}~\bibnamefont
  {Ignatovich}},\ }\href@noop {} {\emph {\bibinfo {title} {The Physics of
  Ultracold Neutrons}}},\ Oxford science publications\ (\bibinfo  {publisher}
  {Clarendon Press},\ \bibinfo {year} {1990})\BibitemShut {NoStop}%
\bibitem [{\citenamefont {Golub}\ and\ \citenamefont
  {B{\"o}ning}(1983)}]{Golub1983}%
  \BibitemOpen
  \bibfield  {author} {\bibinfo {author} {\bibfnamefont {R.}~\bibnamefont
  {Golub}}\ and\ \bibinfo {author} {\bibfnamefont {K.}~\bibnamefont
  {B{\"o}ning}},\ }\href {\doibase 10.1007/BF01308763} {\bibfield  {journal}
  {\bibinfo  {journal} {Zeitschrift f{\"u}r Physik B Condensed Matter}\
  }\textbf {\bibinfo {volume} {51}},\ \bibinfo {pages} {95} (\bibinfo {year}
  {1983})}\BibitemShut {NoStop}%
\bibitem [{\citenamefont {Muhrer}\ \emph {et~al.}(2000)\citenamefont {Muhrer},
  \citenamefont {Ferguson}, \citenamefont {Russell},\ and\ \citenamefont
  {Pitcher}}]{Muhrer2000}%
  \BibitemOpen
  \bibfield  {author} {\bibinfo {author} {\bibfnamefont {G.}~\bibnamefont
  {Muhrer}}, \bibinfo {author} {\bibfnamefont {P.}~\bibnamefont {Ferguson}},
  \bibinfo {author} {\bibfnamefont {G.}~\bibnamefont {Russell}}, \ and\
  \bibinfo {author} {\bibfnamefont {E.}~\bibnamefont {Pitcher}},\ }\href@noop
  {} {\bibfield  {journal} {\bibinfo  {journal} {Fourth International Topical
  Meeting on Nuclear Applications of Accelerator Technology (Washington, D.C.,
  Nov 12- 16)}\ ,\ \bibinfo {pages} {120}} (\bibinfo {year}
  {2000})}\BibitemShut {NoStop}%
\end{thebibliography}%

\end{document}